\documentclass[useAMS,usenatbib,a4paper]{mn2e}
\usepackage{amsmath}
\usepackage{amssymb}
\usepackage{amsfonts}
\usepackage {multirow}
\usepackage {graphicx}
\usepackage {natbib}
\usepackage{txfonts}
\usepackage{graphicx,amsmath}
\usepackage[hyperindex,breaklinks=true, colorlinks, citecolor=blue]{hyperref}  % for active links 
\usepackage{enumerate}
\usepackage{tablefootnote}
\usepackage{eurosym}
\usepackage[usenames,dvipsnames]{xcolor}

\def\setsymbol#1#2{\expandafter\def\csname #1\endcsname{#2}}
\def\getsymbol#1{\csname #1\endcsname}

\def\Planck{\textit{Planck}}

\newbox\tablebox    \newdimen\tablewidth
\def\leaderfil{\leaders\hbox to 5pt{\hss.\hss}\hfil}
\def\tablenote#1 #2\par{\begingroup \parindent=0.8em
    \abovedisplayshortskip=0pt\belowdisplayshortskip=0pt
    \noindent
    $$\hss\vbox{\hsize\tablewidth \hangindent=\parindent \hangafter=1 \noindent
    \hbox to \parindent{$^#1$\hss}\strut#2\strut\par}\hss$$
    \endgroup}

\def\L2{\ifmmode L_2\else $L_2$\fi}

\def\DeltaT{\ifmmode \Delta T\else $\Delta T$\fi}
\def\deltat{\ifmmode \Delta t\else $\Delta t$\fi}
\def\fknee{\ifmmode f_{\rm knee}\else $f_{\rm knee}$\fi}
\def\Fmax{\ifmmode F_{\rm max}\else $F_{\rm max}$\fi}
\def\solar{\ifmmode{\rm M}_{\mathord\odot}\else${\rm M}_{\mathord\odot}$\fi}
\def\Msolar{\ifmmode{\rm M}_{\mathord\odot}\else${\rm M}_{\mathord\odot}$\fi}
\def\Lsolar{\ifmmode{\rm L}_{\mathord\odot}\else${\rm L}_{\mathord\odot}$\fi}
\def\inv{\ifmmode^{-1}\else$^{-1}$\fi}
\def\mo{\ifmmode^{-1}\else$^{-1}$\fi}
\def\sup#1{\ifmmode ^{\rm #1}\else $^{\rm #1}$\fi}
\def\expo#1{\ifmmode \times 10^{#1}\else $\times 10^{#1}$\fi}
\def\,{\thinspace}
\def\lsim{\mathrel{\raise .4ex\hbox{\rlap{$<$}\lower 1.2ex\hbox{$\sim$}}}}
\def\gsim{\mathrel{\raise .4ex\hbox{\rlap{$>$}\lower 1.2ex\hbox{$\sim$}}}}

\def\simprop{\mathrel{\raise .4ex\hbox{\rlap{$\propto$}\lower 1.2ex\hbox{$\sim$}}}}
\def\deg{\ifmmode^\circ\else$^\circ$\fi}
\def\pdeg{\ifmmode $\setbox0=\hbox{$^{\circ}$}\rlap{\hskip.11\wd0 .}$^{\circ}
          \else \setbox0=\hbox{$^{\circ}$}\rlap{\hskip.11\wd0 .}$^{\circ}$\fi}
\def\arcs{\ifmmode {^{\scriptstyle\prime\prime}}
          \else $^{\scriptstyle\prime\prime}$\fi}
\def\arcm{\ifmmode {^{\scriptstyle\prime}}
          \else $^{\scriptstyle\prime}$\fi}
\newdimen\sa  \newdimen\sb
\def\parcs{\sa=.07em \sb=.03em
     \ifmmode \hbox{\rlap{.}}^{\scriptstyle\prime\kern -\sb\prime}\hbox{\kern -\sa}
     \else \rlap{.}$^{\scriptstyle\prime\kern -\sb\prime}$\kern -\sa\fi}
\def\parcm{\sa=.08em \sb=.03em
     \ifmmode \hbox{\rlap{.}\kern\sa}^{\scriptstyle\prime}\hbox{\kern-\sb}
     \else \rlap{.}\kern\sa$^{\scriptstyle\prime}$\kern-\sb\fi}
\def\ra[#1 #2 #3.#4]{#1\sup{h}#2\sup{m}#3\sup{s}\llap.#4}
\def\dec[#1 #2 #3.#4]{#1\deg#2\arcm#3\arcs\llap.#4}
\def\deco[#1 #2 #3]{#1\deg#2\arcm#3\arcs}
\def\rra[#1 #2]{#1\sup{h}#2\sup{m}}

\def\dots{\relax\ifmmode \ldots\else $\ldots$\fi}
\def\WHzsr{\ifmmode $W\,Hz\mo\,sr\mo$\else W\,Hz\mo\,sr\mo\fi}
\def\mHz{\ifmmode $\,mHz$\else \,mHz\fi}
\def\GHz{\ifmmode $\,GHz$\else \,GHz\fi}
\def\mKs{\ifmmode $\,mK\,s$^{1/2}\else \,mK\,s$^{1/2}$\fi}
\def\muKs{\ifmmode \,\mu$K\,s$^{1/2}\else \,$\mu$K\,s$^{1/2}$\fi}
\def\muKRJs{\ifmmode \,\mu$K$_{\rm RJ}$\,s$^{1/2}\else \,$\mu$K$_{\rm RJ}$\,s$^{1/2}$\fi}
\def\muKHz{\ifmmode \,\mu$K\,Hz$^{-1/2}\else \,$\mu$K\,Hz$^{-1/2}$\fi}
\def\MJysr{\ifmmode \,$MJy\,sr\mo$\else \,MJy\,sr\mo\fi}
\def\MJysrmK{\ifmmode \,$MJy\,sr\mo$\,mK$_{\rm CMB}\mo\else \,MJy\,sr\mo\,mK$_{\rm CMB}\mo$\fi}
\def\microns{\ifmmode \,\mu$m$\else \,$\mu$m\fi}

\def\muK{\ifmmode \,\mu$K$\else \,$\mu$\hbox{K}\fi}
\def\microK{\ifmmode \,\mu$K$\else \,$\mu$\hbox{K}\fi}
\def\muW{\ifmmode \,\mu$W$\else \,$\mu$\hbox{W}\fi}
\def\kms{\ifmmode $\,km\,s$^{-1}\else \,km\,s$^{-1}$\fi}
\def\kmsMpc{\ifmmode $\,\kms\,Mpc\mo$\else \,\kms\,Mpc\mo\fi}
\providecommand{\sorthelp}[1]{}

\newcommand{\tC}{\boldsymbol{\rm C}}

\def\reff@jnl#1{{\rm#1\/}}

\def\aj{\reff@jnl{AJ}}                  % Astronomical Journal
\def\araa{\reff@jnl{ARA\&A}}            % Annual Review of Astron and Astrophys
\def\apj{\reff@jnl{ApJ}}                % Astrophysical Journal
\def\apjl{\reff@jnl{ApJ}}               % Astrophysical Journal, Letters
\def\apjs{\reff@jnl{ApJS}}              % Astrophysical Journal, Supplement
\def\ao{\reff@jnl{Appl.Optics}}         % Applied Optics
\def\apss{\reff@jnl{Ap\&SS}}            % Astrophysics and Space Science
\def\aap{\reff@jnl{A\&A}}               % Astronomy and Astrophysics
\def\aapr{\reff@jnl{A\&A~Rev.}}         % Astronomy and Astrophysics Reviews
\def\aaps{\reff@jnl{A\&AS}}             % Astronomy and Astrophysics, Supplement
\def\azh{\reff@jnl{AZh}}                        % Astronomicheskii Zhurnal
\def\baas{\reff@jnl{BAAS}}              % Bulletin of the AAS
\def\jcap{\reff@jnl{JCAP}}              % Journal of Cosmology and Astroparticle Physics
\def\jrasc{\reff@jnl{JRASC}}            % Journal of the RAS of Canada
\def\memras{\reff@jnl{MmRAS}}           % Memoirs of the RAS
\def\mnras{\reff@jnl{MNRAS}}            % Monthly Notices of the RAS
\def\pra{\reff@jnl{Phys.Rev.A}}         % Physical Review A: General Physics
\def\prb{\reff@jnl{Phys.Rev.B}}         % Physical Review B: Solid State
\def\prc{\reff@jnl{Phys.Rev.C}}         % Physical Review C
\def\prd{\reff@jnl{Phys.Rev.D}}         % Physical Review D
\def\prl{\reff@jnl{Phys.Rev.Lett}}      % Physical Review Letters
\def\pasp{\reff@jnl{PASP}}              % Publications of the ASP
\def\pasj{\reff@jnl{PASJ}}              % Publications of the ASJ
\def\qjras{\reff@jnl{QJRAS}}            % Quarterly Journal of the RAS
\def\skytel{\reff@jnl{S\&T}}            % Sky and Telescope
\def\solphys{\reff@jnl{Solar~Phys.}}    % Solar Physics
\def\sovast{\reff@jnl{Soviet~Ast.}}     % Soviet Astronomy
 \def\ssr{\reff@jnl{Space~Sci.Rev.}}    % Space Science Reviews
\def\zap{\reff@jnl{ZAp}}                % Zeitschrift fuer Astrophysik
\def\nat{\reff@jnl{Nature}}             % Nature 
\def\procspie{\reff@jnl{Proceedings of the SPIE}}             % Proc. of SPIE

\def\Planck{\textit{Planck}}
\def\core{\textit{CORE}}
\def\litebird{\textit{LiteBIRD}}
\def\pixie{\textit{PIXIE}}
\def\pico{\textit{PICO}}

\def\ben{\begin{enumerate}}
\def\een{\end{enumerate}}
\def\bi{\begin{itemize}}
\def\ei{\end{itemize}}
\def\be{\begin{equation}}
\def\ee{\end{equation}}
\def\bea{\begin{eqnarray}}
\def\eea{\end{eqnarray}}
\def\ba{\begin{align}}
\def\ea{\end{align}}

\def\bdw{\boldsymbol{w }}
\def\bda{\boldsymbol{a }}

\def\bdn{\boldsymbol{n }}

\def\bdx{\boldsymbol{x }}
\def\bdw{\boldsymbol{w }}

\newcommand{\healpix}{{\tt HEALPix}}

\makeatletter
\newcommand\footnoteref[1]{\protected@xdef\@thefnmark{\ref{#1}}\@footnotemark}
\makeatother

\title[Extracting $\mu$-anisotropies]
{Extracting foreground-obscured $\mu$-distortion anisotropies to constrain primordial non-Gaussianity}
\author[M.\,Remazeilles and J.\,Chluba]{M.\,Remazeilles\thanks{E-mail:~\url{mathieu.remazeilles@manchester.ac.uk}} and J.\,Chluba\thanks{E-mail:~\url{jens.chluba@manchester.ac.uk}} 
\\
Jodrell Bank Centre for Astrophysics, Alan Turing Building, School of Physics and Astronomy, The University of Manchester, \\
Oxford Road, Manchester, M13 9PL, U.K.\\
}

\begin{document}

\date{\vspace{-0mm}{Accepted 2018 --. Received 2018 February 26}}

\maketitle

%%%%%%%%%%%%%%%%%%%%%%%%%%%%%%%%%%%%%%%%%%%%%%%%%%%%
%%%
%%% Abstract
%%%
\begin{abstract}
Correlations between cosmic microwave background (CMB) temperature, polarization and spectral distortion anisotropies can be used as a probe of primordial non-Gaussianity. Here, we perform a reconstruction of $\mu$-distortion anisotropies in the presence of Galactic and extragalactic foregrounds, applying the so-called {\it Constrained ILC} component separation method to simulations of proposed CMB space missions (\textit{PIXIE}, \textit{LiteBIRD}, \textit{CORE}, \textit{PICO}). Our sky simulations include Galactic dust, Galactic synchrotron, Galactic free-free, thermal Sunyaev-Zeldovich effect, as well as primary CMB temperature and $\mu$-distortion anisotropies, the latter being added as correlated field. The Constrained ILC method allows us to null the CMB temperature anisotropies in the reconstructed $\mu$-map (and vice versa), in addition to mitigating the contaminations from astrophysical foregrounds and instrumental noise. We compute the cross-power spectrum between the reconstructed (CMB-free) $\mu$-distortion map and the ($\mu$-free) CMB temperature map, after foreground removal and component separation. Since the cross-power spectrum is proportional to the primordial non-Gaussianity parameter, $f_{\rm NL}$, on scales $k\simeq 740\,\rm Mpc^{-1}$, this allows us to derive $f_{\rm NL}$-detection limits for the aforementioned future CMB experiments. Our analysis shows that foregrounds degrade the theoretical detection limits (based mostly on instrumental noise) by more than one order of magnitude, with {\it PICO} standing the best chance at placing upper limits on scale-dependent non-Gaussianity.
We also discuss the dependence of the constraints on the channel sensitivities and chosen bands. Like for $B$-mode polarization measurements, extended coverage at frequencies $\nu\lesssim 40\,{\rm GHz}$ and $\nu\gtrsim 400\,{\rm GHz}$ provides more leverage than increased channel sensitivity.
\end{abstract}


%%%%%%%%%%%%%%%%%%%%%%%%%%%%%%%%%%%%%%%%%%%%%%%%%%%%
\section{Introduction}
\label{sec:intro}
%%%%%%%%%%%%%%%%%%%%%%%%%%%%%%%%%%%%%%%%%%%%%%%%%%%%%%
Interactions between cosmic microwave background (CMB) photons and matter generate temperature and polarization anisotropies \citep{Sunyaev1970, Peebles1970} as well as spectral distortions of the CMB radiation \citep{Zeldovich1969, Sunyaev1970mu}. Precisely characterizing the CMB thus enables us to probe different stages in the evolution of the Universe, thereby learning about the main cosmological parameters and the nature of dark matter and dark energy \citep[e.g.,][]{Spergel2003, Planck2015params}. 

In the late Universe ($z < 10-100$), large-scale structures modify the primordial CMB light by gravitational effects and Thomson scattering, while the hot gas inside galaxy clusters distorts the CMB radiation through the Sunyaev-Zeldovich (SZ) effect \citep{Zeldovich1969, Carlstrom2002}, generating $y$-distortion anisotropies over a wide range of scales. 
A similar type of spectral distortion can also be created in the pre-recombination era ($z>10^3$) when no structures were present yet, but most excitingly, in the early Universe ($z > 10^4$), other types of spectral distortions can be formed. One prominent example is the $\mu$-distortion resulting from energy injection to the CMB \citep[e.g.,][]{Sunyaev1970mu, Burigana1991, Hu1993, Chluba2012}, e.g., by annihilating or decaying particles \citep{Sarkar1984, McDonald2001, Hu1993b, Chluba2013fore}, primordial black holes \citep{Carr2010, Yacine2017, Poulin2017} or the dissipation of small-scale acoustic modes \citep{Sunyaev1970diss, Daly1991, Hu1994}. Other types of distortions, with more rich spectral structure, can be formed through photon injection \citep{Chluba2015GreensII} or in the partially Comptonized regime \citep{Chluba2012, Khatri2012mix, Chluba2013Green}. Although so far undiscovered, $\mu$-distortions will open a new window to the early Universe, potentially shedding light on the nature of dark matter, particle and inflation physics, all from behind the last-scattering surface well into the pre-recombination-era at $z > 10^3$ \citep[e.g.,][]{Chluba2012, Sunyaev2013, deZotti2015, Chluba2016}. 

Aside from {\it average} CMB distortions, primordial distortion fluctuations (sometimes referred to as spectral-spatial anisotropies) can be created through anisotropic heating mechanisms, as discussed by \citet{Chluba2012_2x2}. Aside from late time effects (e.g., related to SZ clusters), in $\Lambda$CDM these anisotropies are usually expected to be small ($\Delta \mu/\mu\lesssim 10^{-5}-10^{-4}$).  Exotic mechanisms could in principle be connected to anisotropic dark matter annihilation during structure formation (density square enhancement), dissipation of large-scale primordial magnetic fields, energy injection by decaying particles in cosmologies with iso-curvature perturbations at small scales or heating by primordial black hole clusters, to name a few possibilities that come to mind.

One mechanism capable of creating sizable primordial distortion anisotropies is through the damping of primordial small-scale acoustic modes in the presence of (enhanced) primordial non-Gaussianity in the ultra-squeezed limit, as first discussed by \citet{Pajer2012}. There it was demonstrated that by measuring intrinsic spatial correlations between $\mu$-distortion and CMB temperature anisotropies we can place new constraints on the primordial non-Gaussianity at small scales with wavenumbers corresponding to ${k\simeq 10^3}-{10^4\,{\rm Mpc^{-1}}}$. 

At face value, the expected theoretical limits are fairly weak, corresponding to $|f_{\rm NL}|\lesssim \mathcal{O}(10^3)$ on the (local-type) non-Gaussianity parameter for {\rm PIXIE}-like experiments  \citep{Pajer2012, Ganc2012}. However, as stressed again later \citep{Pajer2013, Biagetti2013, Emami2015}, these apply to currently unconstrained scales, thus allowing to place novel bounds on the {\it scale-dependence} of non-Gaussianity, for which theoretical motivation can be given \citep[e.g.,][]{Dimastrogiovanni2016}. One of the most important aspects here is the huge lever arm between typical CMB scales, $k\simeq 10^{-2}\,{\rm Mpc^{-1}}$, and those relevant to the creation of the $\mu$-distortion anisotropies, $k\simeq 740\,{\rm Mpc^{-1}}$, which in principle can lead to greatly enhanced distortion anisotropies while being consistent with CMB temperature anisotropy limits $|f_{\rm NL}|\lesssim 5$ \citep{Planck2013ng}. In addition, correlations between temperature and $y$-distortion anisotropies can be created from modes dissipating at scales $k\simeq 10\,{\rm Mpc^{-1}}$ \citep{Emami2015}. Similarly, correlations of distortion signals with polarization anisotropies can be used to further disentangle different contributions \citep{Ota2016, Ravenni2017}. Higher order statistics can also be affected \citep{Bartolo2016, Shiraishi2016}. Finally, the overall level of the anisotropic distortion signal also depends on the average value of the dissipation-induced distortion \citep{Chluba_mu_2017}, which in principle can be strongly enhanced in non-standard early-Universe models \citep[e.g.,][]{Chluba2012inflaton}.

Here, we wish to address the question about the detectability of these distortion signals in the presence of real world limitations caused by foregrounds. For the average CMB distortion signals it was recently shown that foregrounds indeed pose a serious challenge, with one of the biggest limitations caused by lack of low-frequency coverage \citep{abitbol_pixie}. 
Also, like for the detection of primordial $B$-mode polarization signals, one always is faced with questions about the robustness of the analysis with respect to biases \citep{Remazeilles2016, Remazeilles2017}.

To obtain unbiased constraints on the non-Gaussianity parameter, $f_{\rm NL}$, from the $T$-$\mu$ cross-angular power spectrum between CMB temperature anisotropies and $\mu$-distortion anisotropies, it is essential to cancel out residual CMB temperature anisotropies in the reconstructed map of $\mu$-distortion anisotropies. This can be robustly achieved by using the {\it Constrained ILC} component separation method \citep{Remazeilles2011}. Utilizing the known spectral signature of $\mu$-distortion and CMB temperature anisotropies, we apply the Constrained ILC component separation method to map the $\mu$-distortion anisotropies while nulling the CMB temperature contamination, and vice versa. 
With this approach, we assess the performance of the proposed CMB space missions \textit{PIXIE} \citep{Pixie2016}, \textit{LiteBIRD} \citep{Litebird2016}, \textit{CORE} \citep{Core2016}, and \textit{PICO} (Shaul Hanany, priv. comm.) in this context. We also discuss the optimization of future CMB experiments in terms of sensitivity and frequency bands to allow a detection of anisotropic $\mu$-distortions for average ${\langle\mu\rangle=2\times 10^{-8}}$ and $|f_{\rm NL}(k\simeq 740\,{\rm Mpc^{-1}})| \lesssim 4500$. For $f_{\rm NL}(k_0)\simeq 5$ at pivot scale, $k_0= 0.05\,{\rm Mpc^{-1}}$, this would impose a limit of\footnote{A small additional enhancement of the average distortion caused by scale-dependent non-Gaussianity \citep{Chluba_mu_2017} was taken into account for this estimate.} $n_{\rm NL}\lesssim 1.6$ on the spectral index of $f_{\rm NL}(k)\simeq f_{\rm NL}(k_0)(k/k_0)^{n_{\rm NL}-1}$ for scale-dependent non-Gaussianity, providing a new way to constrain non-standard early-universe model (e.g., multi-field inflation). 

We do not consider higher order statistics of the $\mu$-distortion field here, but the method can be easily extended. This might become relevant if indeed large values of $f_{\rm NL}(k)>10^4$ are indicated by future data, but a more careful assessment is beyond the scope of this paper. We also do not include any effects from the {\it residual} ($r$-type) distortion signal created by heating at $10^4\lesssim z\lesssim 2 \times 10^5$ \citep[e.g.,][]{Chluba2013PCA}. Just like for primordial $\mu$ and $y$, this would lead to $r-T$ correlations dominated by acoustic damping of perturbations with $k\simeq 50\,{\rm Mpc}^{-1}$, but at a level that is about one order of magnitude smaller than the primordial $\mu$ or $y$ distortion signals. In refined analysis, this signal could cause another contamination that should to be considered.

The paper is organized as follows. In Sect.~\ref{sec:simu} we describe our simulations of $\mu$-distortions and foregrounds for different CMB experiments. In Sect.~\ref{sec:comp_sep}, we first discuss the details of the component separation method employed for the analysis (Sect.~\ref{subsec:method}). We then present our results on the reconstruction of the $\mu$-$T$ correlation signal and detection limits on $f_{\rm NL}$ after foreground removal for the different CMB satellite concepts (Sect.~\ref{subsec:results}). In Sect.~\ref{sec:discussion}, we discuss some important issues to be addressed in order to optimize for the detection of $\mu$-distortions. We conclude in Sect.~\ref{sec:conc}.

\vspace{3mm}
\section{Simulations}
\label{sec:simu}
%%%%%%%%%%%%%%%%%%%%%%%%%%%%%%%%%%%%%%%%%%%%%%%%%%%%%%
In this section we outline the various ingredients of our simulations, starting with the distortion anisotropies (Sect.~\ref{sec:sim_mu}), then discussing foregrounds (Sect.~\ref{sec:sim_fg}) and closing with the various mission specifications (Sect.~\ref{sec:exp}).

\vspace{3mm}
\subsection{Correlated $\boldsymbol{\mu}$ distortion and CMB temperature anisotropies}
\label{sec:sim_mu}
%%%%%%%%%%%%%%%%%%%%%%%%%%%%%%%%%%%%%%%%%%%%%%%%%%%%%%
To simulate the CMB $\mu$-distortion anisotropies, we need to provide a description of its auto- and cross-power spectra. We mainly use the compact expressions given by \citet{Emami2015} and \citet{Chluba_mu_2017}, which link the various power spectra to the $f_{\rm NL}$ parameter and average $\mu$-distortion, $\left<\mu\right>$, in a simple manner. 
To capture the $\ell$-dependence of the temperature and $\mu$-distortion cross-power spectrum, $C_\ell^{\,\mu\times T}$, corrections beyond the Sachs-Wolfe limit originally used by \citet{Pajer2012} have to be included. These were first considered by \cite{Ganc2012} and at large angular scales ($2 \lesssim \ell \lesssim 200$) can be represented using analytical expressions \citep{Chluba_mu_2017}. 
A more accurate computation of the $\mu$-$T$ cross-power spectrum (extending to small scales) was recently carried out by \cite{Ravenni2017}. These results were published while we were completing this work, so that most of the results derived here are based on the modelling of \citet{Chluba_mu_2017}. However, both models have very similar amplitude and scale-dependence at large angular scales (${\ell \lesssim 200}$) probed in our analysis, so that our results on foreground removal and signal reconstruction are consistent for both models (see Sect.~\ref{subsec:ravenni}). 

Depending on the model for the $\mu$-$T$ cross-power spectrum, we simulate full-sky maps of $\mu$-distortion anisotropies and CMB temperature anisotropies as correlated fields using the following prescription: the auto-power spectrum of CMB temperature anisotropies in the Sachs-Wolfe (SW) limit is given by
%----------------------
\begin{align}
C_\ell^{\,TT,\,{\rm SW}} \simeq {2\pi\over 25}{T^2_{\rm CMB}\,\Delta_0^2\over \ell(\ell+1)}
\end{align}
%----------------------
in thermodynamic temperature units, where the CMB blackbody temperature is $T_{\rm CMB} = 2.7255$\,K and the amplitude of the curvature perturbation power spectrum is $\Delta_0^2 = 2.4\times 10^{-9}$ under the assumption of scale-invariance ($n_{\rm S}=1$).

In the description of \cite{Chluba_mu_2017}, the cross-power spectrum between $\mu$-distortion anisotropies and CMB temperature anisotropies on angular scales ${2 \leq \ell \lesssim 200}$ is then given by:
%----------------------
\begin{align}
\label{eq:theory}
C_\ell^{\,\mu \times T} = 12\ C_\ell^{\,TT,\,{\rm SW}}\,\rho(\ell)\,f_{\rm NL}\,\langle\mu\rangle,
\end{align}
%----------------------
where the monopole of the $\mu$-distortion is set to ${\langle\mu\rangle = 2\times 10^{-8}}$ \citep{Chluba2016} and $f_{\rm NL}$ is the primordial non-Gaussianity parameter. The scale-dependence of the correlation is approximated by:
%----------------------
\begin{align}
\rho(\ell) &= 1.08\bigl(1-0.022\ell-1.72\times10^{-4}\ell^2\cr
              &\qquad+2.00\times10^{-6}\ell^3-4.56\times 10^{-9}\ell^4\bigr).
\end{align}
%----------------------
We adopt the opposite sign convention to that of \emph{WMAP} \citep{Komatsu2001ng}: $f_{\rm NL}=-f_{\rm NL}^{\rm WMAP}$. Thus, for positive $f_{\rm NL}$, $\mu$ and $T$ are correlated at the largest angular scales. In Sect.~\ref{subsec:fnl_neg}, we show that this is not a limitation, as the results directly apply to negative values of $f_{\rm NL}$ too, making this choice unimportant.

Equation~\eqref{eq:theory} clearly illustrates the dependence of the cross-correlation on $f_{\rm NL}$ and $\langle\mu\rangle$, showing that the obtained limits from anisotropy measurements alone only constrain the product $f_{\rm NL}\,\langle\mu\rangle$ \citep{Chluba_mu_2017}. To break the degeneracy, an absolute measurement using a {\it PIXIE}-type experiment is required. Thus, a combination of imager (providing angular resolution) and spectrometer (providing spectral coverage) might be one viable avenue forward towards clear detections and constraints.

%%%%%%%%%%%%%%%%%%%%%%%%%%%%%%%%%%%%%%%%%%%%%%%%%%%%%%
\begin{figure}
  \begin{center}
    \includegraphics[width=0.99\columnwidth]{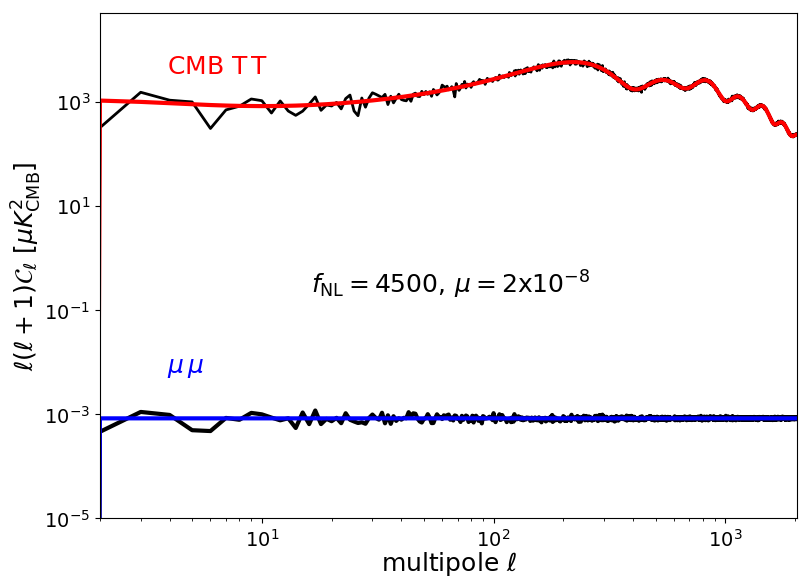}
    \\[3mm]
    \includegraphics[width=0.99\columnwidth]{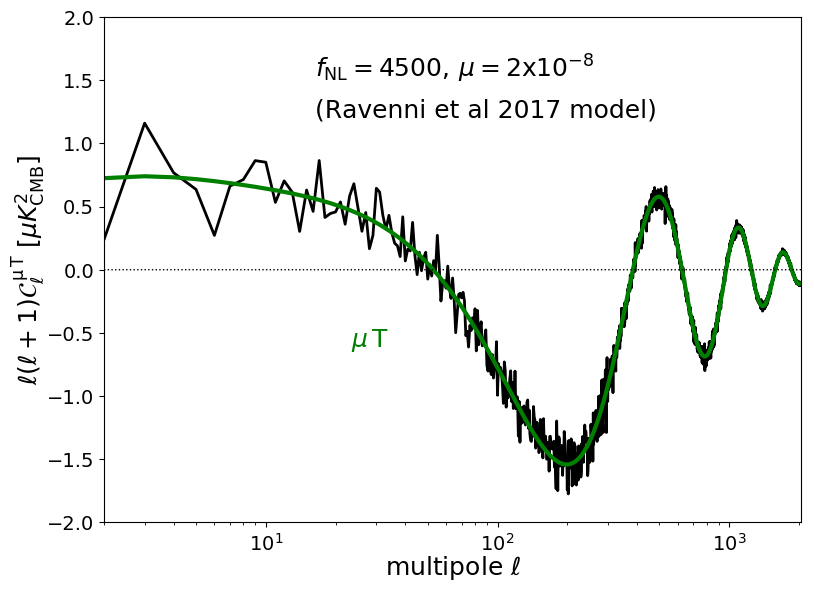}
  \end{center}
\caption{Angular power spectra of anisotropies from theory (red, blue, and green lines) and simulations (black lines), for average $\langle\mu\rangle=2\times 10^{-8}$ and $f_{\rm NL}=4500$: CMB $TT$ and $\mu\mu$ distortion auto-power spectra (\emph{upper panel}), and $\mu\times T$ cross-power spectrum (\emph{lower panel}). The upper panel illustrates the large dynamic range between CMB temperature anisotropies (red line) and $\mu$-distortion anisotropies (blue line).}
\label{Fig:cls}
\end{figure}
%%%%%%%%%%%%%%%%%%%%%%%%%%%%%%%%%%%%%%%%%%%%%%%%%%%%%%

In the more complete computation of \cite{Ravenni2017}, the $\mu$-$T$ cross-power spectrum at all angular scales is given by:
%----------------------
\begin{align}
\label{eq:ravenni}
C_\ell^{\,\mu \times T} &= 4\pi\,\left({12\over 5}\right)\,\int\,{k^2\,dk\over 2\pi^2}\,\mathcal{T}^{T}_{\ell}\left(k\right)\,j_\ell\left(k\,r_{\rm ls}\right)\,P(k)\,\cr
                  &\qquad \times\int\,{q_1^2\,dq_1\over 2\pi^2}\,f^{\mu}(q_1,q_1,k)\,P(q_1),
\end{align}
%----------------------
where $\mathcal{T}^{T}_{\ell}(k)$ denotes the radiation transfer function for CMB temperature, $f^{\mu}(q_1,q_1,k)$ is the transfer function for the $\mu$-distortion, $j_\ell$ is the spherical Bessel function of the first kind, $r_{\rm ls}$ is the comoving distance to the last-scattering surface, and $P(k)$ is the primordial power spectrum. The resultant cross power spectrum is illustrated in Fig.~\ref{Fig:cls}, exhibiting several oscillations towards small angular scales. However, the added signal-to-noise is limited at $\ell\gtrsim 200$ even for {\it PICO} (see Sect.~\ref{subsec:ravenni}), so that the approximation Eq.~\eqref{eq:theory} is sufficient for our main estimates.

The $\mu$-distortion auto-power spectrum in the Gaussian limit is negligible in comparison to the noise-level of future CMB imagers \citep{Pajer2012, Ganc2012}. For the non-Gaussian contribution we use \citep{Emami2015}
%----------------------
\begin{align}
C_\ell^{\,\mu \times \mu} = 144\,C_\ell^{\,TT,\,\rm SW}\,f_{\rm NL}^2\,\langle\mu\rangle^2,
\end{align}
%----------------------
which is in good agreement with the original estimates \citep{Pajer2012}. For illustration, we adopt different values of primordial non-Gaussianity in our simulations: 
%----------------------
\begin{align}
f_{\rm NL} = 4.5\times 10^3\,; 10^4\;; 10^5.
\end{align}
%----------------------
The first value is close to the estimated detection limit of a {\it PIXIE}-like experiment without considering foregrounds in more detail \citep{Chluba_mu_2017}. The last value, putting us into the large signal-to-noise regime, is mainly chosen to validate the method as it reaches the limits of perturbation theory.

\vspace{5mm}
The covariance matrix of CMB temperature anisotropies and $\mu$-distortions anisotropies is then given by:
%----------------------
\begin{align}
\label{eq:covariance}
{\rm C} =  \left(
\begin{array}{cc}
C_\ell^{\,TT} & C_\ell^{\,\mu\times T}  \\
C_\ell^{\,\mu\times T} & C_\ell^{\,\mu\times \mu}  \\
\end{array}
\right).
\end{align}
%----------------------
By computing the singular-value decomposition (SVD) of the covariance matrix, ${\rm C}$, we can first generate two independent Gaussian random fields, then reproject them in the appropriate basis in order to obtain a simulated CMB map and a simulated $\mu$-distortion map that are correlated according to Eq.~\eqref{eq:covariance}.

The spectral energy distribution (SED) of CMB temperature anisotropies is the derivative of the blackbody spectrum with respect to temperature, thus in thermodynamic units corresponding to a constant spectrum across frequency, $\nu$:
%----------------------
\begin{align}
\label{eq:spectra0}
a_{\rm T}(\nu) =  T_{\rm CMB}.
\end{align}
%---------------------- 
This expression neglects higher order terms $\propto (\Delta T/T)^2$, which introduce a $y$-type spectral dependence \citep[e.g.,][]{Chluba2004} that are negligible in our discussion.
Conversely, the SED of $\mu$-distortion anisotropies, $a_\mu$, scales across frequencies as:
%----------------------
\begin{align}
\label{eq:spectra}
a_\mu(\nu) =  {T_{\rm CMB}\over x} \left({x\over 2.19} - 1 \right),\quad x\equiv{h\nu\over kT_{\rm CMB}},
\end{align}
%----------------------
in thermodynamic temperature units. The distinct spectral signatures of CMB temperature and $\mu$-distortion anisotropies should allow us to separate the two signals by using multi-frequency observations from CMB satellite experiments. We use Eqs.~\eqref{eq:spectra0} and \eqref{eq:spectra} in our simulations to integrate the CMB and $\mu$-distortion template maps over the frequency bands of different CMB experiments.

Figure~\ref{Fig:cls} shows both the auto- and cross-angular power spectra of the simulated CMB and $\mu$ maps (black lines), plotted against the theory power spectra (coloured lines) for the model of \cite{Ravenni2017}. CMB temperature anisotropies are a significant foreground to $\mu$-distortion anisotropies, dominating the signal at all angular scales by more than six orders of magnitude for $f_{\rm NL}\lesssim 10^3$. Conversely, the $\mu$-$T$ cross-correlation signal is only about three orders of magnitude lower than the CMB $TT$ power spectrum, therefore providing a potentially more accessible target for future CMB experiments \citep{Pajer2012}.

%%%%%%%%%%%%%%%%%%%%%%%%%%%%%%%%%%%%%%%%%%%%%%%%%%%%%%
\begin{figure}
  \begin{center}
    \includegraphics[width=0.99\columnwidth]{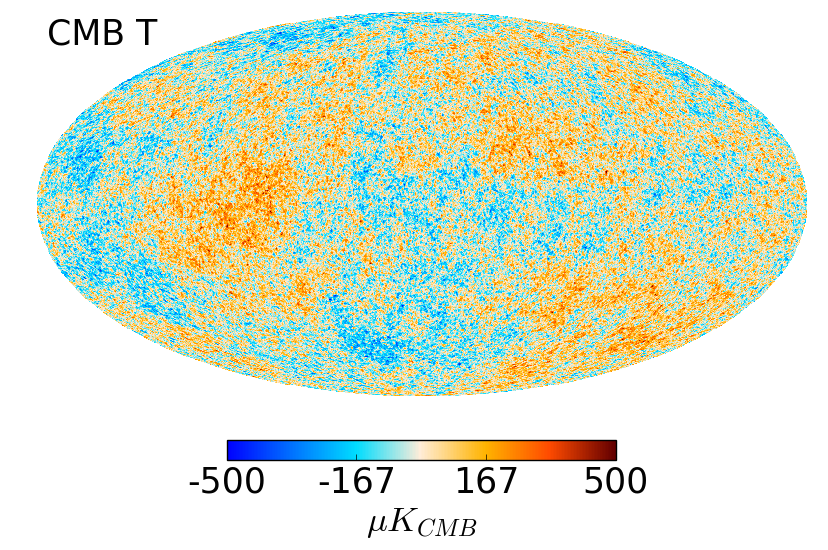}
    \\[3mm]
    \includegraphics[width=0.99\columnwidth]{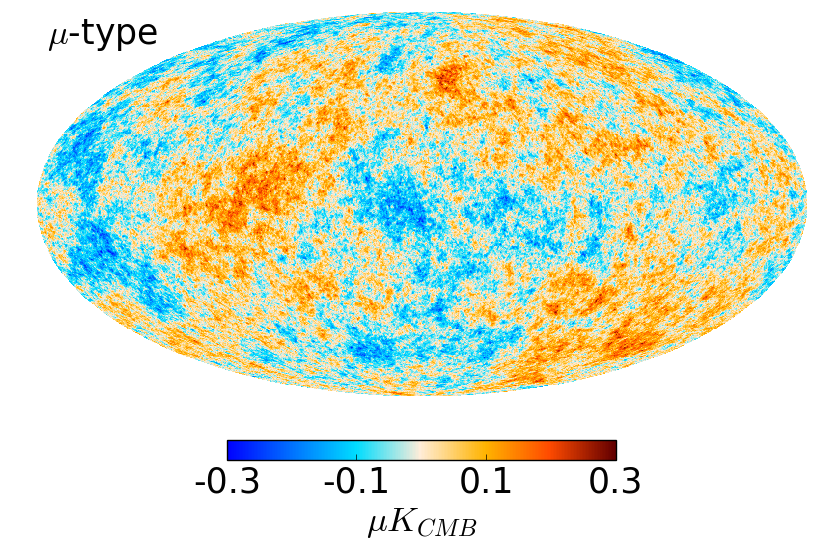}
  \end{center}
\caption{Simulation of the \emph{correlated} maps of CMB temperature anisotropies (\emph{top}) and $\mu$-distortion anisotropies (\emph{bottom}) at infinite resolution. Here $\langle\mu\rangle=2\times 10^{-8}$ and ${f_{\rm NL}=4500}$.}
\label{Fig:cmb_n_mu}
\end{figure}
%%%%%%%%%%%%%%%%%%%%%%%%%%%%%%%%%%%%%%%%%%%%%%%%%%%%%%
The resulting maps of correlated CMB temperature and $\mu$-distortion anisotropies are shown in Fig.~\ref{Fig:cmb_n_mu} for $\langle\mu\rangle=2\times 10^{-8}$, which is close to the value expected within standard $\Lambda$CDM \citep{Chluba2016}, and $f_{\rm NL}=4.5\times 10^3$.  The top panel shows typical degree-scale fluctuations of CMB temperature anisotropies over the sky, while the bottom panel shows that bulk of the $\mu$-distortion fluctuations are present at large angular scales, giving the impression of a low-resolution version of the temperature map.

%%%%%%%%%%%%%%%%%%%%%%%%%%%%%%%%%%%%%%%%%%%%%%%%%%%%%%
\begin{figure*}
  \begin{center}
    \includegraphics[width=0.98\columnwidth]{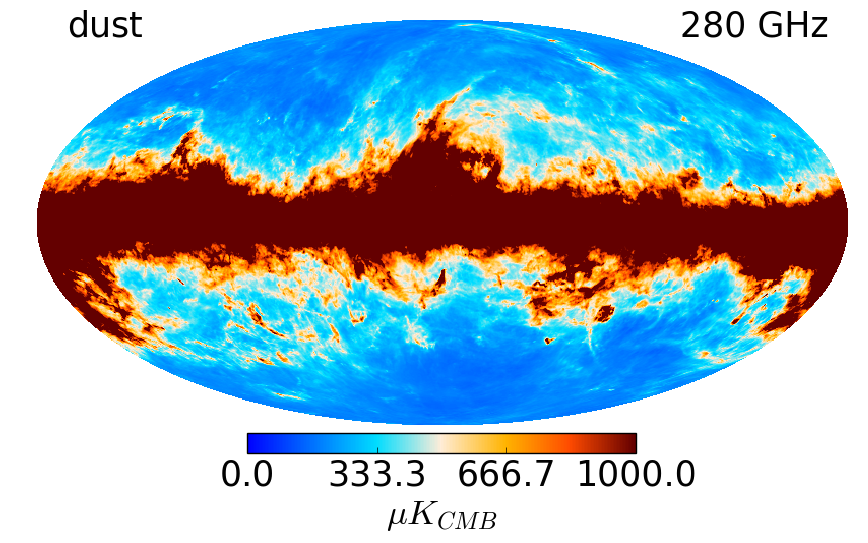}\hspace{4mm}
\includegraphics[width=0.98\columnwidth]{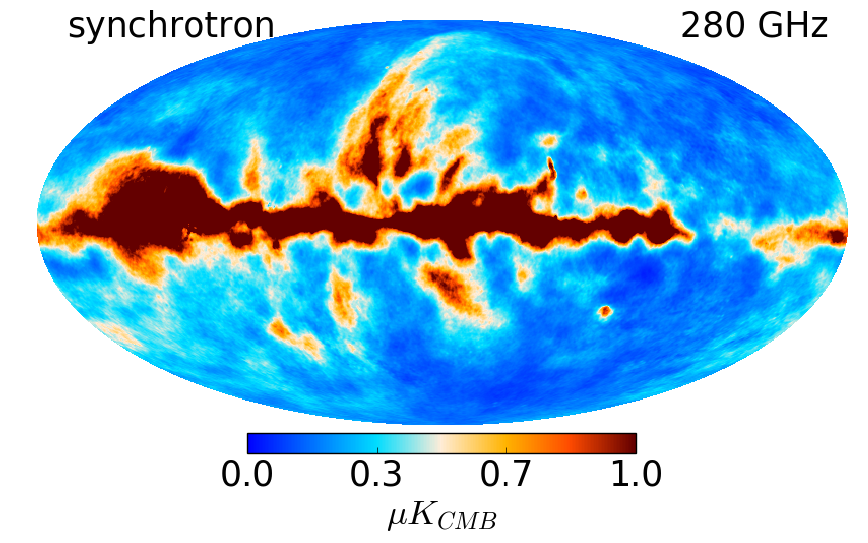}
\\[3mm]
    \includegraphics[width=0.98\columnwidth]{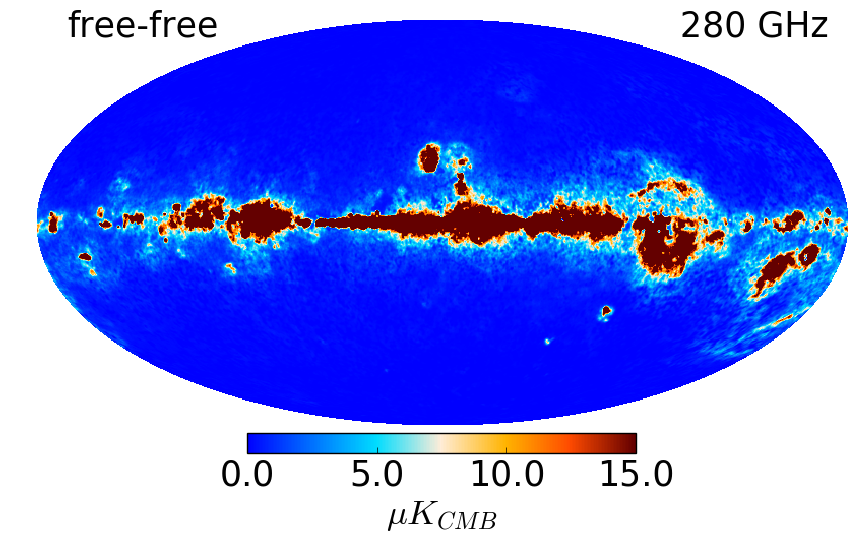}\hspace{4mm}
\includegraphics[width=0.98\columnwidth]{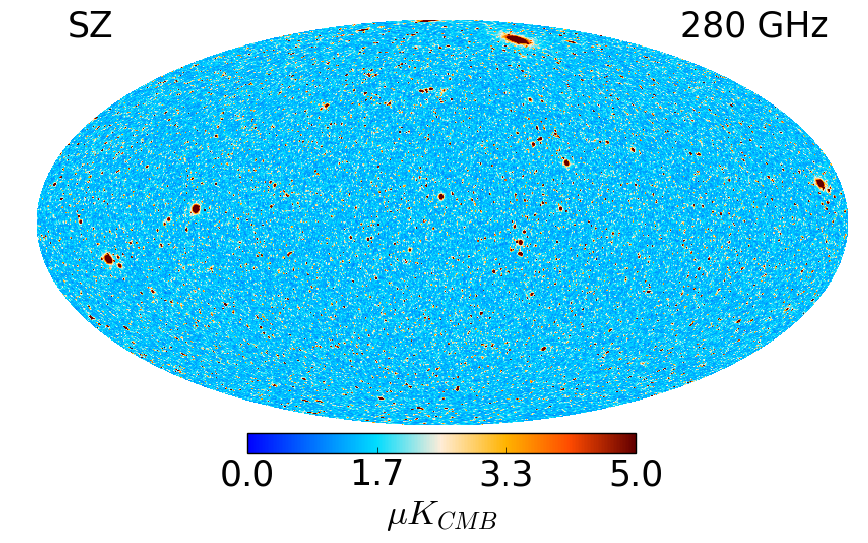}
\\[3mm]
    \includegraphics[width=0.98\columnwidth]{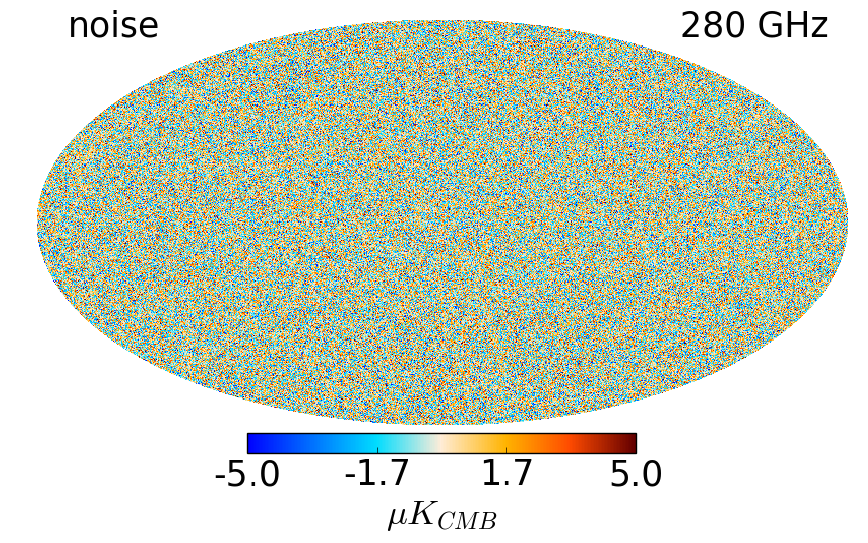}\hspace{4mm}
 \includegraphics[width=0.98\columnwidth]{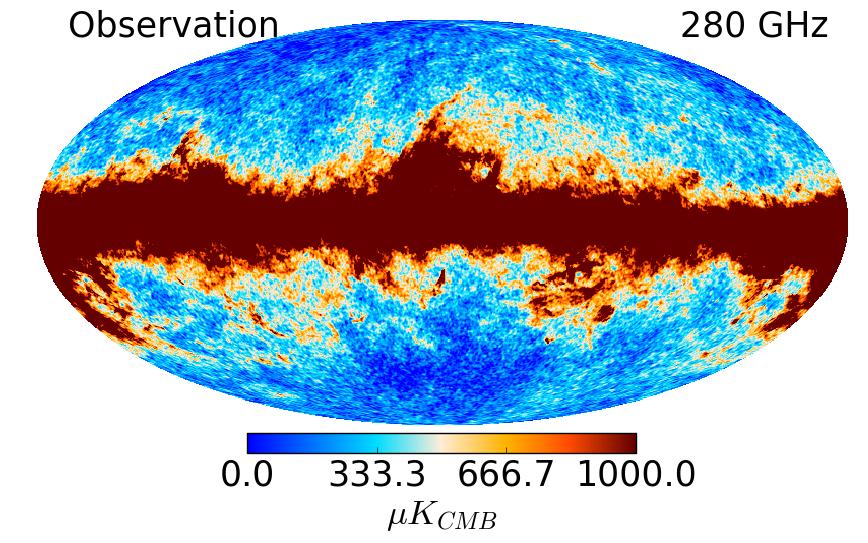}
 \end{center}
\caption{Simulation of Galactic and extragalactic foregrounds observed in the $280$\,GHz frequency band of \textit{LiteBIRD}: thermal dust (\emph{top left}); synchrotron (\emph{top right}); free-free (\emph{middle left}); SZ (\emph{middle right}); \textit{LiteBIRD} instrumental noise (\emph{bottom left}); total observation map at $280$\,GHz (\emph{bottom right}), which includes foregrounds, CMB anisotropies, $\mu$-distortion anisotropies, and noise.}
\label{Fig:gal_fg}
\end{figure*}
%%%%%%%%%%%%%%%%%%%%%%%%%%%%%%%%%%%%%%%%%%%%%%%%%%%%%%

\vspace{2mm}
\subsection{Foregrounds}
\label{sec:sim_fg}
%%%%%%%%%%%%%%%%%%%%%%%%%%%%%%%%%%%%%%%%%%%%%%%%%%%%%%
We use the \textsc{PSM} \citep[Planck Sky Model;][]{Delabrouille2013} software to simulate foregrounds and instrumental noise. We include both Galactic and extragalactic foregrounds in our sky simulations: thermal dust emission, synchrotron radiation, Galactic free-free emission, and thermal Sunyaev-Zeldovich (SZ) effect ($y$-distortion) from galaxy clusters. We neglect potential effects of line-of-sight and beam averaging on the SEDs of the different components \citep{Chluba2017foregrounds}. We also do not include any intrinsic $y$-$T$ correlations and focus only on the $\mu$-$T$ signal described in Sect.~\ref{sec:sim_mu}. As shown in \citet{Ravenni2017}, $y$-$T$ correlations and also correlations of distortions with CMB polarization signals can help us to separate different contributions, but we leave a more detailed analysis to future work. 

One risk of ignoring $y$-$T$ correlations in the analysis of real data would be that residual SZ emission in the reconstructed $\mu$-distortion map might bias the measurement of the $\mu$-$T$ correlation signal. Here, masking and the explicit scale-dependence of the SZ-$T$ and SZ-$E$ correlations can be utilized to separate contributions \citep{Creque2016, Ravenni2017}. The primordial $y$-$T$ correlation signal itself is about one order of magnitude smaller than the $\mu$-$T$ signal \citep[e.g.,][]{Ravenni2017}, so that residual $y$-$T$ contamination will also be much lower than the measured $\mu$-$T$ signal, while in contrast spurious residual $TT$ correlations might be larger than the signal itself. Thus, the main enemy here is rather residual CMB temperature anisotropies in the reconstructed $\mu$-distortion map, which, if not canceled, will add spurious $T$-$T$ correlations at a larger level than residual primordial $y$-$T$ correlations to the measured $\mu$-$T$ signal (see Fig.~\ref{Fig:std_vs_cst}).

Galactic thermal dust emission (top left panel in Fig.~\ref{Fig:gal_fg}) arise from small dust grains of various sizes in the interstellar medium (silicates and carbonaceous grains, molecules of polycyclic aromatic hydrocarbon) that are heated by the emission from stars and re-emit photons at infra-red wavelengths. This is the dominant astrophysical foreground at high frequencies ($> 100$\,GHz) in CMB observations. We use the publicly released \Planck\ GNILC dust all-sky map at $353$\,GHz \citep{Planck_PIP_XLVIII} as a template for the simulation of the Galactic thermal dust emission. The GNILC dust map does not suffer from contamination by cosmic infrared background anisotropies thanks to filtering by the GNILC algorithm \citep{Remazeilles2011b}. The dust template is then integrated over the frequency bands of the considered CMB experiment assuming a modified blackbody emission law 
%----------------------
\begin{align}
I_\nu^{\rm dust} = I_{353\,{\rm GHz}}^{\rm GNILC}\, \left({\nu\over 353}\right)^{\beta_{\rm d}} \, {B_\nu\left(T_{\rm d}\right)\over B_{353}\left(T_{\rm d}\right)},
\end{align}
%----------------------
with variable emissivity, $\beta_{\rm d}$, and temperature, $T_{\rm d}$, over the sky, $B_\nu(T_{\rm d})$ being the Planck's law for blackbody radiation. The released \Planck\ GNILC maps of dust temperature and emissivity, with average values over the sky of $\langle\beta_{\rm d}\rangle = 1.6$ and $\langle T_{\rm d}\rangle = 19.4$\,K, are used for the spectral scaling of the dust template map.

High-energy cosmic ray electrons spiraling Galactic magnetic fields are responsible for Galactic synchrotron radiation. Since those magnetic fields extend outside the Galaxy, synchrotron emission is present even at high Galactic latitudes in the sky (top right panel in Fig.~\ref{Fig:gal_fg}), and is the main astrophysical foreground at radio frequencies ($< 100$\,GHz) in CMB observations. As a template of Galactic synchrotron emission, we use the reprocessed \cite{Haslam1982} 408\,MHz all-sky map of \cite{Remazeilles2015}, in which extragalactic radio sources and other systematic effects have been subtracted. The 408\,MHz map is scaled across the frequency bands of CMB experiments through a power-law emission law in Rayleigh-Jeans brightness temperature units 
%----------------------
\begin{align}
I_\nu^{\rm synch.} = I_{408\,{\rm MHz}}\, \left({\nu\over 408}\right)^{\beta_{\rm s}}\,,
\end{align}
%----------------------
with a variable spectral index, $\beta_{\rm s}$, over the sky. The synchrotron spectral index map is taken from \mbox{\cite{Miville2008}}, which has an average value of $\langle\beta_{\rm s}\rangle=-3$ over the sky. Similar variations of the spectral index are expected along the line of sight, leading to higher order curvature terms \citep{Chluba2017foregrounds}, which are not included here.

Free electrons loose energy through Coulomb interactions with heavy ions, which results in a Bremsstrahlung emission of photons in ${\rm H}_{\rm II}$ regions of the Galactic plane. This emission is also termed as Galactic free-free emission, which, while fainter than synchrotron and thermal dust emissions, is still a significant foreground in star-forming regions of the Galaxy at low frequencies $\lesssim 100$\, GHz for CMB observations (middle left panel in Fig.~\ref{Fig:gal_fg}). Galactic free-free emission is simulated from the ${\rm H}_\alpha$ emission map corrected for dust extinction \citep{Dickinson2003}, and scaled across frequency bands through a power-law emission law in brightness temperature units, with a uniform spectral index of $\beta_{\rm ff}=-2.1$ over the sky:
%----------------------
\begin{align}
I_\nu^{\rm ff} = 90\,{\rm mK}\,\left({T_{\rm e}\over {\rm K}}\right)^{-0.35}\,\left({\nu\over{\rm GHz}}\right)^{\beta_{\rm ff}}\,\left({{\rm H}_\alpha\over{\rm cm}^{-6}\,{\rm pc}}\right),
\end{align}
%----------------------
where $T_{\rm e}$ is the electronic temperature in K, $\nu$ is the frequency in GHz, and the ${\rm H}_\alpha$ emission measure is in ${\rm cm}^{-6}\,{\rm pc}$ units (electron density squared along the line of sight).

The hot gas of electrons residing in galaxy clusters scatter CMB photons, generating $y$-distortions of the CMB blackbody emission at the location of galaxy clusters (middle right panel in Fig.~\ref{Fig:gal_fg}). This is known as the thermal Sunyaev-Zeldovich (SZ) effect \citep{Sunyaev1972}. While dominant at small angular scales, clustering generates thermal SZ emission on all angular scales in the sky in the low-redshift Universe, and is recognised as an important foreground to early $\mu$-distortion anisotropies \citep{Khatri2015}. 
Thermal SZ emission from galaxy clusters is generated in our simulations by mapping the MCXC catalogue of galaxy clusters \citep{Piffaretti2011} from ROSAT \citep{ROSAT2004} and the SDSS catalogue of galaxy clusters \citep{Koester2007}. The model of $y$-Compton parameter flux is derived from the universal cluster pressure profile of \citep{Arnaud2010}, while the scaling of the thermal SZ $y$-map across frequency bands in the non-relativistic limit is given by \citep{Sunyaev1972}:
%----------------------
\begin{align}
I_\nu^{\rm SZ} = y\,T_{\rm CMB}\,\left(x\, \coth\left({x\over 2}\right) - 4\right)\,,\quad x\equiv {h\nu \over kT_{\rm CMB}}\,,
\end{align}
%---------------------- 
in thermodynamic temperature units. We do not include extra-galactic compact radio and infra-red sources in our simulations because we are interested in quite large angular scales ($\ell \lesssim 200$) to extract the $\mu$-$T$ correlation signal.

Figure~\ref{Fig:gal_fg} provides an overview of all the simulated maps of Galactic and extragalactic foregrounds in the $280$\,GHz frequency band of the \textit{LiteBIRD} experiment, as well as the instrumental noise at this frequency, and the total observation sky map at $280$\,GHz, which consists of foreground emissions, instrumental noise, and CMB temperature and $\mu$-distortion anisotropies.

%%%%%%%%%%%%%%%%%%%%%%%%%%%%%%%%%%%%%%%%%%%%%%%%%%%%%%
\begin{table}
\caption{Instrumental specifications for \pixie\ (A. Kogut, priv. comm.). The aggregated sensitivity in temperature is $4.7$\,$\mu$K.arcmin.}
  \label{tab:pixie}
 \centering
  \begin{tabular}{lll}
\hline\hline
Frequency   & Beam FWHM & $I$ noise r.m.s \\
$[\rm GHz]$   & $[\rm arcmin]$  & $[\rm \mu K.arcmin]$ \\
\hline 
30 & 96 & 219.8 \\
60 & 96 & 59.0 \\
90 & 96 & 29.3 \\
120 & 96 & 19.3 \\
150 & 96 & 14.9 \\
180 & 96 & 13.0 \\
210 & 96 & 12.4 \\
240 & 96 & 12.6 \\
270 & 96 & 13.5 \\
300 & 96 & 15.2  \\
330 & 96 & 17.7 \\
360 & 96 & 21.3 \\
390 & 96 & 26.4 \\
420 & 96 & 33.5 \\
450 & 96  & 43.3 \\
480 & 96  & 57.3 \\
510 & 96  & 76.4 \\
540 & 96  & 103.5 \\
570 & 96  & 142.1 \\
600 & 96  & 197.7 \\
630 & 96  & 277.9 \\
660 & 96  & 393.7 \\
690 & 96  & 564.3 \\
720 & 96  & 810.3 \\
750 & 96  & 1175.2 \\
780 & 96  & 1718.3 \\
810 & 96  & 2528.6 \\
840 & 96  & 3742.0 \\
870 & 96  & 5557.9 \\
900 & 96  & 8315.6 \\
930 & 96  & 12473.4 \\
960 & 96  & 18837.3 \\
990 & 96  & 28510.5 \\
1020 & 96  & 43274.9 \\
1050 & 96  & 66185.2 \\
1080 & 96  & 101399.0 \\
1110 & 96  & 155705.0 \\
1140 & 96  & 240133.0 \\
1170 & 96  & 371231.0 \\
1200 & 96  & 576999.0 \\
\hline
   \end{tabular}
\end{table}
%%%%%%%%%%%%%%%%%%%%%%%%%%%%%%%%%%%%%%%%%%%%%%%%%%%%%%

\subsection{Instrumental specifications}
\label{sec:exp}

Motivated by the success of the \emph{Planck} space mission \citep{planck2015_overview} in mapping CMB temperature anisotropies with high precision over the full sky, a certain number of future high-sensitive CMB satellite experiments are now being considered around the world, mainly to detect the primordial CMB $B$-mode polarization at very large angular scales \citep{Remazeilles2016,Remazeilles2017}. Future CMB satellites will also allow to probe temperature anisotropies with unprecedented sensitivity at those very large angular scales, where bulk of the $\mu$-$T$ correlation signal lies \citep{Finelli2016}. 

%%%%%%%%%%%%%%%%%%%%%%%%%%%%%%%%%%%%%%%%%%%%%%%%%%%%%%
\begin{table}
\caption{Instrumental specifications for \litebird. The aggregated sensitivity in temperature is $1.7$\,$\mu$K.arcmin.}
  \label{tab:litebird}
 \centering
  \begin{tabular}{lll}
\hline\hline
Frequency   & Beam FWHM & $I$ noise r.m.s \\
$[\rm GHz]$   & $[\rm arcmin]$  & $[\rm \mu K.arcmin]$ \\
\hline 
40 & 69 & 26.0 \\
50 & 56 & 16.7 \\
60 & 48 & 13.8 \\
68 & 43 & 11.2 \\
78 & 39 & 9.4 \\
89 & 35 & 8.1 \\
100 & 29 & 6.4 \\
119 & 25 & 5.3 \\
140 & 23 & 4.1 \\
166 & 21 & 4.5  \\
195 & 20 & 4.0 \\
235 & 19 & 5.3 \\
280 & 24 & 9.2 \\
337 & 20 & 13.5 \\
402 & 17  & 26.1 \\
\hline
   \end{tabular}
\end{table}
%%%%%%%%%%%%%%%%%%%%%%%%%%%%%%%%%%%%%%%%%%%%%%%%%%%%%%

\pixie\ \citep{Kogut2011PIXIE, Pixie2016} is a US CMB space mission concept which was proposed to NASA (National Aeronautics and Space Administration) in 2011 and 2016. \pixie\ is based on a Fourier transform spectrometer and would allow us to probe the full sky in 400 spectral bands ranging from $30$ to $6000$\,GHz, with only four bolometric detectors per channel. Through absolute spectroscopy, \pixie\ can perform absolute calibration of the CMB blackbody spectrum, and in that sense is the only proposed CMB experiment capable of measuring the absolute monopole of spectral distortions, e.g. $\langle\mu\rangle$. \pixie\ has a relatively low angular resolution of $\delta \theta \simeq 96'$ in all channels, and its aggregated sensitivity depends on the splitting of the $30$--$6000$ GHz spectral band into individual frequency bands allowing for variable bandwidths. 

In our \pixie\ simulation, we use 40 frequency channels from $30$\,GHz to $1200$\,GHz with $30$\,GHz bandwidths, resulting in an effective overall sensitivity of $\simeq 7$\,$\mu$K.arcmin in polarization. In polarization mode, two beams observe the sky, while in intensity mode for spectral distortion monopoles, only one beam observes the sky resulting in lower sensitivity. However, we are interested in anisotropies of spectral distortions in our analysis, so that the two beams for polarization can be used in differential mode and averaged, giving effectively an aggregated sensitivity of $\simeq 4.7$ \,$\mu$K.arcmin in intensity for temperature anisotropies. Table ~\ref{tab:pixie} summarizes the instrumental specifications of \pixie\ in this chosen configuration (A. Kogut, private communication).   

\litebird\ \citep{Litebird2016} is a proposed Japanese CMB satellite experiment, selected for Phase-A study by JAXA (Japan Aerospace Exploration Agency). In its current proposed design \citep{Litebird2018}, \litebird\ must observe the full sky through 15 frequency channels ranging from $40$ to $402$ GHz in order to remove foreground contamination, with a small cross-Dragone telescope of $40$\,cm diameter, thus providing an overall beam resolution of $\delta \theta\simeq 20'$ in CMB channels. The \litebird\ satellite concept is an imager with 2622 detectors composed of Transition Edge Sensor (TES) bolometers, providing a combined sensitivity from all frequency channels of about $2.4$\,$\mu$K.arcmin in polarization ($1.7$ \,$\mu$K.arcmin in intensity).\footnote{The sensitivities and resolutions of the proposed CMB satellites used in our simulations are just indicative and subject to discussion during the design study of the experiment.} The instrumental specifications of \litebird\ used for our simulations are summarized in Table~\ref{tab:litebird}.

%%%%%%%%%%%%%%%%%%%%%%%%%%%%%%%%%%%%%%%%%%%%%%%%%%%%%%
\begin{table}
\caption{Instrumental specifications for \core. The aggregated sensitivity in temperature is $1.2$\,$\mu$K.arcmin.}
  \label{tab:core}
 \centering
  \begin{tabular}{lll}
\hline\hline
Frequency   & Beam FWHM & $I$ noise r.m.s \\
$[\rm GHz]$   & $[\rm arcmin]$  & $[\rm \mu K.arcmin]$ \\
\hline 
60 & 17.87 & 7.5 \\
70 & 15.39 & 7.1 \\
80 & 13.52 & 6.8 \\
90 & 12.08 & 5.1 \\
100 & 10.92 & 5.0 \\
115 & 9.56 & 5.0 \\
130 & 8.51 & 3.9 \\
145 & 7.68 & 3.6 \\
160 & 7.01 & 3.7 \\
175 & 6.45 & 3.6  \\
195 & 5.84 & 3.5 \\
220 & 5.23 & 3.8 \\
255 & 4.57 & 5.6 \\
295 & 3.99 & 7.4 \\
340 & 3.49  & 11.1 \\
390 & 3.06  & 22.0 \\
450 & 2.65  & 45.9 \\
520 & 2.29  & 116.6 \\
600 & 1.98  & 358.3 \\
\hline
   \end{tabular}
\end{table}
%%%%%%%%%%%%%%%%%%%%%%%%%%%%%%%%%%%%%%%%%%%%%%%%%%%%%%

%%%%%%%%%%%%%%%%%%%%%%%%%%%%%%%%%%%%%%%%%%%%%%%%%%%%%%
\begin{table}
\caption{Instrumental specifications for \pico\ (S. Hanany, priv. comm.). The aggregated sensitivity in temperature is $0.8$\,$\mu$K.arcmin.}
  \label{tab:pico}
 \centering
  \begin{tabular}{lll}
\hline\hline
Frequency   & Beam FWHM & $I$ noise r.m.s \\
$[\rm GHz]$   & $[\rm arcmin]$  & $[\rm \mu K.arcmin]$ \\
\hline 
      21 & 40.9 &   35.4 \\
      25 & 34.1 &   23.3 \\
      30 & 28.4 &   15.8 \\
      36 & 23.7 &   10.6 \\
      43 & 19.7 &    6.4 \\
      52 & 16.4 &    4.9 \\
      62 & 13.7 &    3.5 \\
      75 & 11.4 &    2.8 \\
      90 &  9.5 &    2.3 \\
     110 &  7.9 &    2.1 \\
     130 &  6.6 &    1.9 \\
     155 &  5.5 &    1.8 \\
     185 &  4.6 &    2.5 \\
     225 &  3.8 &    3.7 \\
     270 &  3.2 &    6.4 \\
     320 &  2.7 &   11.3 \\
     385 &  2.2 &   22.6 \\
     460 &  1.8 &   53.0 \\
     555 &  1.5 &  155.6 \\
     665 &  1.3 &  777.8 \\
     800 &  1.1 & 7071.1 \\
\hline
   \end{tabular}
\end{table}
%%%%%%%%%%%%%%%%%%%%%%%%%%%%%%%%%%%%%%%%%%%%%%%%%%%%%%

\core\ \citep{Core2016} is a European CMB space mission concept that was proposed to ESA (European Space Agency) in 2016 as a medium-class (M) mission. \core\ is an imager allowing to observe the full sky within 19 frequency bands ranging from $60$ to $600$\,GHz for foreground subtraction, and with a cross-Dragone telescope diameter of $120$\,cm diameter, providing a beam resolution of $\delta \theta\simeq 10'$ in CMB channels. The focal plane of the \core\ satellite would be composed of 2100 sensitive Kinetic Inductance Detectors (KIDs), providing an aggregated sensitivity from all channels of $1.7$\,$\mu$K.arcmin in polarization ($1.2$ \,$\mu$K.arcmin in intensity). While not selected by ESA, a set of ten papers of the \core\ collaboration covering the large science legacy that \core\ would provide has been published (see \cite{Core2016} and references therein), showing the strong interest of a large community in such a future CMB satellite experiment. The instrumental specifications of \core\ are listed in Table~\ref{tab:core}.

\pico\ (S. Hanany, private communication) is a US CMB space mission concept currently being discussed for proposal to NASA, which would observe the sky in 21 frequency bands ranging from $21$ to $800$\, GHz with 12060 TES detectors, thus providing a combined sensitivity of $1.1$ \,$\mu$K.arcmin in polarization ($0.8$ \,$\mu$K.arcmin in intensity). \pico\ is thus the most sensitive CMB satellite experiment among the ones investigated in this work, and has the largest number of frequency bands and broadest frequency range for foreground subtraction among the CMB imagers considered. The \pico\ satellite concept will be composed of an open-Dragone telescope with a diameter of $140$\,cm, providing a beam resolution of $\delta \theta\simeq 7'$ in CMB channels. The possibility of adding a small spectrometer that will complement the imager for absolute calibration is still being discussed. The proposed configuration of \pico\ for our simulations is summarized in Table~\ref{tab:pico}.

For the aforementioned instrumental configurations, the simulated sky components are integrated over the dedicated frequency bands, co-added, and convolved of by a Gaussian beam of full-width half-maximum (FWHM) depending on the frequency channel. Instrumental white noise maps are co-added to the sky frequency maps, depending of the sensitivity of each CMB experiment in each frequency band. These simulations built the starting point of our analysis and can be shared upon request.

\section{Reconstruction of anisotropic $\boldsymbol{\mu}$-distortions}
\label{sec:comp_sep}
The detection of $\mu$-distortion anisotropies will rely on component separation methods to subtract foreground contamination and reconstruct the signal. Most component separation techniques, either parametric fitting approaches \citep[e.g.][]{Eriksen2008} or blind variance minimization methods \citep[e.g.][]{Delabrouille2009,Fernandez2012}, aim at minimizing the global foreground contamination. However, while the residual contamination is minimized, it is not completely nulled in the reconstructed signal. Also specific residuals might be more of an issue for measuring the $\mu$-$T$ correlation signal than others. For instance, any non-zero residual of CMB temperature anisotropies in the reconstructed $\mu$-distortion map adds spurious $T$-$T$ correlations in the $\mu$-$T$ cross-correlation measurement, therefore largely biasing the signal (see further discussion in Sect.~\ref{subsec:vs}). Indeed, the spectral template of temperature anisotropies correlates with $\mu$-distortions at a similar level as for example $y$-signals \citep{Chluba2013PCA}. Thus, instead of just minimizing the \emph{global} foreground contamination, it is beneficial to \emph{null} the unwanted CMB temperature anisotropy component in the reconstructed $\mu$-distortion map, at the expense of slightly more contamination by other foregrounds and noise.

\subsection{Component separation methodology}
\label{subsec:method}

To achieve this, we apply the Constrained ILC component separation method \citep{Remazeilles2011} on the set of simulations. Because this technique allows one to reconstruct the $\mu$-distortion anisotropies, while simultaneously nulling CMB temperature anisotropies (and vice versa), it will allow us to minimize potential biases created by temperature fluctuations. It also automatically removes temperature fluctuations related to the integrated-SW of clusters \citep{Creque2016, Ravenni2017}.
From the foreground-cleaned, CMB-free, $\mu$-distortion map, $\widehat{\mu}$, and the $\mu$-free CMB map, $\widehat{T}$, we will then compute the cross-power spectrum, $\widehat{C}_\ell^{\,\mu\times T}$, between the two maps. The reconstructed cross-power spectrum will allow us to derive forecasts for the detection of the local-type non-Gaussianity parameter, $f_{\rm NL}$, in the presence of foregrounds for the different CMB satellite experiments like \textit{PIXIE}, \textit{LiteBIRD}, \textit{CORE}, and \textit{PICO}.

For a given frequency band $i$, the sky observation map $x_i$ can be
modelled as the combination of different emission components:
%--------------------
\begin{align}
\label{eq:obs}
  x_i(\hat{\mathbf{\theta}})\, =\, a_{\mu,i}\, s_{\rm \mu} (\hat{\mathbf{\theta}})\,
  +\,a_{{\rm T},i}\, s_{\rm CMB} (\hat{\mathbf{\theta}})\,
  +n_i(\hat{\mathbf{\theta}}),
\end{align}
%--------------------
where $s_{\rm \mu} (\hat{\mathbf{\theta}})$ is the $\mu$-distortion anisotropy at pixel $\hat{\mathbf{\theta}}$, $s_{\rm CMB} (\hat{\mathbf{\theta}})$ is the CMB temperature anisotropy in the same direction, and $n_i(p)$ is a ``nuisance'' term including instrumental noise and Galactic foregrounds in the frequency channel~$i$. The $\mu$-distortion and CMB temperature anisotropies scale with frequency through distinct emission laws that are parameterized by the vectors $\bda_{\mu}$ and $\bda_{\rm T}$ (Eq.~\ref{eq:spectra}), with $n_f$ dimensions accounting for the number of frequency bands of the CMB experiment. 

Like the standard {\tt NILC} method \citep{Delabrouille2009,Remazeilles2013}, the Constrained ILC method \mbox{\citep{Remazeilles2011}} constructs a minimum-variance weighted linear combination of the sky maps: ${\hat{s}_{\mu} (\hat{\mathbf{\theta}}) = \bdw^t \bdx(\hat{\mathbf{\theta}}) = \sum_{i=1}^{n_f} w_i x_i(\hat{\mathbf{\theta}})}$ ($\bdw^t$ is the transpose of $\bdw$), under the condition that the scalar product of the weight vector, $\bdw$, and $\mu$-distortion SED vector, $\bda_{\mu}$, is equal to unity, i.e. ${\sum_{i=1}^{n_f} w_i a_{\mu,i} = 1}$. This guarantees the full conservation of the $\mu$-distortion anisotropies in the reconstruction. 

However, the Constrained ILC \mbox{\citep{Remazeilles2011}} generalizes the standard {\tt NILC} method by offering an additional constraint for the ILC weights to be orthogonal to the CMB emission law, $\bda_{\rm T}$, while guaranteeing the conservation of the $\mu$-distortion component. The Constrained ILC estimate of the CMB-free map of $\mu$-distortion anisotropies, $\hat{s}_{\mu} (\hat{\mathbf{\theta}})\, =\, \bdw^t \bdx(\hat{\mathbf{\theta}})$, is thus a solution of the minimization problem:
%--------------------
\begin{subequations}
\begin{align}
 \label{eq:varcon}
  \min_{\bdw}\, {\rm E}\left[\hat{s}_{\mu} ^2 \right], 
  \\
 \label{eq:mucon}
  \bdw^{t} \bda_{\mu} = 1, 
   \\
  \label{eq:cmbcon}
  \bdw^{t} \bda_{\rm T} = 0. 
\end{align}
\end{subequations}
%--------------------
Benefiting from the knowledge of the spectral shapes of $\mu$-distortion and CMB temperature anisotropies, the weights of the Constrained ILC are thus adjusted to yield simultaneously unit response to the $\mu$-distortion emission law $\bda_{\mu}$ (Eq.~\ref{eq:mucon}) and zero response to the CMB emission law $\bda_{\rm T}$ (Eq.~\ref{eq:cmbcon}). 

The method is {\it blind} in the sense that no parametrization or assumption is made on the foregrounds, but just for the $\mu$ and temperature signals. The two-dimensional constraint (Eqs.~\ref{eq:mucon}, \ref{eq:cmbcon}) allows us to null CMB temperature anisotropies, while preserving $\mu$-distortion anisotropies. The residual contamination from Galactic foregrounds and instrumental noise is also controlled through the minimum-variance condition (Eq.~\ref{eq:varcon}). The exact expression for the Constrained ILC weights was derived in \mbox{\citet{Remazeilles2011}} by solving the minimization problem Eqs.~(\ref{eq:varcon})--(\ref{eq:cmbcon}), which for a CMB-free reconstruction of $\mu$-distortion anisotropies is given by:
%---------------
\begin{align}
\label{eq:2D-ILC}
  \hat{s}^{\rm T\mbox{-}free}_{\rm \mu}(\hat{\mathbf{\theta}}) &=
  { \left( \bda_{\rm T}^{t} {\tC}^{-1} \bda_{\rm T} \right)
  \bda_{\mu}^{t} {{\tC}}^{-1} - \left( \bda_{\mu}^{t}
  {\tC}^{-1} \bda_{\rm T} \right) \bda_{\rm T}^{t}
  {{\tC}}^{-1}
  \over
  \left(\bda_{\mu}^{t} {\tC}^{-1} \bda_{\mu}
  \right)\left(\bda_{\rm T}^{t} {\tC}^{-1} \bda_{\rm T} \right) -
  \left( \bda_{\mu}^{t} {\tC}^{-1} \bda_{\rm T} \right)^2 }\,
  \bdx(\hat{\mathbf{\theta}}).
\end{align}
%---------------
Here ${\tC}^{\rm i j} = \left<\,x_{i}\, x_{j}\,\right>$
are the coefficients of the frequency-frequency covariance matrix of the sky channel maps. Similarly, a $\mu$-free estimate of CMB anisotropies can be derived by exchanging $\bda_{\mu}$ and $\bda_{\rm T}$ in the formula Eq. ~(\ref{eq:2D-ILC}). 

%%%%%%%%%%%%%%%%%%%%%%%%%%%%%%%%%%%%%%%%%%%%%%%%%%%%%%
\begin{figure}
  \begin{center}
    \includegraphics[width=0.96\columnwidth]{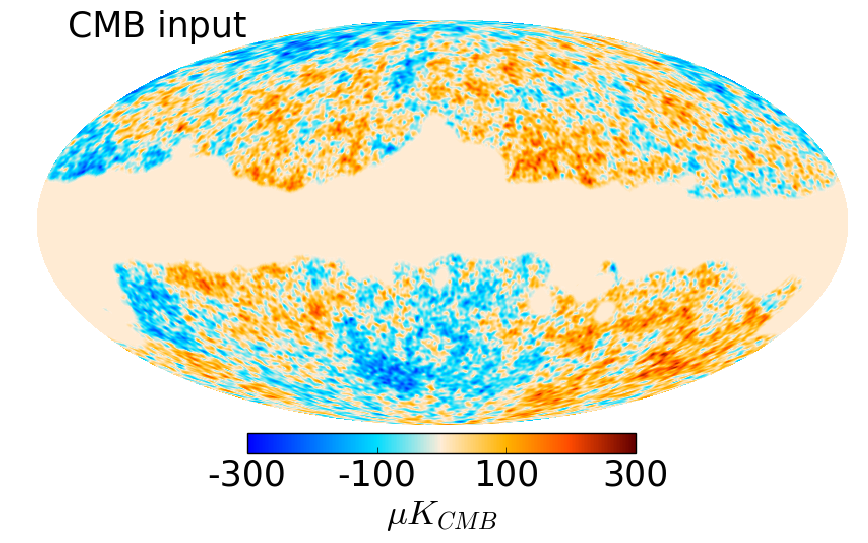}
    \\[1.5mm]
    \includegraphics[width=0.96\columnwidth]{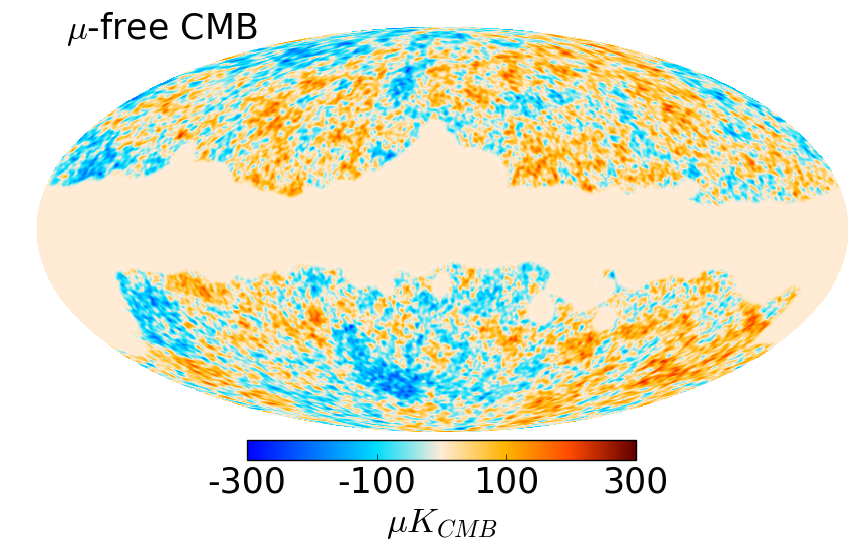}
  \end{center}
\caption{Reconstruction of the CMB temperature anisotropies with the Constrained ILC component separation method at $69'$ resolution for \textit{LiteBIRD}.  \emph{Top panel}: input CMB map realization. \emph{Bottom panel}: Constrained ILC CMB map reconstruction. The Constrained ILC CMB map benefits from the absence of any residual contamination by $\mu$-distortions.}
\label{Fig:ilc_cmb}
\end{figure}
%%%%%%%%%%%%%%%%%%%%%%%%%%%%%%%%%%%%%%%%%%%%%%%%%%%%%%

One can also easily verify that applying the Constrained ILC weights (Eq.~\ref{eq:2D-ILC}) to the frequency maps (Eq.~\ref{eq:obs}) yields:
%---------------
\begin{align}
\label{eq:math}
  \hat{s}^{\rm T\mbox{-}free}_{\rm \mu} &=
  { \left( \bda_{\rm T}^{t} {\tC}^{-1} \bda_{\rm T} \right)
  \bda_{\mu}^{t} {{\tC}}^{-1}\bda_{\mu} - \left( \bda_{\mu}^{t}
  {\tC}^{-1} \bda_{\rm T} \right) \bda_{\rm T}^{t}
  {{\tC}}^{-1}\bda_{\mu}
  \over
  \left(\bda_{\mu}^{t} {\tC}^{-1} \bda_{\mu}
  \right)\left(\bda_{\rm T}^{t} {\tC}^{-1} \bda_{\rm T} \right) -
  \left( \bda_{\mu}^{t} {\tC}^{-1} \bda_{\rm T} \right)^2 }\,
  s_{\rm \mu} \cr
&+ { \left( \bda_{\rm T}^{t} {\tC}^{-1} \bda_{\rm T} \right)
  \bda_{\mu}^{t} {{\tC}}^{-1}\bda_{\rm T} - \left( \bda_{\mu}^{t}
  {\tC}^{-1} \bda_{\rm T} \right) \bda_{\rm T}^{t}
  {{\tC}}^{-1}\bda_{\rm T}
  \over
  \left(\bda_{\mu}^{t} {\tC}^{-1} \bda_{\mu}
  \right)\left(\bda_{\rm T}^{t} {\tC}^{-1} \bda_{\rm T} \right) -
  \left( \bda_{\mu}^{t} {\tC}^{-1} \bda_{\rm T} \right)^2 }\,
  s_{\rm CMB} \cr
&+ { \left( \bda_{\rm T}^{t} {\tC}^{-1} \bda_{\rm T} \right)
  \bda_{\mu}^{t} {{\tC}}^{-1} - \left( \bda_{\mu}^{t}
  {\tC}^{-1} \bda_{\rm T} \right) \bda_{\rm T}^{t}
  {{\tC}}^{-1}
  \over
  \left(\bda_{\mu}^{t} {\tC}^{-1} \bda_{\mu}
  \right)\left(\bda_{\rm T}^{t} {\tC}^{-1} \bda_{\rm T} \right) -
  \left( \bda_{\mu}^{t} {\tC}^{-1} \bda_{\rm T} \right)^2 }\,
  \bdn\cr
&=  1\, s_{\rm \mu} + 0\, s_{\rm CMB} + \bdw^{t} \bdn.
\end{align}
%---------------
In other words, the $\mu$-distortion signal, $s_{\rm \mu}$, is reconstructed without any bias by this projection (first term of Eq.~\ref{eq:math}), while the CMB temperature signal is not just mitigated but cancelled out owing to the orthogonal weighting (second term of Eq.~\ref{eq:math}). Residual foregrounds and instrumental noise are minimized by the minimum-variance weighting (third  term of Eq.~\ref{eq:math}). 

It should be noted that additional constraints can in principle be implemented in the Constrained ILC pipeline, for example to also cancel the contaminations by $y$-distortion anisotropies. However, any added constraint usually comes with the cost of further increasing the noise in the reconstructed $\mu$-distortion map, since less information is available for reducing the noise variance. To address the limitations caused by this aspect, we plan more comprehensive simulations also including information from CMB polarization.

Lastly, the Constrained ILC method also operates on a needlet (spherical wavelet) frame \citep{Narcowich2006} because the localization properties of the wavelets allows the weights of components to re-adjust depending on the local conditions of foreground contamination both over the sky and over the angular scales.

%%%%%%%%%%%%%%%%%%%%%%%%%%%%%%%%%%%%%%%%%%%%%%%%%%%%%%
\begin{figure*}
  \begin{center}
    \includegraphics[width=0.99\columnwidth]{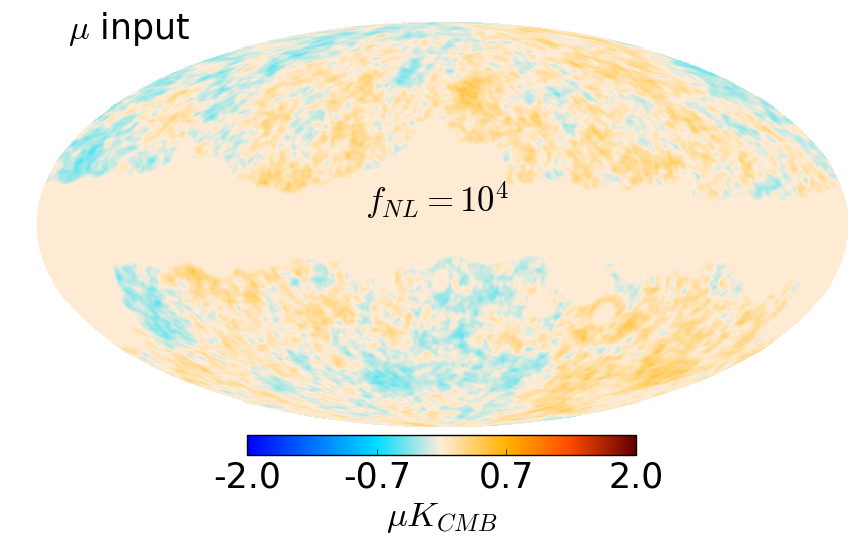}\hspace{3mm}
    \includegraphics[width=0.99\columnwidth]{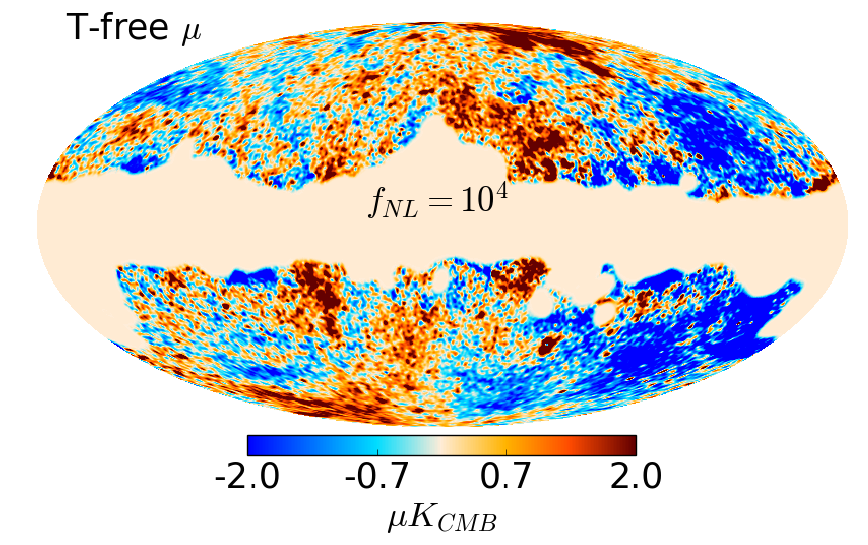}
        \\[3mm]
    \includegraphics[width=0.99\columnwidth]{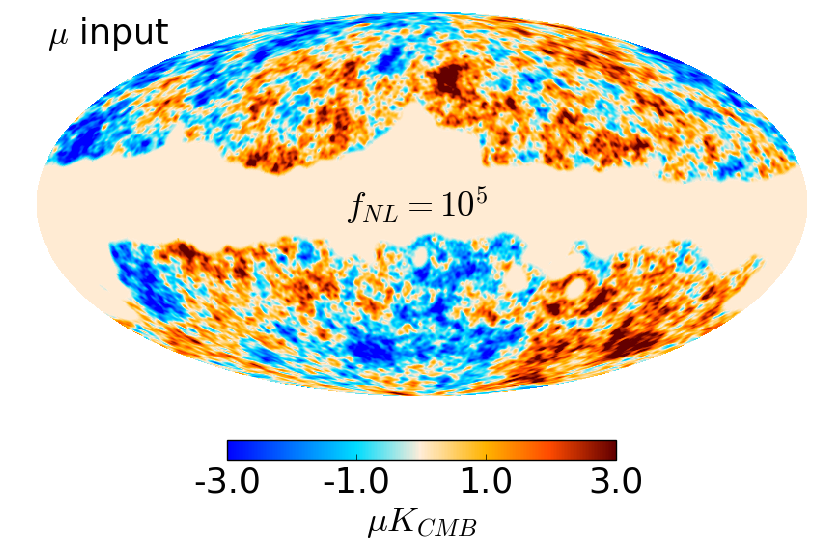}\hspace{3mm}
    \includegraphics[width=0.99\columnwidth]{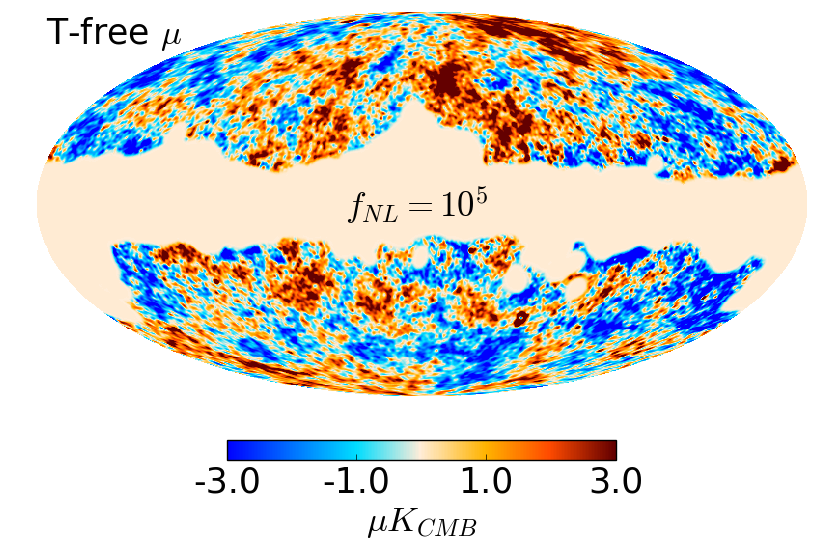}   
     \\[3mm]
    \includegraphics[width=0.99\columnwidth]{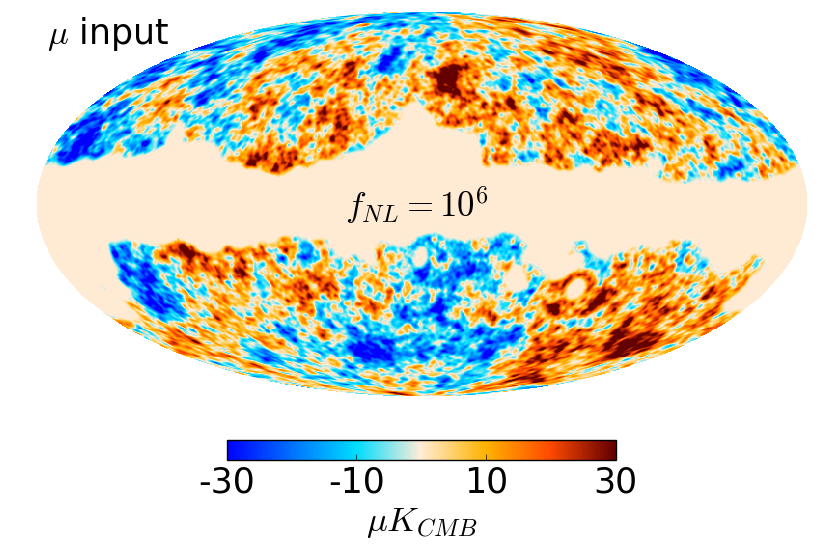}\hspace{3mm}
    \includegraphics[width=0.99\columnwidth]{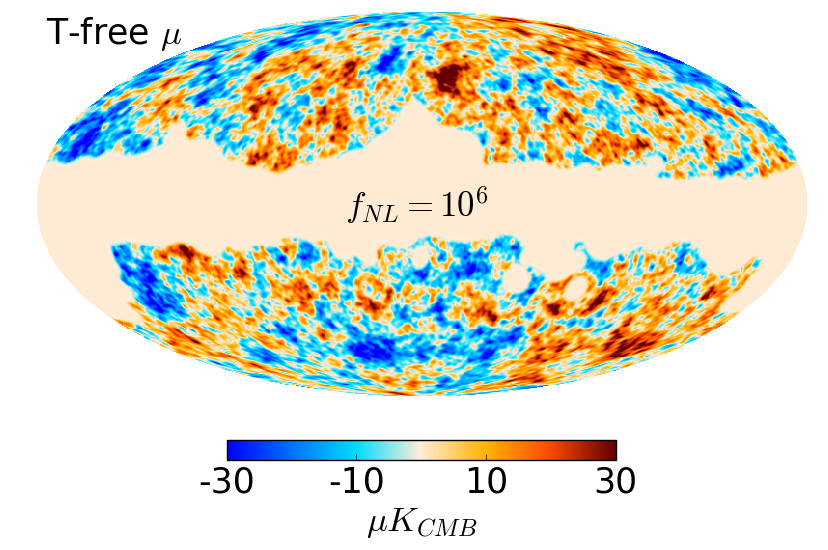}
  \end{center}
\caption{CMB-free reconstruction of $\mu$-distortion anisotropies ($\langle\mu\rangle=2\times 10^{-8}$) with the Constrained ILC component separation method at $69'$ resolution corresponding to \textit{LiteBIRD}, and for increasing values of the non-Gaussianity parameter: $f_{\rm NL}=10^4$ (\emph{top}), $f_{\rm NL}=10^5$ (\emph{middle}), $f_{\rm NL}=10^6$ (\emph{bottom}). \emph{Left panels}: input $\mu$-map realizations. \emph{Right panels}: Constrained ILC $\mu$-map reconstructions. The case with $f_{\rm NL}=10^6$ is mainly shown as a demonstration that the method can recover the anisotropic $\mu$-distortion signal.}
\label{Fig:ilc}
\end{figure*}
%%%%%%%%%%%%%%%%%%%%%%%%%%%%%%%%%%%%%%%%%%%%%%%%%%%%%%

%%%%%%%%%%%%%%%%%%%%%%%%%%%%%%%%%%%%%%%%%%%%%%%%%%%%%%
\begin{figure*}
  \begin{center}
   \includegraphics[width=0.89\columnwidth]{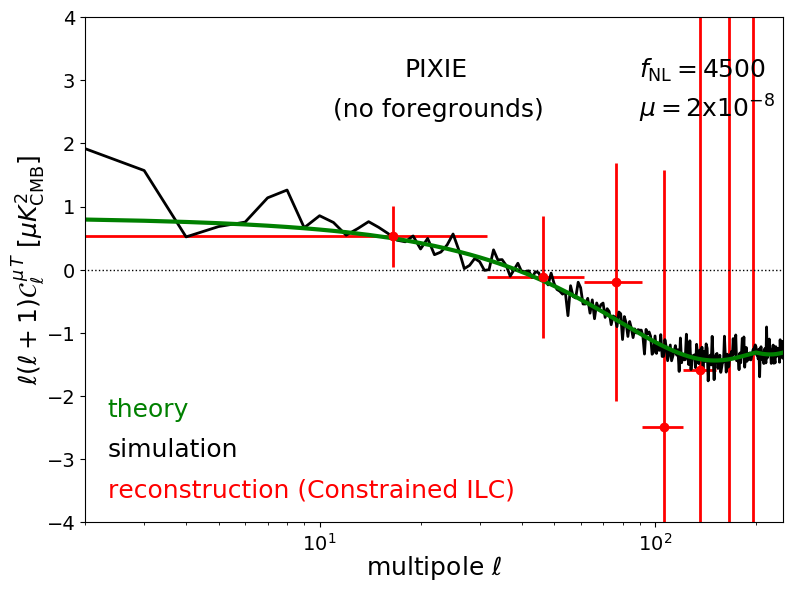}\hspace{4mm}
   \includegraphics[width=0.89\columnwidth]{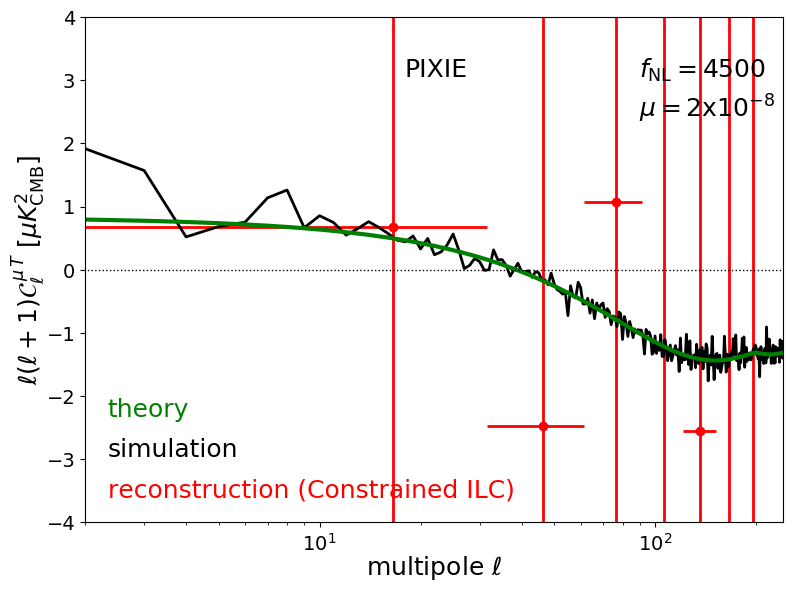}
   \\
   \includegraphics[width=0.89\columnwidth]{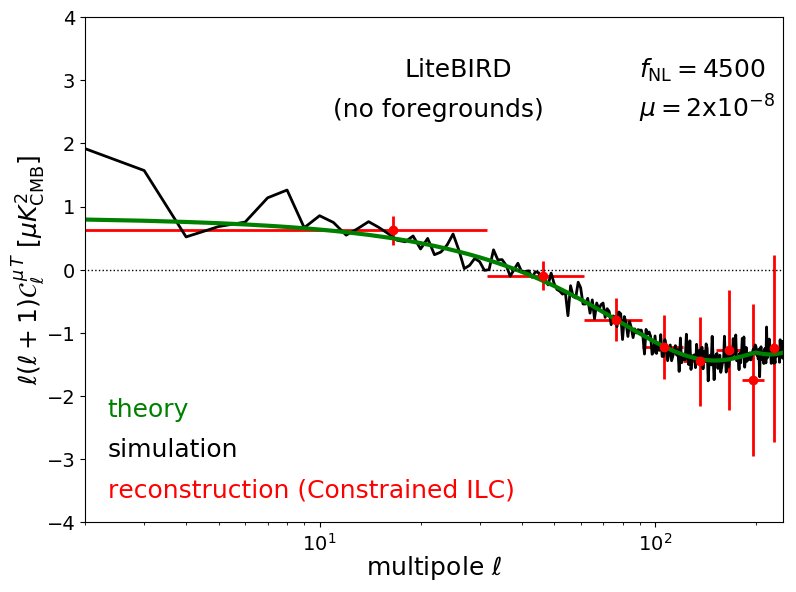}\hspace{4mm}
   \includegraphics[width=0.89\columnwidth]{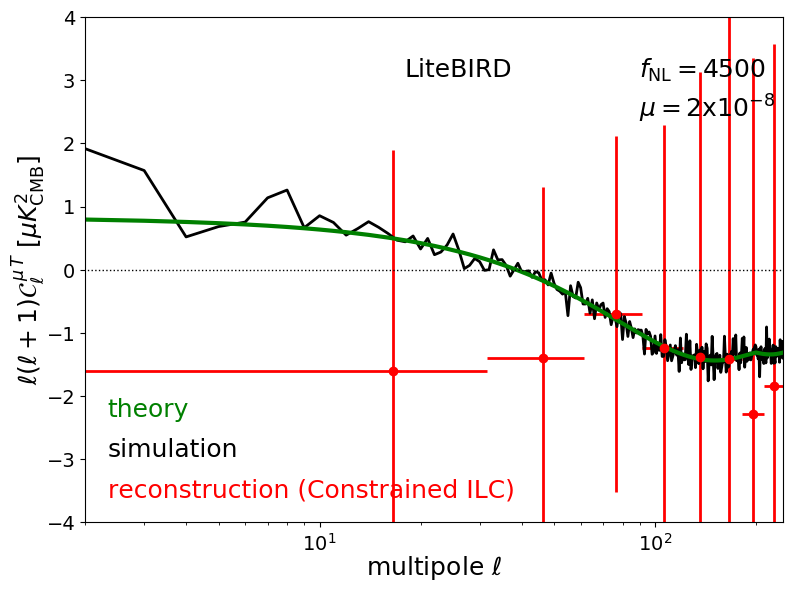}
   \\
   \includegraphics[width=0.89\columnwidth]{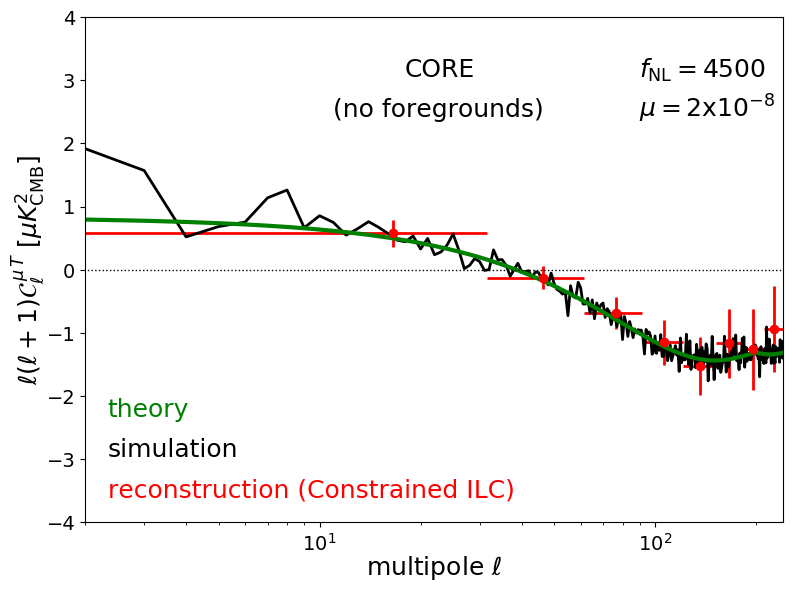}\hspace{4mm}
   \includegraphics[width=0.89\columnwidth]{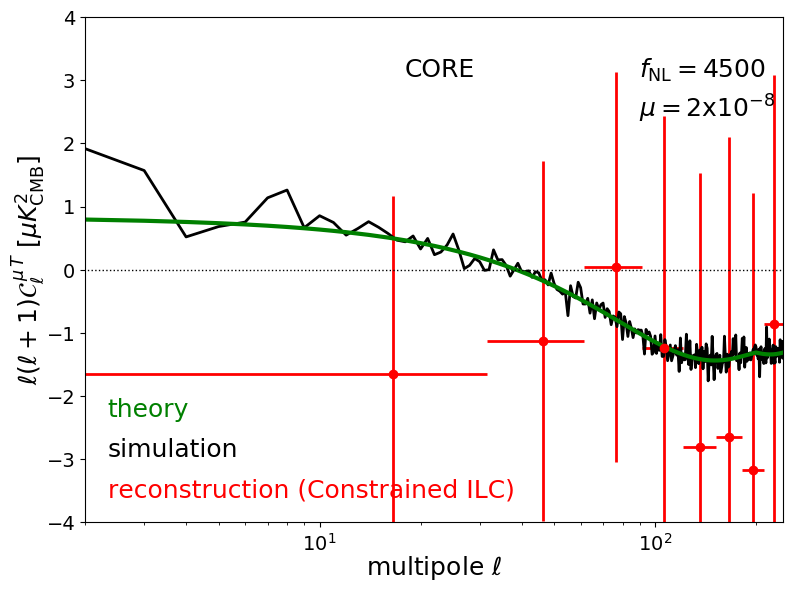}
   \\
   \includegraphics[width=0.89\columnwidth]{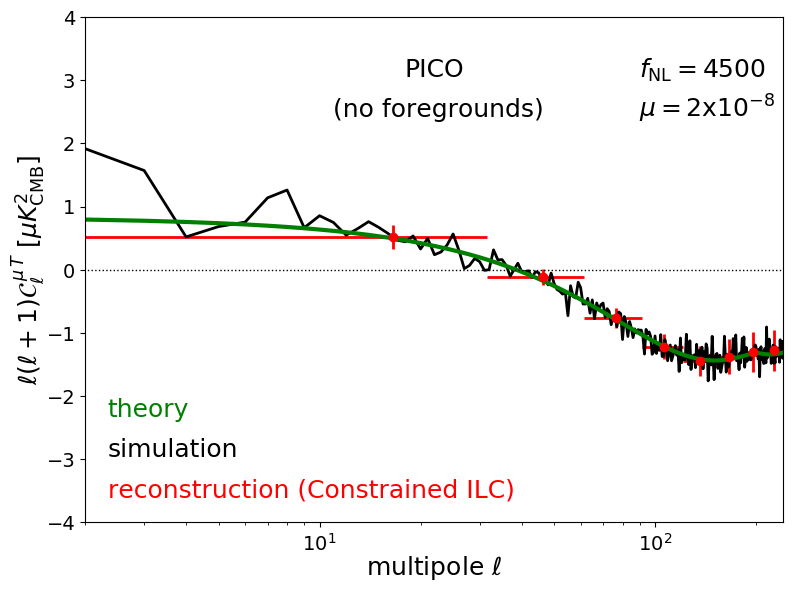}\hspace{4mm}
   \includegraphics[width=0.89\columnwidth]{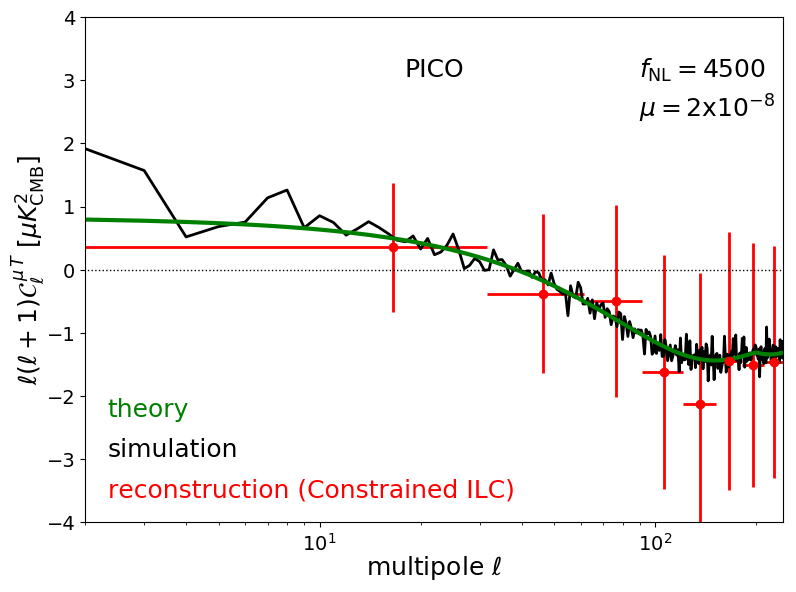}
  \end{center}
\caption{Measurement of the cross-power spectrum, $\widehat{C}_\ell^{\,\mu\times T}$, for $f_{\rm NL}=4500$ \emph{in the absence of foregrounds} (left panels) and {\it with foregrounds} (right panels) for \textit{PIXIE} (first row), \textit{LiteBIRD} (second row), \emph{CORE} (third row), and {\it PICO} (last row): theory (\emph{green}), input realization (\emph{black}),  Constrained ILC reconstruction (\emph{red}). The beam resolution adopted for the reconstruction is $96'$ for \textit{PIXIE}, $69'$ for \textit{LiteBIRD}, and $60'$ for \textit{CORE} and \textit{PICO}. }
\label{Fig:ilc_clx_fnl3_nofg_fg}
\end{figure*}
%%%%%%%%%%%%%%%%%%%%%%%%%%%%%%%%%%%%%%%%%%%%%%%%%%%%%%

\vspace{-3mm}
\subsection{Main results}
\label{subsec:results}
%%%%%%%%%%%%%%%%%%%%%%%%%%%%%%%%%%%%%%%%%%%%%%%%%%%%%%
The result of foreground removal and reconstruction of a \mbox{``$\mu$-free''} CMB temperature anisotropy map by the Constrained ILC method is shown in Fig.~\ref{Fig:ilc_cmb} for \emph{LiteBIRD}. We can clearly see that the method nicely recovers the input temperature sky map, as expected from the huge signal-to-noise in this case.

Similarly, Fig.~\ref{Fig:ilc} shows the reconstruction of the \mbox{``CMB-free''} map of $\mu$-distortion anisotropies for the specifications of \textit{LiteBIRD} (as an example), and for different values of the non-Gaussianity parameter: $f_{\rm NL}=10^4$ (top panels), $f_{\rm NL}=10^5$ (middle panels) and $f_{\rm NL}=10^6$ (bottom panels), the latter being chosen mainly for illustration. While guaranteed by the Constrained ILC method, the total absence of residual CMB temperature ($\leftrightarrow$ $\mu$-distortion) anisotropies in the reconstructed $\mu$-map ($\leftrightarrow$ CMB map) is not obvious by looking at the maps, but will be clearly established when considering the angular power spectra (see e.g. Sect.~\ref{subsec:vs}). We see from Fig.~\ref{Fig:ilc} how foreground removal for faint $\mu$-distortion anisotropies becomes challenging for decreasing values of $f_{\rm NL}$, despite the 15 frequency bands of \emph{LiteBIRD}. For instance, the $\mu$-map for $f_{\rm NL}=10^4$ is clearly still contaminated by strong residual foregrounds, while we obtain accurate reconstruction of the $\mu$-distortion anisotropies for $f_{\rm NL} \gtrsim 10^5$. However, even for $f_{\rm NL}=10^4$ the $\mu$-distortion anisotropies could still be detected by cross-correlation with CMB temperature anisotropies, as we will show below.

%%%%%%%%%%%%%%%%%%%%%%%%%%%%%%%%%%%%%%%%%%%%%%%%%%%%%%
\begin{figure*}
  \begin{center}
   \includegraphics[width=0.89\columnwidth]{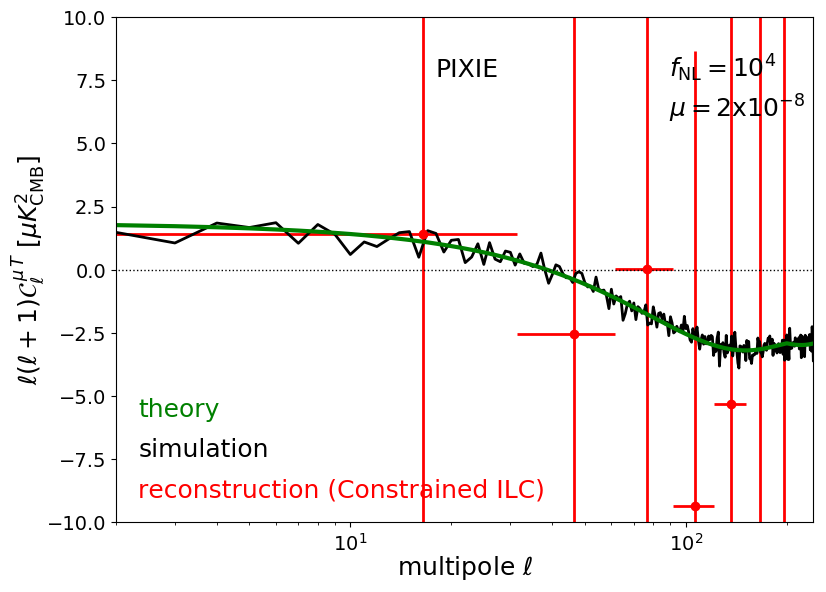}\hspace{4mm}
   \includegraphics[width=0.89\columnwidth]{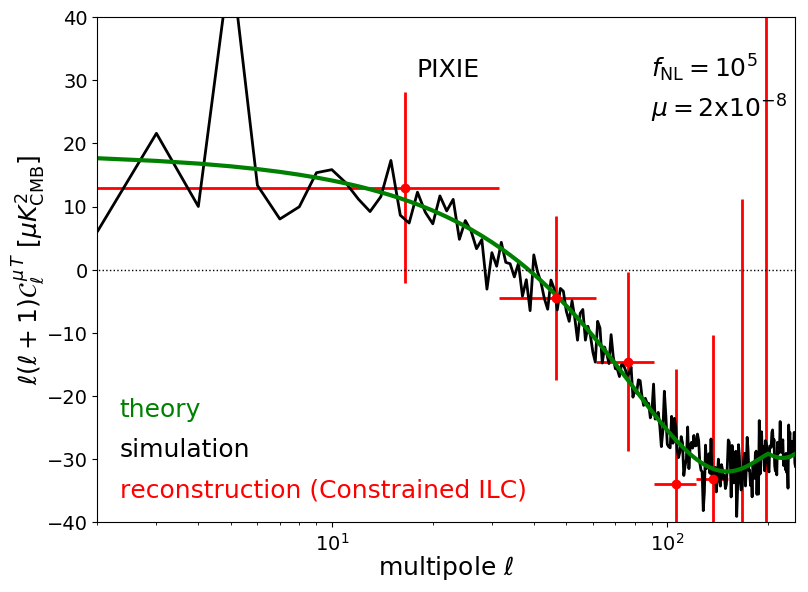}
   \\
   \includegraphics[width=0.89\columnwidth]{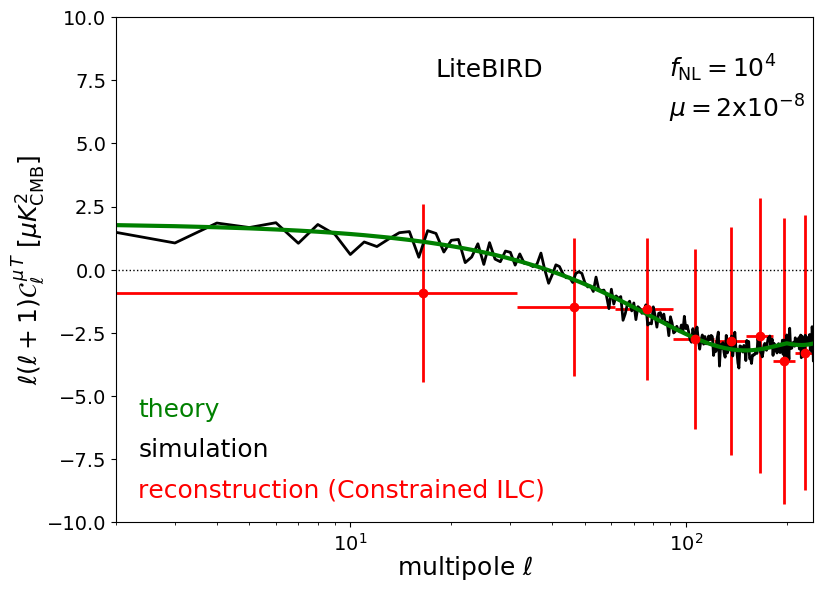}\hspace{4mm}
   \includegraphics[width=0.89\columnwidth]{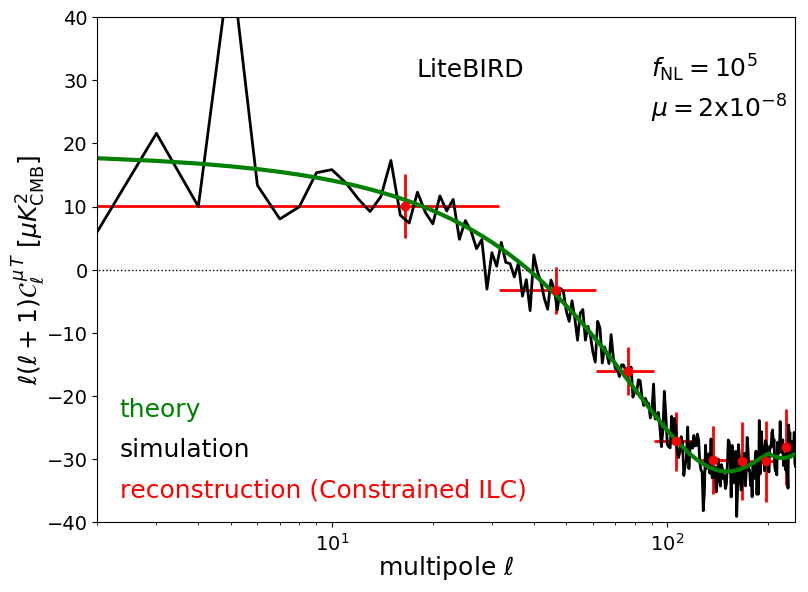} 
   \\
   \includegraphics[width=0.89\columnwidth]{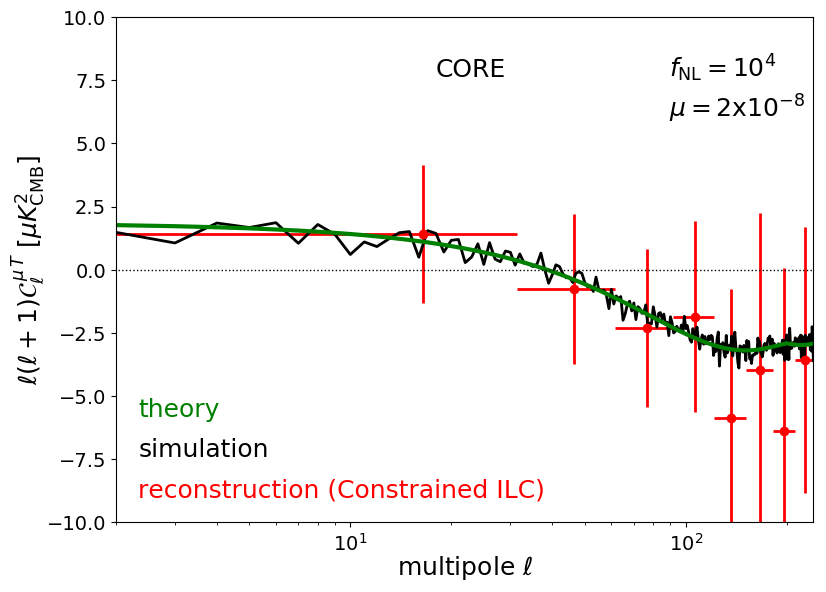}\hspace{4mm}
   \includegraphics[width=0.89\columnwidth]{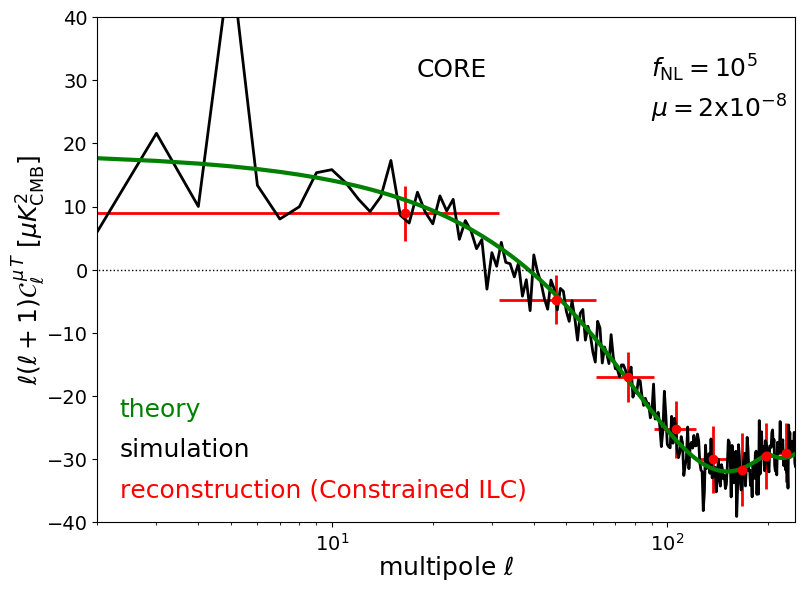}
   \\
   \includegraphics[width=0.89\columnwidth]{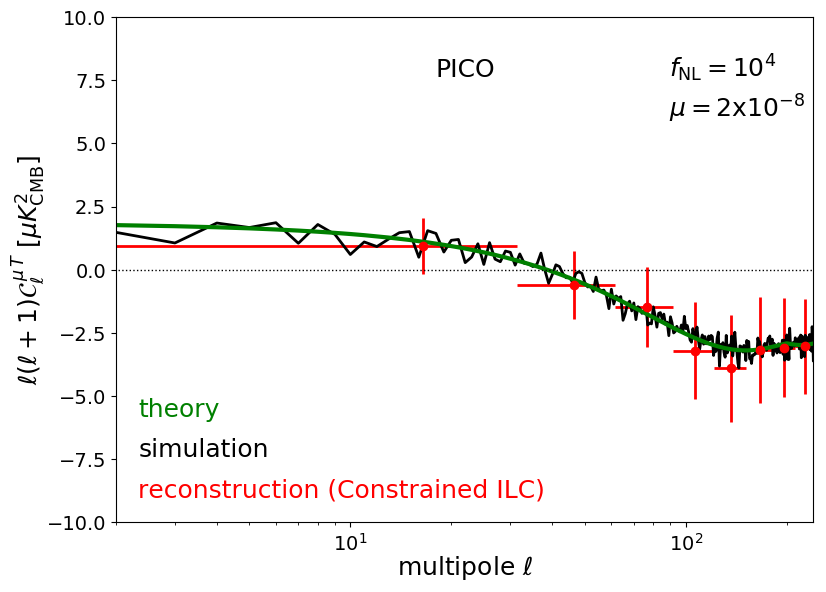}\hspace{4mm}
   \includegraphics[width=0.89\columnwidth]{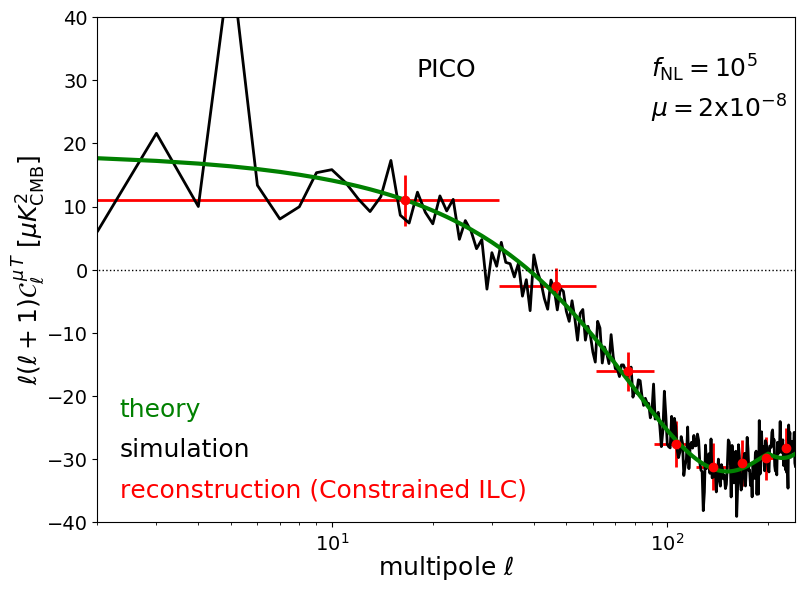}
  \end{center}
\caption{Similar to Fig.~\ref{Fig:ilc_clx_fnl3_nofg_fg} with foregrounds included and for $f_{\rm NL}=10^4$ (left panels) and $f_{\rm NL}=10^5$ (right panels).}
\label{Fig:ilc_clx_fnl4}
\end{figure*}
%%%%%%%%%%%%%%%%%%%%%%%%%%%%%%%%%%%%%%%%%%%%%%%%%%%%%%

As mentioned several times, one advantage of the Constrained ILC reconstruction is that the resulting $\mu$-distortion map is free from any residual CMB temperature anisotropies, and vice versa, so that we can measure the cross-power spectrum, $C_\ell^{\,\mu\times T}$, between $\mu$-distortion and CMB temperature anisotropies without suffering from spurious $T$-$T$ correlations arising from CMB residuals in the $\mu$ map, and thus obtain unbiased estimates of $f_{\rm NL}$. 
Figures~\ref{Fig:ilc_clx_fnl3_nofg_fg} and \ref{Fig:ilc_clx_fnl4} present our results from the calculation of the cross-power spectrum  between the reconstructed CMB and $\mu$-distortion maps (red points) after component separation with the Constrained ILC, for the different CMB experiments and different values of $f_{\rm NL}$. The multipole bin width is $\Delta\ell = 30$, and the uncertainty on $\widehat{C}_\ell^{\,\mu\times T}$ has been computed in each bin using \citep{Tristram2005}:
%---------------
\begin{align}
\label{eq:sigma}
 \sigma_\ell (\widehat{C}_\ell^{\,\mu\times T}) = \sqrt{ {\widehat{C}_\ell^{\,\mu \mu}\,\widehat{C}_\ell^{\,TT} + \left(\widehat{C}_\ell^{\,\mu\times T}\right)^2 \over (2\ell+1)f_{\rm sky}} },
\end{align}
%---------------
therefore including residual foregrounds leaking in the power spectra of the reconstructed CMB temperature and $\mu$-distortion maps. The fraction of the sky used for component separation is ${f_{\rm sky}=0.66}$. The reconstructed $\widehat{C}_\ell^{\,\mu\times T}$ can be compared to the cross-power spectrum of the input CMB and $\mu$ realizations of the simulation (black lines) and and the theory $C_\ell^{\,\mu\times T}$ (green lines).

%%%%%%%%%%%%%%%%%%%%%%%%%%%%%%%%%%%%%%%%%%%%%%%%%%%%%%
\begin{table*}
\caption{Detection forecasts on $f_{\rm NL}(k\simeq 740\,\textrm{Mpc}^{-1})$ after component separation, based on multipoles $2 \leq \ell \leq 200$.}
  \label{tab:fnl}
 \centering
  \begin{tabular}{lcccc}
\hline\hline
$f_{\rm NL}$ (fiducial)   & $10^5$ & $10^4$ & $4500$ & $4500$ \\
                        &        &        &        & w/o foregrounds \\
\hline 
\emph{PIXIE}             & $(1.11 \pm 0.40)\times 10^5$   & $(2.17 \pm 3.90)\times 10^4$   & $(1.5\pm 3.9)\times 10^4$  &  $4778 \pm 3868$                \\
                        & $2.5\sigma$  & --  & -- &  $1.2\sigma$ \\
\emph{LiteBIRD}         &  $(0.98 \pm 0.08)\times 10^5$  &  $(0.91 \pm 0.68)\times 10^4$ & $4272 \pm 6788$ &  $4753 \pm 930$               \\
                        & $12.5\sigma$  & $1.5\sigma$  & -- & $4.8\sigma$ \\
\emph{CORE}             &  $(0.97 \pm 0.08)\times 10^5$  &  $(1.35 \pm 0.74)\times 10^4$ & $5692 \pm 6397$ &  $4336 \pm 653$               \\
                        & $12.5\sigma$  & $1.4\sigma$  & -- & $6.9\sigma$ \\
\emph{PICO}             &  $(0.99 \pm 0.06)\times 10^5$  &  $(1.07 \pm 0.30)\times 10^4$ & $5094 \pm 2929$ &  $4480 \pm 371$               \\
                        & $17.8\sigma$  & $3.3\sigma$  & $1.5\sigma$ & $12.1\sigma$ \\
\hline
   \end{tabular}
\end{table*}
%%%%%%%%%%%%%%%%%%%%%%%%%%%%%%%%%%%%%%%%%%%%%%%%%%%%%%

From the reconstructed $\mu$-$T$ cross-power spectrum after component separation, we can derive Fisher forecasts on the primordial non-Gaussianity parameter, $f_{\rm NL}$ in the presence of foregrounds. The 1$\sigma$ uncertainty on $f_{\rm NL}$ is computed using the Fisher information as:
%----------------------
\begin{align}
\sigma\left( f_{\rm NL} \right)= \left[ \sum_{\ell} \left({1\over \sigma_\ell\left(\widehat{C}_\ell^{\,\mu\times T}\right)}\,{\partial C_\ell^{\,\mu\times T} \over \partial f_{\rm NL}} \right)^2\, \right]^{-1/2}\,,
\end{align}
%----------------------
where $\sigma_\ell\left(\widehat{C}_\ell^{\,\mu\times T}\right)$ are the $1\sigma$ error bars on the reconstructed cross-power spectrum (assuming a diagonal covariance matrix across multipole bins), which include foreground residuals from component separation, and $C_\ell^{\,\mu\times T}(f_{\rm NL})$ is the theoretical cross-power spectrum from Eq.~\eqref{eq:theory}, which is linear in the parameter $f_{\rm NL}$.

In addition to the uncertainty limit, $\sigma\left(f_{\rm NL}\right)$, we quantify the bias on $f_{\rm NL}$ due to foreground residuals by computing the maximum-likelihood estimate:
%----------------------
\begin{align}
\widehat{f}_{\rm NL} = \frac{\sum_{\ell} \widehat{C}_\ell^{\,\mu\times T} C_\ell^{\,\mu\times T}(f_{\rm NL}=1)/\sigma_\ell^2}{\sum_{\ell} \left[C_\ell^{\,\mu\times T}(f_{\rm NL}=1)\right]^2/\sigma_\ell^2},
\end{align}
%----------------------
where $C_\ell^{\,\mu\times T}(f_{\rm NL}=1) \equiv \partial C_\ell^{\,\mu\times T} / \partial f_{\rm NL}$ is the theoretical cross-power spectrum from Eq.~\eqref{eq:theory} for the fiducial value of ${f_{\rm NL} = 1}$. Our forecasts on $f_{\rm NL}(k\simeq 740\,\textrm{Mpc}^{-1})$ are summarized in Tables~\ref{tab:fnl} and \ref{tab:fnl2} for the four CMB satellite configurations. 

\vspace{-2mm}
\subsubsection{Results without foregrounds}
In \cite{Chluba_mu_2017}, the detection limit on $f_{\rm NL}$ from the measurement of anisotropic $\mu$-distortions by future CMB satellites have been estimated to be of order $f_{\rm NL} \lesssim 4500$ for $\langle\mu\rangle = 2\times 10^{-8}$ \emph{in the absence of foregrounds}. In order to verify this assertion, let us first consider \pixie, \litebird, \core, and \pico\ simulations \emph{without foregrounds}, therefore including only instrumental noise and correlated CMB temperature and $\mu$-distortion anisotropies in each frequency band, with fiducial values $f_{\rm NL} = 4500$ and $\langle\mu\rangle = 2\times 10^{-8}$. The left panels of Fig.~\ref{Fig:ilc_clx_fnl3_nofg_fg} show the reconstructed $\mu$-$T$ cross-power spectrum after component separation in this case. Indeed, using the Constrained ILC method, without foregrounds we do recover the scale-dependence and amplitude of the cross-power spectrum $C_\ell^{\,\mu\times T}$ for all CMB satellite experiments. Due to limited angular resolution, \pixie\ quickly looses leverage for modes at $\ell>10^2$, while these are still accessible for the other configurations.

The derived limits on $f_{\rm NL}$ in the absence of foregrounds are listed in the last column of Table~\ref{tab:fnl} when integrating over multipoles $2\leq \ell \leq 200$. The most sensitive experiment, \pico, could measure $f_{\rm NL}=4500$ with $12\sigma$ significance in the absence of foregrounds, while the least sensitive experiment, \pixie, would detect $f_{\rm NL}=4500$ at $1.2\sigma$, consistent with earlier estimations \citep{Chluba_mu_2017}. In the absence of foregrounds, \litebird\ and \core\ could detect $f_{\rm NL}=4500$ at $5\sigma$ and $7\sigma$ significance, respectively.

\vspace{-2mm}
\subsubsection{Results with foregrounds}
In the presence of foregrounds, the bias and uncertainty on the reconstructed $\mu$-distortion signal increase dramatically. The right panels of Fig.~\ref{Fig:ilc_clx_fnl3_nofg_fg} show the reconstructed $\mu$-$T$ cross-power spectrum after foreground cleaning for all CMB experiments assuming $f_{\rm NL}=4500$. In this case, the measurement of the $\mu$-$T$ correlation signal is of poorer quality at all angular scales, with a reconstructed $\widehat{C}_\ell^{\,\mu\times T}$ compatible with zero for \pixie, \litebird\ and \core. Conversely, due to a higher sensitivity and large number of frequencies, \pico\ is the only CMB experiment among those considered here which is able to detect a signal for $f_{\rm NL} = 4500$ in the presence of foregrounds (bottom right panel of Fig.~\ref{Fig:ilc_clx_fnl3_nofg_fg}), with a measurement of $f_{\rm NL} = 4500$ at $\simeq 2\sigma$ significance (third column of Tables \ref{tab:fnl} and \ref{tab:fnl2}). 

For increasing $f_{\rm NL} \gtrsim 10^4$ (Fig.~\ref{Fig:ilc_clx_fnl4}), \litebird, \core, and \pico\ clearly detect the $\mu$-$T$ correlation signal even after foreground cleaning with Constrained ILC. Because of increased sensitivity and number of frequency bands, \pico\ performs better than \core\ and \litebird\ in the presence of foregrounds. Typically, \pico\ could allow to constrain $f_{\rm NL}=10^4$ at $\simeq 3\sigma$ (fourth column of Table \ref{tab:fnl}) after foreground cleaning (third column of Table \ref{tab:fnl}), while \litebird\ and \core\ could detect $f_{\rm NL}=10^4$ at $\simeq 1.5 \sigma$. Conversely, mainly due to lower sensitivity, \pixie\ cannot detect $f_{\rm NL} \lesssim 10^4$ after foreground cleaning with the Constrained ILC. 

Lastly, if primordial non-Gaussianity is as large as $f_{\rm NL}=10^5$ on scales $k\simeq 740 \textrm{Mpc}^{-1}$, then the four CMB experiments would be able to measure the $\mu$-$T$ correlation signal (Fig.~\ref{Fig:ilc_clx_fnl4}), with \pixie\ detecting $f_{\rm NL}=10^5$ with $2.5\sigma$ significance, \litebird\ and \core\ with $\simeq 12\sigma$ significance, and \pico\ with $\simeq 18\sigma$ significance (second column of Table \ref{tab:fnl}). Again due to its lower angular resolution, \pixie\ looses modes at $\ell>10^2$ (see upper row of Fig.~\ref{Fig:ilc_clx_fnl4}).

It should be noted that the relative sensitivity between different CMB experiments in the presence of foregrounds depends on the foreground complexity assumed in the simulations. For much more complex foregrounds than in our simulations, \litebird, \core\, and \pico\ may face additional difficulties in capturing the full signal complexity for detecting $\mu$-distortions. In this case, spectrometer concepts like \pixie\ could make a difference, being able to provide additional information thanks to the large number of available frequency bands, though at lower overall channel sensitivity. Conversely, in the case of basic foregrounds, where the foreground subtraction can be controlled decently with $15$ to $20$ frequencies, it is the overall sensitivity that makes a difference. 

%%%%%%%%%%%%%%%%%%%%%%%%%%%%%%%%%%%%%%%%%%%%%%%%%%%%%%
\begin{figure}
  \begin{center}
    \includegraphics[width=0.96\columnwidth]{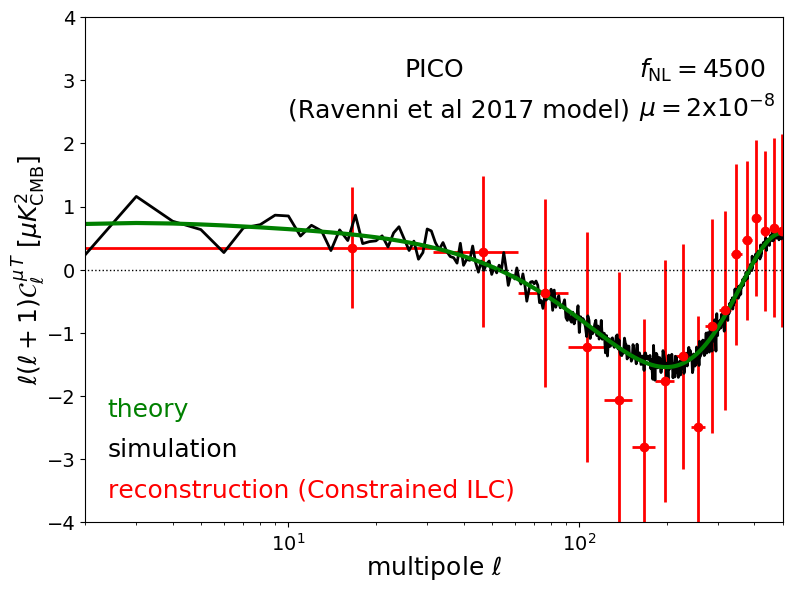}
    \\[1mm]
    \includegraphics[width=0.96\columnwidth]{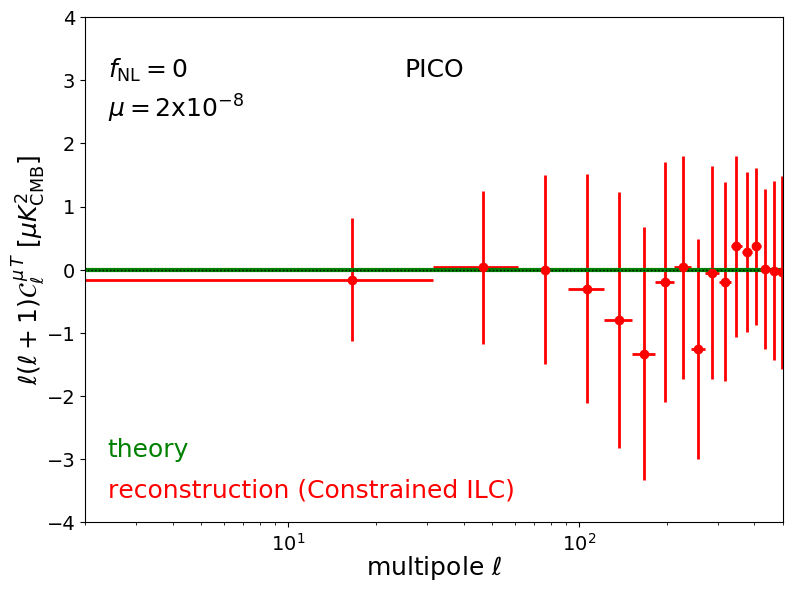}
    \\[1mm]
    \includegraphics[width=0.96\columnwidth]{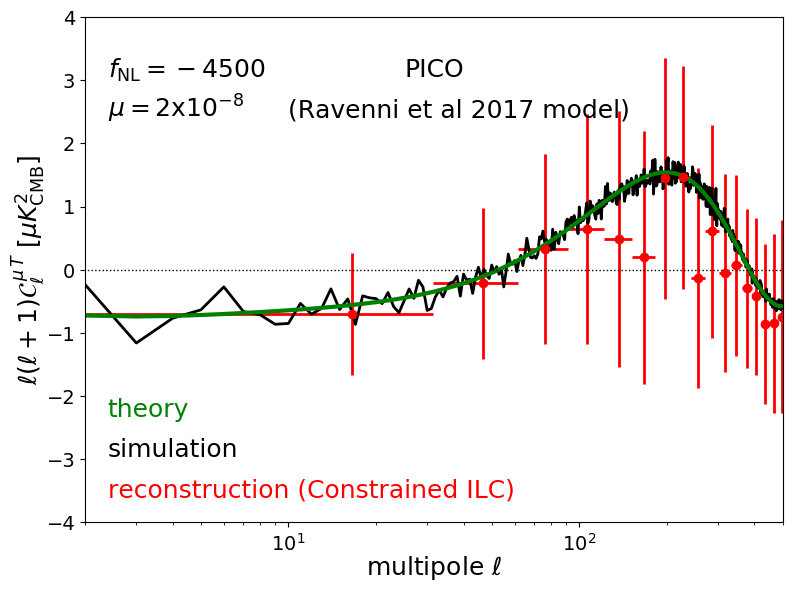}
  \end{center}
\caption{Simulated \textit{PICO} measurement of the cross-power spectrum, $\widehat{C}_\ell^{\,\mu\times T}$, for $f_{\rm NL}=\{4500, 0, -4500\}$ based on the full model of \protect\cite{Ravenni2017} after foreground cleaning with the Constrained ILC: theory (\emph{green}), input realization (\emph{black}),  Constrained ILC reconstruction (\emph{red}). The adopted beam resolution for the reconstruction is $41'$.}
\label{Fig:ravenni}
\end{figure}
%%%%%%%%%%%%%%%%%%%%%%%%%%%%%%%%%%%%%%%%%%%%%%%%%%%%%%

\subsection{Results on the updated model of $\boldsymbol{\mu$-$T}$ correlations}
\label{subsec:ravenni}
%%%%%%%%%%%%%%%%%%%%%%%%%%%%%%%%%%%%%%%%%%%%%%%%%%%%%%
As mentioned above, while this work was being completed, a more accurate modelling of the $\mu$-$T$ cross-power spectrum on a wider range of angular scales has been obtained by \cite{Ravenni2017}. In this new model, the crossing point at which $\mu$-distortion anisotropies anti-correlate with CMB temperature anisotropies is at a slightly smaller angular scale ($\ell\simeq 50$ versus $\ell\simeq 40$) compared to the model used in this work. At large scales, the overall amplitude of the signal is the same in the two models, however, the improved model allows one to include more modes beyond $\ell\simeq 200$, which we do find to improve the constraints at the level of $\simeq 20-30\%$.

To complete our analysis, we performed additional sky simulations in which the $\mu$-distortion map and the CMB map are simulated as correlated fields according to this improved model. In this simulation, the foregrounds and the noise remain unchanged with respect to the other simulations used in this work. We only considered the most challenging case investigated in this work, i.e. $f_{\rm NL}=4500$, and we adopt the instrumental configuration of \pico, for which we have obtained the best results throughout this work. 

%%%%%%%%%%%%%%%%%%%%%%%%%%%%%%%%%%%%%%%%%%%%%%%%%%%%%%
\begin{table}
\caption{Detection limits for \pico\ on $f_{\rm NL}(k\simeq 740\,\textrm{Mpc}^{-1})$ after component separation, based on the multipole range $2 \leq \ell \leq 500$ using the model of \citet{Ravenni2017} to describe the $\mu-T$ cross-correlation. Foregrounds are included in all cases and the fiducial $f_{\rm NL}$ parameter was varied.}
  \label{tab:fnl2}
 \centering
  \begin{tabular}{lccc}
  \hline\hline
$f_{\rm NL}$ (fiducial)   & $-4500$ & $0$ & $4500$  \\
\hline
\emph{PICO}             
&  $-2996 \pm 2112$  &  $1325 \pm 2114$ & $5698 \pm 2121$  \\
& $2\sigma$  & --  & $2\sigma$  
\\
\hline
   \end{tabular}
\end{table}
%%%%%%%%%%%%%%%%%%%%%%%%%%%%%%%%%%%%%%%%%%%%%%%%%%%%%%
Figure~\ref{Fig:ravenni} (upper panel) shows the reconstruction of $C_\ell^{\,\mu\times T}$ by the Constrained ILC for the \pico\ simulation based on the new input model. Clearly, the reconstruction is of similar quality as presented in Fig.~\ref{Fig:ilc_clx_fnl3_nofg_fg}, but extending to slightly smaller angular scales. The constraints for different fiducial values of $f_{\rm NL}$ are summarized in Table~\ref{tab:fnl2}. 
The resulting $1\sigma$ uncertainty of \pico\ for using the model of \cite{Ravenni2017} is $\sigma(f_{\rm NL}=4500) = 3059$, when including modes at $2\leq \ell \leq 200$. Comparing this to the previous constraint, i.e., $\sigma(f_{\rm NL}=4500) = 2929$ from Table~\ref{tab:fnl}, shows that the conclusion is not affected significantly. Also adding modes with $200\leq \ell \leq 500$, we find $\sigma(f_{\rm NL}=4500) = 2121$.  It highlights that the additional gain from modes at $\ell>200$ is about $\simeq 30\%$, improving the expected detection significance for $f_{\rm NL}=4500$ from $\simeq 1.5\sigma$ to $\simeq 2\sigma$ after foreground cleaning.

It has been shown by \cite{Ravenni2017} that correlations between $\mu$-distortions and CMB $E$-mode polarization anisotropies could add constraining power to $f_{\rm NL}$ measurements because of more available modes. However, this assertion has to be confronted to the presence of strongly polarized foregrounds, for which foreground cleaning for polarization might be more challenging than for intensity. This will be investigated in a future work.

\vspace{-3mm}
\subsubsection{Constraints on $f_{\rm NL} \leq 0$}
\label{subsec:fnl_neg}
%%%%%%%%%%%%%%%%%%%%%%%%%%%%%%%%%%%%%%%%%%%%%%%%%%%%%%
As mentioned above, we also checked the detection significance for $f_{\rm NL}<0$. For $f_{\rm NL}=-4500$, The reconstructed cross-power spectrum is shown in the lower panel of Fig.~\ref{Fig:ravenni}, and Table~\ref{tab:fnl2} provides the expected \pico\ constraint. Comparing the result with that for $f_{\rm NL}=4500$ shows that the reconstruction is fairly insensitive to the sign of $f_{\rm NL}$: the estimated errors are comparable and the biases are consistent with statistical fluctuations. Note that $f_{\rm NL}=-4500$ corresponds to $f^{\rm WMAP}_{\rm NL}=4500$ in the standard {\it WMAP} definition, so that our sign convention is not expected to affect the main conclusions significantly.

We also ran a simulation without including any $\mu$-signal in the simulations. The reconstruction is shown in the middle panel of Fig.~\ref{Fig:ravenni}, clearly being consistent with $f_{\rm NL}=0$ at all $\ell$. The obtained constraint is $|f_{\rm NL}|<2114$ (see Table~\ref{tab:fnl2}), which is consistent with the $1\sigma$ error in the simulations with non-zero signal, $|f_{\rm NL}|=4500$.

\vspace{-3mm}
%%%%%%%%%%%%%%%%%%%%%%%%%%%%%%%%%%%%%%%%%%%%%%%%%%%%%%
\section{Discussion}
\label{sec:discussion}
%%%%%%%%%%%%%%%%%%%%%%%%%%%%%%%%%%%%%%%%%%%%%%%%%%%%%%

\subsection{Standard ILC versus Constrained ILC}
\label{subsec:vs}
%%%%%%%%%%%%%%%%%%%%%%%%%%%%%%%%%%%%%%%%%%%%%%%%%%%%%%
As mentioned earlier, most of the component separation techniques aim at minimizing the variance of the global foreground contamination in the reconstructed signal, e.g., as for standard ILC methods like {\tt NILC} \citep{Delabrouille2009}. However, in the context of reconstructing the $\mu$-$T$ correlation signal, CMB temperature anisotropies are a main foreground to the anisotropic $\mu$-distortion signal, and this CMB foreground could in addition be spatially correlated to the $\mu$-distortion signal due to primordial non-Gaussianity, so that residuals of CMB temperature anisotropies in the reconstructed $\mu$-distortion map after component separation will add residual $T$-$T$ correlations in the $\mu$-$T$ correlation measurement. Even if CMB temperature anisotropies are minimized in the $\mu$-distortion map by standard component separation techniques, the spurious $T$-$T$ correlation can be large enough to bias the measurement of the $\mu$-$T$ correlation signal. It is therefore essential to \emph{eliminate} CMB temperature anisotropies in the component separation process rather than minimizing the global foreground contamination. This strong constraint is made possible by the Constrained ILC method \citep{Remazeilles2011} through the knowledge of the SED of both CMB temperature and $\mu$-distortion anisotropies.

\begin{figure}
  \begin{center}
    \includegraphics[width=\columnwidth]{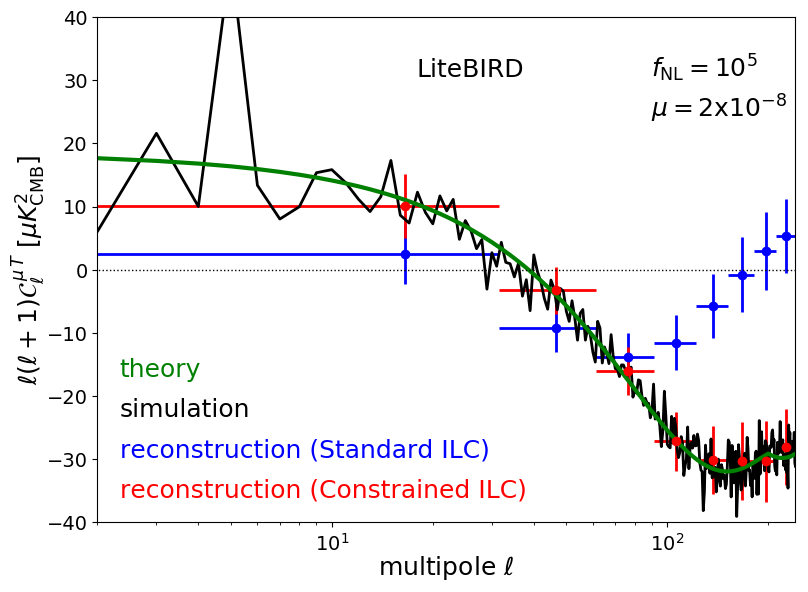} 
  \end{center}
\caption{Standard ILC (\emph{blue}) versus Constrained ILC (\emph{red}), on the example of \litebird\ and $f_{\rm NL}=10^5$. The constraint of orthogonality of the Constrained ILC weights to the SED of CMB anisotropies guarantees the absence of any CMB contamination in the reconstructed $\mu$-distortion map. The Constrained ILC approach is essential to avoid spurious/unphysical $\mu$-$T$ correlations that are inherent to standard component separation methods because of residual CMB contamination in the reconstructed signal. }
\label{Fig:std_vs_cst}
\end{figure}

To illustrate our point, in Fig.~\ref{Fig:std_vs_cst} we compare the reconstruction of the cross-power spectrum $\widehat{C}_\ell^{\,\mu\times T}$ between the standard ILC method (\emph{blue}) and the Constrained ILC method (\emph{red}), in the case of \litebird\ for $f_{\rm NL} = 10^5$. Without the extra orthogonality condition on the CMB SED, the standard ILC reconstruction of $\widehat{C}_\ell^{\,\mu\times T}$ shows a significant bias at all angular scales, due to spurious $T$-$T$ correlations between the CMB temperature anisotropies and the residuals of CMB temperature anisotropies in the standard ILC $\mu$-distortion map. In contrast, we see how the additional constraint of orthogonality to the CMB SED in the Constrained ILC approach is essential for recovering an unbiased $\mu$-$T$ correlation signal on all angular scales.  
Without explicitly nulling the CMB temperature contribution, we obtain $f_{\rm NL}=\left(0.27\pm 0.08\right)\times 10^5$, which is noticeably biased ($\simeq -9\sigma$) with respect to the fiducial value $f_{\rm NL}=10^5$, while with the Constrained ILC method the bias is consistent with statistical fluctuations (compare Table~\ref{tab:fnl}). We thus recommend the use of \emph{constrained} approaches, like the Constrained ILC component separation method proposed in this work, to achieve extraction of anisotropic $\mu$-distortions.

Here, we briefly mention the analysis performed by \cite{Khatri2015} using \emph{Planck} data. There, a standard parametric fitting method was applied to construct a map of $\mu$-distortion anisotropies. No orthogonality condition for the CMB temperature was added, which does seem problematic in light of our considerations. A detailed assessment of this issue is beyond the scope of this paper, however, we plan to reanalyze the \emph{Planck} data using a Constrained ILC method to obtain an unbiased $\mu$-distortion map, which at the sensitivity of \emph{Planck} is expected to be consistent with noise. For this analysis, the role of the integrated SW effect from clusters as well as possible primordial $y-T$ correlations also have to be carefully addressed. 

%%%%%%%%%%%%%%%%%%%%%%%%%%%%%%%%%%%%%%%%%%%%%%%%%%%%%%
%%%%%%%%%%%%%%%%%%%%%%%%%%%%%%%%%%%%%%%%%%%%%%%%%%%%%%

\subsection{Averaging effects}
\label{subsec:averaging}
%%%%%%%%%%%%%%%%%%%%%%%%%%%%%%%%%%%%%%%%%%%%%%%%%%%%%%
The extraction of ever fainter cosmological signals, such as $\mu$-type spectral distortions or primordial CMB $B$-modes, will be particularly sensitive to shortcomings of the foreground modelling \citep[see e.g.][]{Remazeilles2016}. 
Even in the ideal (but unrealistic) situation, in which we know the SEDs of all foregrounds perfectly and use the exact same model in a parametric fitting approach, still some biases on the reconstructed cosmological signal may occur because of averaging effects. Recently, \citet{Chluba2017foregrounds} demonstrated that the actual SED of the foregrounds \emph{in the maps} will differ from the physical SED of the foregrounds \emph{on the sky} because of finite-size pixelization and beam convolution, which both lead to an average of the spectral parameters from different lines-of-sight. Due to averaging, say different power-law SEDs (with varying spectral indices for different lines-of-sight), the effective SED within a pixel is no longer described by a power-law. Similarly, the effect of averaging along one line-of-sight is unavoidable.

Therefore, the effective foreground SED in each pixel differs from the physical SED in each line-of-sight. As shown by \cite{Remazeilles2017}, the spurious curvature created by averaging effects, if ignored in the parametric modelling, is significant enough to bias the reconstruction of the faint primordial CMB $B$-mode signal at a level of $r \simeq 10^{-3}$, which is the sensitivity target of current CMB satellite concepts. Thus, due to the very large dynamic range between the foregrounds and the signals ($\mu$-distortions and CMB $B$-modes), the impact of foreground mismodelling can no longer be ignored. Beyond simple pixel/beam averaging effects, the spherical harmonic transforms that are performed on the maps by most component separation algorithms effectively also is a weighted average of different SEDs \citep{Chluba2017foregrounds}, thus adding even further complications on foreground cleaning.

With these subtle averaging effects in mind, it is useful to implement blind component separation approaches, such as the Constrained ILC, in which there is no parametrization of the foreground SEDs in the process, thus making it less susceptible to mismodelling. However, for a detection of faint cosmological signals such as $\mu$-type distortion and primordial B-modes one will always want to compare and/or combine independent component separation methods, both parametric and blind, to ensure the robustness of the obtained result.

\vspace{-3mm}
\subsection{Optimization: more detectors or more frequencies?}
\label{subsec:sensitivity}
%%%%%%%%%%%%%%%%%%%%%%%%%%%%%%%%%%%%%%%%%%%%%%%%%%%%%%
In this section, we open a discussion on the optimization of the configuration of a CMB experiment to improve the quality of the $C_\ell^{\,\mu\times T}$ measurement in the presence of foregrounds. The main question we address is: \emph{do we need more frequencies or more detectors ($\leftrightarrow$ sensitivity)?} A similar optimization is also highly relevant to ongoing and planned $B$-mode searches.
In this regard, we considered a ``super-\litebird'' experiment having the same frequency range and distribution ($40$--$402$\,GHz) as \litebird\, but 100 times more detectors per frequency, therefore an overall sensitivity increased by a factor 10 (i.e. $< 0.2$\,$\mu$K.arcmin). While applied here to spectral distortions, the conclusions are instructive for $B$-mode searches.

\begin{figure}
  \begin{center}
    \includegraphics[width=\columnwidth]{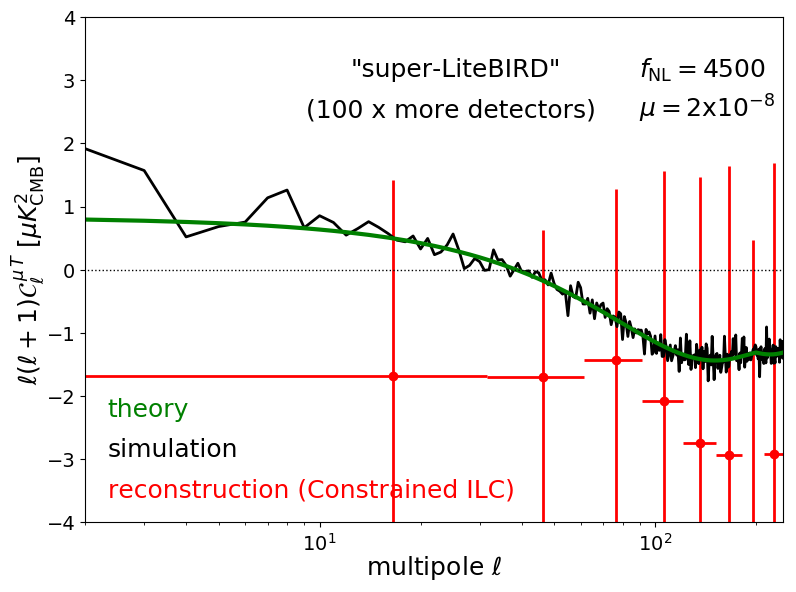}~
  \end{center}
\caption{The reconstructed $C_\ell^{\,\mu\times T}$ for $f_{\rm NL} \simeq 4500$ assuming a {\it super}-\litebird\ concept with 100 times more detectors.}
\label{Fig:detectors}
\end{figure}

In Fig.~\ref{Fig:detectors}, we show the reconstructed cross-power spectrum $C_\ell^{\,\mu\times T}$ for $f_{\rm NL} \simeq 4500$, with a ``super-configuration'' of \litebird. When increasing the number of detectors by a factor of 100, the uncertainties on $C_\ell^{\,\mu\times T}$ are slightly reduced by about 8\% when compared to the baseline configuration presented in Fig.~\ref{Fig:ilc_clx_fnl3_nofg_fg}. Nevertheless the reconstructed signal is still consistent with zero, even for this large boost in sensitivity. This suggests that astrophysical foregrounds, not the instrumental noise, set the ultimate limit to which the $\mu$-$T$ correlation signal can be measured.
With more frequencies ($21$--$800$\,GHz) but lower overall sensitivity ($0.8$\,$\mu$K.arcmin), \pico\ (bottom right panel of Fig.~\ref{Fig:ilc_clx_fnl3_nofg_fg}) actually obtains a higher level of detection of the $\mu$-distortion signal than a \textit{super-LiteBIRD} concept. This indeed demonstrates that the uncertainty after foregrounds cleaning by the Constrained ILC will decrease more significantly if we increase the number of frequency bands and the frequency range rather than the number of detectors. 

We note that \pixie\ has the largest number of frequency bands among all the proposed CMB experiments, nevertheless \pixie\ results are of poorer quality compared to \core, \litebird, and \pico. One reason is that the overall sensitivity of \pixie\ is significantly lower (e.g., more than 4 times lower than \core). This results in a degradation of the ILC foreground cleaning due to the trade-off between minimizing the variance of foregrounds and the variance of noise. 
However, it is also the lack of angular resolution that diminishes the constraining power of \pixie\ even at large angular scales. Considering a simulation where \pixie\ has $40'$ resolution (instead of $96'$) in each frequency band, while keeping the same sensitivity ($4.7$\,$\mu$K.arcmin) and frequency coverage, we find that the constraint on $f_{\rm NL}=10^4$ from multipoles $2\leq\ell\leq 200$ is improved by more than $50\%$, leading to $\sigma\left(f_{\rm NL}=10^4\right) = 1.78\times 10^4$ (compare with Table~\ref{tab:fnl}). \pixie\ would thus benefit from higher resolution in each spectral band to allow the use of the spatially-correlated information during component separation and access valuable additional modes at $\ell\simeq 100-200$.

%%%%%%%%%%%%%%%%%%%%%%%%%%%%%%%%%%%%%%%%%%%%%%%%%%%%%%
%%%%%%%%%%%%%%%%%%%%%%%%%%%%%%%%%%%%%%%%%%%%%%%%%%%%%%

Last, we are interested in knowing which part of the frequency range (low- or high-frequencies) is the most important for the reconstruction of the $\mu$-distortion signal by the Constrained ILC. We thus investigate two descoped configurations of \pico, one without low-frequency channels (i.e. 17 frequencies from $43$ to $800$\,GHz) and the other one without high-frequency channels (i.e. 18 frequencies from $21$ to $460$\,GHz). We find that, in the absence of frequencies above $460$ GHz the detection of $f_{\rm NL}=4500$ by \pico\ would be degraded by about 7\%, while in the absence of frequencies below $43$ GHz, the detection would be degraded by about 30\%, lowering the detection of $f_{\rm NL}=4500$ to $1\sigma$ significance ($2\leq \ell\leq 200$).  
Therefore, for the given sensitivities of \pico\ we conclude that frequencies below $43$\, GHz are more essential than frequencies above $460$\, GHz for a detection of $\mu$-$T$ correlation signals. 
 
\subsection{Calibration uncertainties and the imperfect knowledge of the CMB monopole temperature}
\label{subsec:calibration}

Although future CMB imagers like \pico\ are in a good position to detect $\mu$-distortion anisotropies for $f_{\rm NL}\gtrsim 4500$, the interpretation has to rely on knowledge of the average sky spectrum and its distortion. Indeed, the $\mu$-$T$ correlation signal is proportional to the product $f_{\rm NL}\langle\mu\rangle$, causing a direct degeneracy between $f_{\rm NL}$ and the monopole distortion $\langle\mu\rangle$ that can only be broken through absolute measurements. In this respect, Fourier transform spectrometers like \pixie\ might be essential for disentangling different effects \citep[see also][]{Chluba_mu_2017}. 

In addition, relative inter-channel calibration errors may impact the component separation. It has been shown by \cite{Dick2010} that 1\% calibration errors on the CMB spectrum in \Planck\ channels can dramatically degrade the reconstruction of CMB temperature anisotropies by an ILC method in the high signal-to-noise regime, because of the mismatch between the actual CMB SED in the miscalibrated experiment and the assumed CMB SED in the ILC weights. For current CMB temperature measurements (0.1\% calibration errors for \Planck), this seems to be sufficiently under control \citep{Dick2010},
 but for the large dynamic range encountered with $\mu$-distortion signals additional studies are required. 
 
We therefore ran several simulations with different levels of calibration uncertainties (covering $\simeq 0.01\%-1\%$) and different levels of non-Gaussianity ($f_{\rm NL}=10^5$ and $f_{\rm NL}=4500$). For calibration errors $\gtrsim$ 0.1\% in \pico\ channels, we find a significant bias/loss on the reconstruction of the $\mu$-$T$ correlated signal, even for a large $f_{\rm NL}=10^5$ value, because of the partial erasing by the ILC of the variance of the CMB temperature map in the high signal-to-noise regime \citep{Dick2010}. The sensitivity of future CMB satellite experiments like \pico\ is about an order of magnitude larger than the \emph{Planck} sensitivity, therefore increasing the high signal-to-noise regime for CMB temperature anisotropies, which makes the CMB $T$ signal reconstruction with \pico\ more sensitive to miscalibration than with \Planck. We thus find that the allowed calibration uncertainty for \pico\ has to be $\lesssim$ 0.01\%, which is the typical level of precision expected for \core\ \citep[e.g.][]{Burigana2017}. The conclusion holds also for lower $f_{\rm NL}=4500$ values, for which we find that 0.01\% calibration errors guarantee the full recovery of the CMB $T$ signal, while only slightly impacting the reconstruction of the $\mu$ signal, with less than $0.6\sigma$ bias on $f_{\rm NL}=4500$, i.e. an error of $\Delta f_{\rm NL} \simeq 10^3$. This is also found by extrapolation from earlier estimate by \cite{Ganc2012}.

The absolute error in the CMB monopole temperature, $T_{\rm CMB}$, which currently is known to $\Delta T_{\rm CMB}\simeq 1\,{\rm mK}$ \citep{Fixsen1996, Fixsen2009}, will furthermore result in an error of the differential SED of CMB temperature anisotropies and therefore could also impact the component separation processes. For the $\mu$-distortion SED, $\Delta T_{\rm CMB}$ implies a tiny relative error ($\lesssim 0.1\%$) on a small signal, which can thus be neglected. However, the much larger CMB temperature anisotropies, will lead to an extra term $\Delta a_{{\rm T},i}\,s_{\rm CMB} (\hat{\mathbf{\theta}})$ in Eq.~\eqref{eq:obs} that could cause spurious (correlated) residuals in the $\mu$-distortion map with the Constrained ILC method. This is because the assumed CMB temperature SED now has a new ($y$-type) frequency dependence, $\Delta a_{{\rm T},i}\simeq [x\coth(x/2)-2]\,\Delta T_{\rm CMB}/T_{\rm CMB}$ with $x=h\nu/kT_{\rm CMB}$, which is not fully eliminated by the constraint, Eq.~\eqref{eq:cmbcon}. For enhanced $\mu$-anisotropies studied here, we estimate this to matter at the level of $\Delta f_{\rm NL} \simeq {\rm few}\times 10^2$.

While this level of precision for $f_{\rm NL}$ from $\mu-T$ correlations is still futuristic, the current uncertainty in the CMB monopole could become a serious challenge for recently discussed methods that are meant to use differential measurements in frequency (without requiring an absolute measure of the CMB monopole flux) to extract average CMB distortion signals \citep{Mukherjee2018}. The CMB monopole temperature could be measured to $\Delta T_{\rm CMB}\simeq \text{few}\times\,{\rm nK}$ precision with a spectrometer like \pixie\ \citep{abitbol_pixie}, which would certainly eliminate this potential problem. For similar reasons could a combination of absolutely calibrated maps (using a \pixie-like spectrometer) with future CMB imagers allow us to greatly reduce calibration uncertainties, potentially opening a way to access small spectral-spatial signals, e.g., caused by resonance scattering terms \citep{Kaustuv2004}.

\vspace{-4mm}
\section{Conclusions}
\label{sec:conc}

Using sky simulations with foregrounds, we demonstrated that future CMB satellite experiments could be able to achieve a detection of $\mu$-distortion anisotropies caused by primordial non-Gaussianity by cross-correlating with CMB temperature anisotropies. In particular, the CMB satellite concept \pico, with its broad frequency range and high sensitivity for accurate foreground removal, among the considered satellite concepts is in the best position to detect primordial $\mu$-$T$ correlations for $|f_{\rm NL}(k\simeq 740\,{\rm Mpc}^{-1})| \gtrsim 4500$. While this is far off the limits obtained with CMB measurements at scales $k\simeq 10^{-2}\,{\rm Mpc}^{-1}$, we emphasize again that this still provides interesting new constraints on the scale-dependence of non-Gaussianity \citep[see also][]{Biagetti2013, Emami2015, Ravenni2017}. Our forecasts on the primordial non-Gaussianity parameter $f_{\rm NL}$ for the four different CMB satellite concepts: \pixie, \litebird, \core\ and \pico\ are given in Tables~\ref{tab:fnl} and \ref{tab:fnl2}. This work thus provides the first realistic predictions for the detection of $\mu$-distortion anisotropies in the presence of foregrounds, showing that \pico\ could place an upper limit $|f_{\rm NL}(k\simeq 740\,{\rm Mpc}^{-1})|< 4200$ ($95\%$ c.l.) on local-type non-Gaussianity in the ultra-squeezed limit.

We demonstrated the necessity of nulling CMB temperature anisotropies in the reconstruction of the $\mu$-distortion anisotropies during the component separation process in order to avoid biasing the measurement of the $\mu$-$T$ correlation signal by residual $T$-$T$ correlations (Sect.~\ref{subsec:vs}). In this regard, we proposed a tailored component separation method, the Constrained ILC, to reconstruct (CMB-free) $\mu$-distortion maps. 
Our simulations did not include additional information from (primordial) $y-T$ and distortion-polarization correlations, which have been shown to improve (foreground-free) detection limits significantly \citep[e.g.,][]{Ravenni2017}. Thus, the limits derived here could still improve noticeably. However, when adding information from polarization new foreground challenges do appear, such that it is hard to anticipate the outcome. We plan to generalize our method to take these effect into account and then answer this question in future work.

We argued that for future CMB imagers a broad frequency coverage and large number of bands (with the decent sensitivity) is preferred over hugely improved sensitivity in a limited frequency range when attempting to extract $\mu$-distortion anisotropies by the Constrained ILC method (see Sect.~\ref{subsec:sensitivity}).  In particular, coverage at {\it low frequencies} seems important, while for the simulations carried out here high-frequency channels can be sacrificed. 
This conclusion depends strongly on the type of the signal and the assumed complexity of foregrounds. For example, when comparing to $B$-mode searches, the (unpolarized) $\mu$-distortion signal has a spectral dependence with focus towards longer wavelength, enhancing the importance of low- over high-frequency channels. However, to obtain similar constraints when allowing for increased complexity of the dust model (e.g., caused by averaging processes or presence of CIB) might still require extended coverage at high frequencies. We note that, although not directly transferable to $B$-mode searches, our analysis is still quite instructive, exploring the CMB signal extraction in the (extremely) low signal-to-noise regime. 

Finally, we also highlighted the need for absolute measurement of the monopole distortion even for the measurement of anisotropies of $\mu$-distortions to break the parameter degeneracy $C^{\,\mu\times T}_\ell \propto f_{\rm NL}\langle\mu\rangle$ of the amplitude of the $\mu$-$T$ cross-power spectrum. Fourier-transform spectrometer concepts similar to \pixie\ would be very useful in this respect. Similarly, the channel inter-calibration and possible errors from imperfect knowledge of the CMB monopole temperature have to be carefully considered to safely extract possible spectral-spatial distortion signals from the very early phases of our Universe (see Sect.~\ref{subsec:calibration}).

\small

%%%%%%%%%%%%%%%%%%%%%%%%%%%%%%%%%%%%%%%%%%%%%%%%%%%%
%%%
%%% Acknowledgements
%%%
\section*{Acknowledgements}

We cordially thank Andrea Ravenni for useful discussions on the modelling of anisotropic $\mu$-and $y$-type spectral distortions and for providing us with his $\mu-T$ cross-power spectra. 
We also thank Eiichiro Komatsu and Enrico Pajer for their useful comments on the manuscript, and the referee for their suggestions that helped improve the paper.
Some of the results in this paper have been derived using the \healpix\ package \citep{Gorski2005}. We also acknowledge the use of the PSM package \citep{Delabrouille2013}, developed by the \Planck\ working group on component separation, for making the simulations used in this work.
MR acknowledges funding from the European Research Council Consolidator Grant ({\it CMBSPEC}, No.~725456). JC is supported by the Royal Society as a Royal Society University Research Fellow at the University of Manchester, UK.

\bibliographystyle{mn2e}
\bibliography{mu_simulation}

\begin{thebibliography}{82}
\expandafter\ifx\csname natexlab\endcsname\relax\def\natexlab#1{#1}\fi

\bibitem[{{Abitbol} {et~al}\mbox{.}(2017){Abitbol}, {Chluba}, {Hill}, \&
  {Johnson}}]{abitbol_pixie}
{Abitbol} M.~H., {Chluba} J., {Hill} J.~C., {Johnson} B.~R., 2017, \mnras

\bibitem[{{Ali-Ha{\"i}moud} \& {Kamionkowski}(2017)}]{Yacine2017}
{Ali-Ha{\"i}moud} Y., {Kamionkowski} M., 2017, \prd, 95, 043534

\bibitem[{{Arnaud} {et~al}\mbox{.}(2010){Arnaud}, {Pratt}, {Piffaretti},
  {B{\"o}hringer}, {Croston}, \& {Pointecouteau}}]{Arnaud2010}
{Arnaud} M., {Pratt} G.~W., {Piffaretti} R., {B{\"o}hringer} H., {Croston}
  J.~H., {Pointecouteau} E., 2010, \aap, 517, A92

\bibitem[{{Bartolo} {et~al}\mbox{.}(2016){Bartolo}, {Liguori}, \&
  {Shiraishi}}]{Bartolo2016}
{Bartolo} N., {Liguori} M., {Shiraishi} M., 2016, \jcap, 3, 029

\bibitem[{{Basu} {et~al}\mbox{.}(2004){Basu}, {Hern{\'a}ndez-Monteagudo}, \&
  {Sunyaev}}]{Kaustuv2004}
{Basu} K., {Hern{\'a}ndez-Monteagudo} C., {Sunyaev} R.~A., 2004, \aap, 416, 447

\bibitem[{{Biagetti} {et~al}\mbox{.}(2013){Biagetti}, {Perrier}, {Riotto}, \&
  {Desjacques}}]{Biagetti2013}
{Biagetti} M., {Perrier} H., {Riotto} A., {Desjacques} V., 2013, \prd, 87,
  063521

\bibitem[{{B{\"o}hringer} {et~al}\mbox{.}(2004){B{\"o}hringer}, {Schuecker},
  {Guzzo}, {Collins}, {Voges}, {Cruddace}, {Ortiz-Gil}, {Chincarini}, {De
  Grandi}, {Edge}, {MacGillivray}, {Neumann}, {Schindler}, \&
  {Shaver}}]{ROSAT2004}
{B{\"o}hringer} H. {et~al.}, 2004, \aap, 425, 367

\bibitem[{{Burigana} {et~al}\mbox{.}(2017){Burigana}, {Carvalho}, {Trombetti},
  {Notari}, {Quartin}, {De Gasperis}, {Buzzelli}, {Vittorio}, {De Zotti}, {de
  Bernardis}, {Chluba}, {Bilicki}, {Danese}, {Delabrouille}, {Toffolatti},
  {Lapi}, {Negrello}, {Mazzotta}, {Scott}, {Contreras}, {Achucarro}, {Ade},
  {Allison}, {Ashdown}, {Ballardini}, {Banday}, {Banerji}, {Bartlett},
  {Bartolo}, {Basak}, {Bersanelli}, {Bonaldi}, {Bonato}, {Borrill}, {Bouchet},
  {Boulanger}, {Brinckmann}, {Bucher}, {Cabella}, {Cai}, {Calvo}, {Castellano},
  {Challinor}, {Clesse}, {Colantoni}, {Coppolecchia}, {Crook}, {D'Alessandro},
  {Diego}, {Di Marco}, {Di Valentino}, {Errard}, {Feeney}, {Fernandez-Cobos},
  {Ferraro}, {Finelli}, {Forastieri}, {Galli}, {Genova-Santos}, {Gerbino},
  {Gonzalez-Nuevo}, {Grandis}, {Greenslade}, {Hagstotz}, {Hanany}, {Handley},
  {Hernandez-Monteagudo}, {Hervias-Caimapo}, {Hills}, {Hivon}, {Kiiveri},
  {Kisner}, {Kitching}, {Kunz}, {Kurki-Suonio}, {Lamagna}, {Lasenby},
  {Lattanzi}, {Lesgourgues}, {Liguori}, {Lindholm}, {Lopez-Caniego}, {Luzzi},
  {Maffei}, {Mandolesi}, {Martinez-Gonzalez}, {Martins}, {Masi}, {McCarthy},
  {Melchiorri}, {Melin}, {Molinari}, {Monfardini}, {Natoli}, {Paiella},
  {Paoletti}, {Patanchon}, {Piat}, {Pisano}, {Polastri}, {Polenta}, {Pollo},
  {Poulin}, {Remazeilles}, {Roman}, {Rubino-Martin}, {Salvati}, {Tartari},
  {Tomasi}, {Tramonte}, {Trappe}, {Tucker}, {Valiviita}, {Van de Weijgaert},
  {van Tent}, {Vennin}, {Vielva}, {Young}, {Zannoni}, \& {for the CORE
  Collaboration}}]{Burigana2017}
{Burigana} C. {et~al.}, 2017, arXiv:1704.05764, accepted by JCAP

\bibitem[{{Burigana} {et~al}\mbox{.}(1991){Burigana}, {Danese}, \& {de
  Zotti}}]{Burigana1991}
{Burigana} C., {Danese} L., {de Zotti} G., 1991, \aap, 246, 49

\bibitem[{{Carlstrom} {et~al}\mbox{.}(2002){Carlstrom}, {Holder}, \&
  {Reese}}]{Carlstrom2002}
{Carlstrom} J.~E., {Holder} G.~P., {Reese} E.~D., 2002, \araa, 40, 643

\bibitem[{{Carr} {et~al}\mbox{.}(2010){Carr}, {Kohri}, {Sendouda}, \&
  {Yokoyama}}]{Carr2010}
{Carr} B.~J., {Kohri} K., {Sendouda} Y., {Yokoyama} J., 2010, \prd, 81, 104019

\bibitem[{{Chluba}(2013{\natexlab{a}})}]{Chluba2013fore}
{Chluba} J., 2013{\natexlab{a}}, \mnras, 436, 2232

\bibitem[{{Chluba}(2013{\natexlab{b}})}]{Chluba2013Green}
{Chluba} J., 2013{\natexlab{b}}, \mnras, 434, 352

\bibitem[{{Chluba}(2015)}]{Chluba2015GreensII}
{Chluba} J., 2015, \mnras, 454, 4182

\bibitem[{{Chluba}(2016)}]{Chluba2016}
{Chluba} J., 2016, \mnras, 460, 227

\bibitem[{{Chluba} {et~al}\mbox{.}(2017{\natexlab{a}}){Chluba},
  {Dimastrogiovanni}, {Amin}, \& {Kamionkowski}}]{Chluba_mu_2017}
{Chluba} J., {Dimastrogiovanni} E., {Amin} M.~A., {Kamionkowski} M.,
  2017{\natexlab{a}}, \mnras, 466, 2390

\bibitem[{{Chluba} {et~al}\mbox{.}(2012{\natexlab{a}}){Chluba}, {Erickcek}, \&
  {Ben-Dayan}}]{Chluba2012inflaton}
{Chluba} J., {Erickcek} A.~L., {Ben-Dayan} I., 2012{\natexlab{a}}, \apj, 758,
  76

\bibitem[{{Chluba} {et~al}\mbox{.}(2017{\natexlab{b}}){Chluba}, {Hill}, \&
  {Abitbol}}]{Chluba2017foregrounds}
{Chluba} J., {Hill} J.~C., {Abitbol} M.~H., 2017{\natexlab{b}}, \mnras, 472,
  1195

\bibitem[{{Chluba} \& {Jeong}(2014)}]{Chluba2013PCA}
{Chluba} J., {Jeong} D., 2014, \mnras, 438, 2065

\bibitem[{{Chluba} {et~al}\mbox{.}(2012{\natexlab{b}}){Chluba}, {Khatri}, \&
  {Sunyaev}}]{Chluba2012_2x2}
{Chluba} J., {Khatri} R., {Sunyaev} R.~A., 2012{\natexlab{b}}, \mnras, 425,
  1129

\bibitem[{{Chluba} \& {Sunyaev}(2004)}]{Chluba2004}
{Chluba} J., {Sunyaev} R.~A., 2004, \aap, 424, 389

\bibitem[{{Chluba} \& {Sunyaev}(2012)}]{Chluba2012}
{Chluba} J., {Sunyaev} R.~A., 2012, \mnras, 419, 1294

\bibitem[{{CORE Collaboration} {et~al}\mbox{.}(2016){CORE Collaboration},
  {Finelli}, {Bucher}, {Ach{\'u}carro}, {Ballardini}, {Bartolo}, {Baumann},
  {Clesse}, {Errard}, {Handley}, {Hindmarsh}, {Kiiveri}, {Kunz}, {Lasenby},
  {Liguori}, {Paoletti}, {Ringeval}, {V{\"a}liviita}, {van Tent}, {Vennin},
  {Allison}, {Arroja}, {Ashdown}, {Banday}, {Banerji}, {Bartlett}, {Basak},
  {Baselmans}, {de Bernardis}, {Bersanelli}, {Bonaldi}, {Borril}, {Bouchet},
  {Boulanger}, {Brinckmann}, {Burigana}, {Buzzelli}, {Cai}, {Calvo},
  {Carvalho}, {Castellano}, {Challinor}, {Chluba}, {Colantoni}, {Crook},
  {D'Alessandro}, {D'Amico}, {Delabrouille}, {Desjacques}, {De Zotti}, {Diego},
  {Di Valentino}, {Feeney}, {Fergusson}, {Fernandez-Cobos}, {Ferraro},
  {Forastieri}, {Galli}, {Garc{\'{\i}}a-Bellido}, {de Gasperis},
  {G{\'e}nova-Santos}, {Gerbino}, {Gonz{\'a}lez-Nuevo}, {Grandis},
  {Greenslade}, {Hagstotz}, {Hanany}, {Hazra}, {Hern{\'a}ndez-Monteagudo},
  {Hervias-Caimapo}, {Hills}, {Hivon}, {Hu}, {Kisner}, {Kitching}, {Kovetz},
  {Kurki-Suonio}, {Lamagna}, {Lattanzi}, {Lesgourgues}, {Lewis}, {Lindholm},
  {Lizarraga}, {L{\'o}pez-Caniego}, {Luzzi}, {Maffei}, {Mandolesi},
  {Mart{\'{\i}}nez-Gonz{\'a}lez}, {Martins}, {Masi}, {McCarthy}, {Matarrese},
  {Melchiorri}, {Melin}, {Molinari}, {Monfardini}, {Natoli}, {Negrello},
  {Notari}, {Oppizzi}, {Paiella}, {Pajer}, {Patanchon}, {Patil}, {Piat},
  {Pisano}, {Polastri}, {Polenta}, {Pollo}, {Poulin}, {Quartin}, {Ravenni},
  {Remazeilles}, {Renzi}, {Roest}, {Roman}, {Rubi{\~n}o-Martin}, {Salvati},
  {Starobinsky}, {Tartari}, {Tasinato}, {Tomasi}, {Torrado}, {Trappe},
  {Trombetti}, {Tucker}, {Tucci}, {Urrestilla}, {van de Weygaert}, {Vielva},
  {Vittorio}, \& {Young}}]{Finelli2016}
{CORE Collaboration} {et~al.}, 2016, ArXiv:1612.08270

\bibitem[{{Creque-Sarbinowski} {et~al}\mbox{.}(2016){Creque-Sarbinowski},
  {Bird}, \& {Kamionkowski}}]{Creque2016}
{Creque-Sarbinowski} C., {Bird} S., {Kamionkowski} M., 2016, \prd, 94, 063519

\bibitem[{{Daly}(1991)}]{Daly1991}
{Daly} R.~A., 1991, \apj, 371, 14

\bibitem[{{De Zotti} {et~al}\mbox{.}(2016){De Zotti}, {Negrello}, {Castex},
  {Lapi}, \& {Bonato}}]{deZotti2015}
{De Zotti} G., {Negrello} M., {Castex} G., {Lapi} A., {Bonato} M., 2016, \jcap,
  3, 047

\bibitem[{{Delabrouille} {et~al}\mbox{.}(2013){Delabrouille}, {Betoule},
  {Melin}, {Miville-Desch{\^e}nes}, {Gonzalez-Nuevo}, {Le Jeune}, {Castex}, {de
  Zotti}, {Basak}, {Ashdown}, {Aumont}, {Baccigalupi}, {Banday}, {Bernard},
  {Bouchet}, {Clements}, {da Silva}, {Dickinson}, {Dodu}, {Dolag}, {Elsner},
  {Fauvet}, {Fa{\"y}}, {Giardino}, {Leach}, {Lesgourgues}, {Liguori},
  {Mac{\'{\i}}as-P{\'e}rez}, {Massardi}, {Matarrese}, {Mazzotta}, {Montier},
  {Mottet}, {Paladini}, {Partridge}, {Piffaretti}, {Prezeau}, {Prunet},
  {Ricciardi}, {Roman}, {Schaefer}, \& {Toffolatti}}]{Delabrouille2013}
{Delabrouille} J. {et~al.}, 2013, \aap, 553, A96

\bibitem[{{Delabrouille} {et~al}\mbox{.}(2009){Delabrouille}, {Cardoso}, {Le
  Jeune}, {Betoule}, {Fay}, \& {Guilloux}}]{Delabrouille2009}
{Delabrouille} J., {Cardoso} J.-F., {Le Jeune} M., {Betoule} M., {Fay} G.,
  {Guilloux} F., 2009, \aap, 493, 835

\bibitem[{{Delabrouille} {et~al}\mbox{.}(2017){Delabrouille}, {de Bernardis},
  {Bouchet}, {Ach{\'u}carro}, {Ade}, {Allison}, {Arroja}, {Artal}, {Ashdown},
  {Baccigalupi}, {Ballardini}, {Banday}, {Banerji}, {Barbosa}, {Bartlett},
  {Bartolo}, {Basak}, {Baselmans}, {Basu}, {Battistelli}, {Battye}, {Baumann},
  {Beno{\^i}t}, {Bersanelli}, {Bideaud}, {Biesiada}, {Bilicki}, {Bonaldi},
  {Bonato}, {Borrill}, {Boulanger}, {Brinckmann}, {Brown}, {Bucher},
  {Burigana}, {Buzzelli}, {Cabass}, {Cai}, {Calvo}, {Caputo}, {Carvalho},
  {Casas}, {Castellano}, {Catalano}, {Challinor}, {Charles}, {Chluba},
  {Clements}, {Clesse}, {Colafrancesco}, {Colantoni}, {Contreras},
  {Coppolecchia}, {Crook}, {D'Alessandro}, {D'Amico}, {da Silva}, {de Avillez},
  {de Gasperis}, {De Petris}, {de Zotti}, {Danese}, {D{\'e}sert}, {Desjacques},
  {Di Valentino}, {Dickinson}, {Diego}, {Doyle}, {Durrer}, {Dvorkin},
  {Eriksen}, {Errard}, {Feeney}, {Fern{\'a}ndez-Cobos}, {Finelli},
  {Forastieri}, {Franceschet}, {Fuskeland}, {Galli}, {G{\'e}nova-Santos},
  {Gerbino}, {Giusarma}, {Gomez}, {Gonz{\'a}lez-Nuevo}, {Grandis},
  {Greenslade}, {Goupy}, {Hagstotz}, {Hanany}, {Handley},
  {Henrot-Versill{\'e}}, {Hern{\'a}ndez-Monteagudo}, {Hervias-Caimapo},
  {Hills}, {Hindmarsh}, {Hivon}, {Hoang}, {Hooper}, {Hu}, {Keih{\"a}nen},
  {Keskitalo}, {Kiiveri}, {Kisner}, {Kitching}, {Kunz}, {Kurki-Suonio},
  {Lagache}, {Lamagna}, {Lapi}, {Lasenby}, {Lattanzi}, {Le Brun},
  {Lesgourgues}, {Liguori}, {Lindholm}, {Lizarraga}, {Luzzi},
  {Mac{\`i}as-P{\'e}rez}, {Maffei}, {Mandolesi}, {Martin}, {Martinez-Gonzalez},
  {Martins}, {Masi}, {Massardi}, {Matarrese}, {Mazzotta}, {McCarthy},
  {Melchiorri}, {Melin}, {Mennella}, {Mohr}, {Molinari}, {Monfardini},
  {Montier}, {Natoli}, {Negrello}, {Notari}, {Noviello}, {Oppizzi},
  {O'Sullivan}, {Pagano}, {Paiella}, {Pajer}, {Paoletti}, {Paradiso},
  {Partridge}, {Patanchon}, {Patil}, {Perdereau}, {Piacentini}, {Piat},
  {Pisano}, {Polastri}, {Polenta}, {Pollo}, {Ponthieu}, {Poulin}, {Pr{\^e}le},
  {Quartin}, {Ravenni}, {Remazeilles}, {Renzi}, {Ringeval}, {Roest}, {Roman},
  {Roukema}, {Rubino-Martin}, {Salvati}, {Scott}, {Serjeant}, {Signorelli},
  {Starobinsky}, {Sunyaev}, {Tan}, {Tartari}, {Tasinato}, {Toffolatti},
  {Tomasi}, {Torrado}, {Tramonte}, {Trappe}, {Triqueneaux}, {Tristram},
  {Trombetti}, {Tucci}, {Tucker}, {Urrestilla}, {V{\"a}liviita}, {Van de
  Weygaert}, {Van Tent}, {Vennin}, {Verde}, {Vermeulen}, {Vielva}, {Vittorio},
  {Voisin}, {Wallis}, {Wandelt}, {Wehus}, {Weller}, {Young}, {Zannoni}, \& {for
  the CORE collaboration}}]{Core2016}
{Delabrouille} J. {et~al.}, 2017, arXiv:1706.04516, accepted by JCAP

\bibitem[{{Dick} {et~al}\mbox{.}(2010){Dick}, {Remazeilles}, \&
  {Delabrouille}}]{Dick2010}
{Dick} J., {Remazeilles} M., {Delabrouille} J., 2010, \mnras, 401, 1602

\bibitem[{{Dickinson} {et~al}\mbox{.}(2003){Dickinson}, {Davies}, \&
  {Davis}}]{Dickinson2003}
{Dickinson} C., {Davies} R.~D., {Davis} R.~J., 2003, \mnras, 341, 369

\bibitem[{{Dimastrogiovanni} \& {Emami}(2016)}]{Dimastrogiovanni2016}
{Dimastrogiovanni} E., {Emami} R., 2016, \jcap, 12, 015

\bibitem[{{Emami} {et~al}\mbox{.}(2015){Emami}, {Dimastrogiovanni}, {Chluba},
  \& {Kamionkowski}}]{Emami2015}
{Emami} R., {Dimastrogiovanni} E., {Chluba} J., {Kamionkowski} M., 2015, \prd,
  91, 123531

\bibitem[{{Eriksen} {et~al}\mbox{.}(2008){Eriksen}, {Jewell}, {Dickinson},
  {et~al.}}]{Eriksen2008}
{Eriksen} H.~K., {Jewell} J.~B., {Dickinson} C., {et~al.}, 2008, \apj, 676, 10

\bibitem[{{Fern{\'a}ndez-Cobos} {et~al}\mbox{.}(2012){Fern{\'a}ndez-Cobos},
  {Vielva}, {Barreiro}, {et~al.}}]{Fernandez2012}
{Fern{\'a}ndez-Cobos} R., {Vielva} P., {Barreiro} R.~B., {et~al.}, 2012,
  \mnras, 420, 2162

\bibitem[{{Fixsen}(2009)}]{Fixsen2009}
{Fixsen} D.~J., 2009, \apj, 707, 916

\bibitem[{{Fixsen} {et~al}\mbox{.}(1996){Fixsen}, {Cheng}, {Gales}, {Mather},
  {Shafer}, \& {Wright}}]{Fixsen1996}
{Fixsen} D.~J., {Cheng} E.~S., {Gales} J.~M., {Mather} J.~C., {Shafer} R.~A.,
  {Wright} E.~L., 1996, \apj, 473, 576

\bibitem[{{Ganc} \& {Komatsu}(2012)}]{Ganc2012}
{Ganc} J., {Komatsu} E., 2012, \prd, 86, 023518

\bibitem[{{G{\'o}rski} {et~al}\mbox{.}(2005){G{\'o}rski}, {Hivon}, {Banday},
  {Wandelt}, {Hansen}, {Reinecke}, \& {Bartelmann}}]{Gorski2005}
{G{\'o}rski} K.~M., {Hivon} E., {Banday} A.~J., {Wandelt} B.~D., {Hansen}
  F.~K., {Reinecke} M., {Bartelmann} M., 2005, \apj, 622, 759

\bibitem[{{Haslam} {et~al}\mbox{.}(1982){Haslam}, {Salter}, {Stoffel}, \&
  {Wilson}}]{Haslam1982}
{Haslam} C.~G.~T., {Salter} C.~J., {Stoffel} H., {Wilson} W.~E., 1982, \aaps,
  47, 1

\bibitem[{{Hu} {et~al}\mbox{.}(1994){Hu}, {Scott}, \& {Silk}}]{Hu1994}
{Hu} W., {Scott} D., {Silk} J., 1994, \apjl, 430, L5

\bibitem[{{Hu} \& {Silk}(1993{\natexlab{a}})}]{Hu1993}
{Hu} W., {Silk} J., 1993{\natexlab{a}}, \prd, 48, 485

\bibitem[{{Hu} \& {Silk}(1993{\natexlab{b}})}]{Hu1993b}
{Hu} W., {Silk} J., 1993{\natexlab{b}}, Physical Review Letters, 70, 2661

\bibitem[{{Khatri} \& {Sunyaev}(2015)}]{Khatri2015}
{Khatri} R., {Sunyaev} R., 2015, \jcap, 9, 026

\bibitem[{{Khatri} \& {Sunyaev}(2012)}]{Khatri2012mix}
{Khatri} R., {Sunyaev} R.~A., 2012, \jcap, 9, 16

\bibitem[{{Koester} {et~al}\mbox{.}(2007){Koester}, {McKay}, {Annis},
  {Wechsler}, {Evrard}, {Bleem}, {Becker}, {Johnston}, {Sheldon}, {Nichol},
  {Miller}, {Scranton}, {Bahcall}, {Barentine}, {Brewington}, {Brinkmann},
  {Harvanek}, {Kleinman}, {Krzesinski}, {Long}, {Nitta}, {Schneider},
  {Sneddin}, {Voges}, \& {York}}]{Koester2007}
{Koester} B.~P. {et~al.}, 2007, \apj, 660, 239

\bibitem[{{Kogut} {et~al}\mbox{.}(2016){Kogut}, {Chluba}, {Fixsen}, {Meyer}, \&
  {Spergel}}]{Pixie2016}
{Kogut} A., {Chluba} J., {Fixsen} D.~J., {Meyer} S., {Spergel} D., 2016, in
  \procspie, Vol. 9904, Society of Photo-Optical Instrumentation Engineers
  (SPIE) Conference Series, p. 99040W

\bibitem[{{Kogut} {et~al}\mbox{.}(2011){Kogut}, {Fixsen}, {Chuss}, {Dotson},
  {Dwek}, {Halpern}, {Hinshaw}, {Meyer}, {Moseley}, {Seiffert}, {Spergel}, \&
  {Wollack}}]{Kogut2011PIXIE}
{Kogut} A. {et~al.}, 2011, \jcap, 7, 25

\bibitem[{{Komatsu} \& {Spergel}(2001)}]{Komatsu2001ng}
{Komatsu} E., {Spergel} D.~N., 2001, \prd, 63, 063002

\bibitem[{{Matsumura} {et~al}\mbox{.}(2016){Matsumura}, {Akiba}, {Arnold},
  {Borrill}, {Chendra}, {Chinone}, {Cukierman}, {de Haan}, {Dobbs}, {Dominjon},
  {Elleflot}, {Errard}, {Fujino}, {Fuke}, {Goeckner-wald}, {Halverson},
  {Harvey}, {Hasegawa}, {Hattori}, {Hattori}, {Hazumi}, {Hill}, {Hilton},
  {Holzapfel}, {Hori}, {Hubmayr}, {Ichiki}, {Inatani}, {Inoue}, {Inoue},
  {Irie}, {Irwin}, {Ishino}, {Ishitsuka}, {Jeong}, {Karatsu}, {Kashima},
  {Katayama}, {Kawano}, {Keating}, {Kibayashi}, {Kibe}, {Kida}, {Kimura},
  {Kimura}, {Kohri}, {Komatsu}, {Kuo}, {Kuromiya}, {Kusaka}, {Lee}, {Linder},
  {Matsuhara}, {Matsuoka}, {Matsuura}, {Mima}, {Mitsuda}, {Mizukami}, {Morii},
  {Morishima}, {Nagai}, {Nagasaki}, {Nagata}, {Nakajima}, {Nakamura},
  {Namikawa}, {Naruse}, {Natsume}, {Nishibori}, {Nishijo}, {Nishino}, {Nitta},
  {Noda}, {Noguchi}, {Ogawa}, {Oguri}, {Ohta}, {Otani}, {Okada}, {Okamoto},
  {Okamoto}, {Okamura}, {Rebeiz}, {Richards}, {Sakai}, {Sato}, {Sato},
  {Segawa}, {Sekiguchi}, {Sekimoto}, {Sekine}, {Seljak}, {Sherwin},
  {Shinozaki}, {Shu}, {Stompor}, {Sugai}, {Sugita}, {Suzuki}, {Suzuki},
  {Tajima}, {Takada}, {Takakura}, {Takano}, {Takei}, {Tomaru}, {Tomita},
  {Turin}, {Utsunomiya}, {Uzawa}, {Wada}, {Watanabe}, {Westbrook}, {Whitehorn},
  {Yamada}, {Yamasaki}, {Yamashita}, {Yoshida}, {Yoshida}, \&
  {Yotsumoto}}]{Litebird2016}
{Matsumura} T. {et~al.}, 2016, Journal of Low Temperature Physics, 184, 824

\bibitem[{{McDonald} {et~al}\mbox{.}(2001){McDonald}, {Scherrer}, \&
  {Walker}}]{McDonald2001}
{McDonald} P., {Scherrer} R.~J., {Walker} T.~P., 2001, \prd, 63, 023001

\bibitem[{{Miville-Desch{\^e}nes} {et~al}\mbox{.}(2008){Miville-Desch{\^e}nes},
  {Ysard}, {Lavabre}, {Ponthieu}, {Mac{\'{\i}}as-P{\'e}rez}, {Aumont}, \&
  {Bernard}}]{Miville2008}
{Miville-Desch{\^e}nes} M.-A., {Ysard} N., {Lavabre} A., {Ponthieu} N.,
  {Mac{\'{\i}}as-P{\'e}rez} J.~F., {Aumont} J., {Bernard} J.~P., 2008, \aap,
  490, 1093

\bibitem[{{Mukherjee} {et~al}\mbox{.}(2018){Mukherjee}, {Silk}, \&
  {Wandelt}}]{Mukherjee2018}
{Mukherjee} S., {Silk} J., {Wandelt} B.~D., 2018, ArXiv e-prints

\bibitem[{{Narcowich} {et~al}\mbox{.}(2006){Narcowich}, {Petrushev}, \&
  {Ward}}]{Narcowich2006}
{Narcowich} F., {Petrushev} P., {Ward} J., 2006, SIAM J. Math. Anal., 38, 574

\bibitem[{{Ota}(2016)}]{Ota2016}
{Ota} A., 2016, \prd, 94, 103520

\bibitem[{{Pajer} \& {Zaldarriaga}(2012)}]{Pajer2012}
{Pajer} E., {Zaldarriaga} M., 2012, Physical Review Letters, 109, 021302

\bibitem[{{Pajer} \& {Zaldarriaga}(2013)}]{Pajer2013}
{Pajer} E., {Zaldarriaga} M., 2013, \jcap, 2, 036

\bibitem[{{Peebles} \& {Yu}(1970)}]{Peebles1970}
{Peebles} P.~J.~E., {Yu} J.~T., 1970, \apj, 162, 815

\bibitem[{{Piffaretti} {et~al}\mbox{.}(2011){Piffaretti}, {Arnaud}, {Pratt},
  {Pointecouteau}, \& {Melin}}]{Piffaretti2011}
{Piffaretti} R., {Arnaud} M., {Pratt} G.~W., {Pointecouteau} E., {Melin} J.-B.,
  2011, \aap, 534, A109

\bibitem[{{Planck Collaboration} {et~al}\mbox{.}(2014){Planck Collaboration},
  {Ade}, {Aghanim}, {Armitage-Caplan}, {Arnaud}, {Ashdown}, {Atrio-Barandela},
  {Aumont}, {Baccigalupi}, {Banday}, \& et~al.}]{Planck2013ng}
{Planck Collaboration} {et~al.}, 2014, \aap, 571, A24

\bibitem[{{Planck Collaboration} {et~al}\mbox{.}(2016){Planck Collaboration},
  {Ade}, {Aghanim}, {Arnaud}, {Ashdown}, {Aumont}, {Baccigalupi}, {Banday},
  {Barreiro}, {Bartlett}, \& et~al.}]{Planck2015params}
{Planck Collaboration} {et~al.}, 2016, \aap, 594, A13

\bibitem[{{\sorthelp{Planck Collaboration 2015A}}{Planck Collaboration
  I}(2016)}]{planck2015_overview}
{\sorthelp{Planck Collaboration 2015A}}{Planck Collaboration I}, 2016, \aap,
  594, A1

\bibitem[{{\sorthelp{Planck Collaboration IntZW}}{Planck Collaboration Int.
  XLVIII}(2016)}]{Planck_PIP_XLVIII}
{\sorthelp{Planck Collaboration IntZW}}{Planck Collaboration Int. XLVIII},
  2016, \aap, 596, A109

\bibitem[{{Poulin} {et~al}\mbox{.}(2017){Poulin}, {Lesgourgues}, \&
  {Serpico}}]{Poulin2017}
{Poulin} V., {Lesgourgues} J., {Serpico} P.~D., 2017, \jcap, 3, 043

\bibitem[{{Ravenni} {et~al}\mbox{.}(2017){Ravenni}, {Liguori}, {Bartolo}, \&
  {Shiraishi}}]{Ravenni2017}
{Ravenni} A., {Liguori} M., {Bartolo} N., {Shiraishi} M., 2017, \jcap, 9, 042

\bibitem[{{Remazeilles} {et~al}\mbox{.}(2013){Remazeilles}, {Aghanim}, \&
  {Douspis}}]{Remazeilles2013}
{Remazeilles} M., {Aghanim} N., {Douspis} M., 2013, \mnras, 430, 370

\bibitem[{{Remazeilles} {et~al}\mbox{.}(2017){Remazeilles}, {Banday},
  {Baccigalupi}, {Basak}, {Bonaldi}, {De Zotti}, {Delabrouille}, {Dickinson},
  {Eriksen}, {Errard}, {Fernandez-Cobos}, {Fuskeland},
  {Herv{\'{\i}}as-Caimapo}, {L{\'o}pez-Caniego}, {Martinez-Gonz{\'a}lez},
  {Roman}, {Vielva}, {Wehus}, {Achucarro}, {Ade}, {Allison}, {Ashdown},
  {Ballardini}, {Banerji}, {Bartolo}, {Bartlett}, {Baumann}, {Bersanelli},
  {Bonato}, {Borrill}, {Bouchet}, {Boulanger}, {Brinckmann}, {Bucher},
  {Burigana}, {Buzzelli}, {Cai}, {Calvo}, {Carvalho}, {Castellano},
  {Challinor}, {Chluba}, {Clesse}, {Colantoni}, {Coppolecchia}, {Crook},
  {D'Alessandro}, {de Bernardis}, {de Gasperis}, {Diego}, {Di Valentino},
  {Feeney}, {Ferraro}, {Finelli}, {Forastieri}, {Galli}, {Genova-Santos},
  {Gerbino}, {Gonz{\'a}lez-Nuevo}, {Grandis}, {Greenslade}, {Hagstotz},
  {Hanany}, {Handley}, {Hernandez-Monteagudo}, {Hills}, {Hivon}, {Kiiveri},
  {Kisner}, {Kitching}, {Kunz}, {Kurki-Suonio}, {Lamagna}, {Lasenby},
  {Lattanzi}, {Lesgourgues}, {Lewis}, {Liguori}, {Lindholm}, {Luzzi}, {Maffei},
  {Martins}, {Masi}, {McCarthy}, {Melin}, {Melchiorri}, {Molinari},
  {Monfardini}, {Natoli}, {Negrello}, {Notari}, {Paiella}, {Paoletti},
  {Patanchon}, {Piat}, {Pisano}, {Polastri}, {Polenta}, {Pollo}, {Poulin},
  {Quartin}, {Rubino-Martin}, {Salvati}, {Tartari}, {Tomasi}, {Tramonte},
  {Trappe}, {Trombetti}, {Tucker}, {Valiviita}, {Van de Weijgaert}, {van Tent},
  {Vennin}, {Vittorio}, {Young}, \& {for the CORE
  collaboration}}]{Remazeilles2017}
{Remazeilles} M. {et~al.}, 2017, arXiv:1704.04501, accepted by JCAP

\bibitem[{{Remazeilles} {et~al}\mbox{.}(2011{\natexlab{a}}){Remazeilles},
  {Delabrouille}, \& {Cardoso}}]{Remazeilles2011}
{Remazeilles} M., {Delabrouille} J., {Cardoso} J.-F., 2011{\natexlab{a}},
  \mnras, 410, 2481

\bibitem[{{Remazeilles} {et~al}\mbox{.}(2011{\natexlab{b}}){Remazeilles},
  {Delabrouille}, \& {Cardoso}}]{Remazeilles2011b}
{Remazeilles} M., {Delabrouille} J., {Cardoso} J.-F., 2011{\natexlab{b}},
  \mnras, 418, 467

\bibitem[{{Remazeilles} {et~al}\mbox{.}(2015){Remazeilles}, {Dickinson},
  {Banday}, {Bigot-Sazy}, \& {Ghosh}}]{Remazeilles2015}
{Remazeilles} M., {Dickinson} C., {Banday} A.~J., {Bigot-Sazy} M.-A., {Ghosh}
  T., 2015, \mnras, 451, 4311

\bibitem[{{Remazeilles} {et~al}\mbox{.}(2016){Remazeilles}, {Dickinson},
  {Eriksen}, \& {Wehus}}]{Remazeilles2016}
{Remazeilles} M., {Dickinson} C., {Eriksen} H.~K.~K., {Wehus} I.~K., 2016,
  \mnras, 458, 2032

\bibitem[{{Sarkar} \& {Cooper}(1984)}]{Sarkar1984}
{Sarkar} S., {Cooper} A.~M., 1984, Physics Letters B, 148, 347

\bibitem[{{Shiraishi} {et~al}\mbox{.}(2016){Shiraishi}, {Bartolo}, \&
  {Liguori}}]{Shiraishi2016}
{Shiraishi} M., {Bartolo} N., {Liguori} M., 2016, \jcap, 10, 015

\bibitem[{{Spergel} {et~al}\mbox{.}(2003){Spergel}, {Verde}, {Peiris},
  {Komatsu}, {Nolta}, {Bennett}, {Halpern}, {Hinshaw}, {Jarosik}, {Kogut},
  {Limon}, {Meyer}, {Page}, {Tucker}, {Weiland}, {Wollack}, \&
  {Wright}}]{Spergel2003}
{Spergel} D.~N. {et~al.}, 2003, \apjs, 148, 175

\bibitem[{{Sunyaev} \& {Khatri}(2013)}]{Sunyaev2013}
{Sunyaev} R.~A., {Khatri} R., 2013, IJMPD, 22, 30014

\bibitem[{{Sunyaev} \& {Zeldovich}(1970{\natexlab{a}})}]{Sunyaev1970diss}
{Sunyaev} R.~A., {Zeldovich} Y.~B., 1970{\natexlab{a}}, \apss, 9, 368

\bibitem[{{Sunyaev} \& {Zeldovich}(1970{\natexlab{b}})}]{Sunyaev1970}
{Sunyaev} R.~A., {Zeldovich} Y.~B., 1970{\natexlab{b}}, \apss, 7, 3

\bibitem[{{Sunyaev} \& {Zeldovich}(1970{\natexlab{c}})}]{Sunyaev1970mu}
{Sunyaev} R.~A., {Zeldovich} Y.~B., 1970{\natexlab{c}}, \apss, 7, 20

\bibitem[{{Sunyaev} \& {Zeldovich}(1972)}]{Sunyaev1972}
{Sunyaev} R.~A., {Zeldovich} Y.~B., 1972, comas, 4, 173

\bibitem[{{Suzuki} {et~al}\mbox{.}(2018){Suzuki}, {Ade}, {Akiba}, {Alonso},
  {Arnold}, {Aumont}, {Baccigalupi}, {Barron}, {Basak}, {Beckman}, {Borrill},
  {Boulanger}, {Bucher}, {Calabrese}, {Chinone}, {Cho}, {Cukierman}, {Curtis},
  {de Haan}, {Dobbs}, {Dominjon}, {Dotani}, {Duband}, {Ducout}, {Dunkley},
  {Duval}, {Elleflot}, {Eriksen}, {Errard}, {Fischer}, {Fujino}, {Funaki},
  {Fuskeland}, {Ganga}, {Goeckner-Wald}, {Grain}, {Halverson}, {Hamada},
  {Hasebe}, {Hasegawa}, {Hattori}, {Hattori}, {Hayes}, {Hazumi}, {Hidehira},
  {Hill}, {Hilton}, {Hubmayr}, {Ichiki}, {Iida}, {Imada}, {Inoue}, {Inoue},
  {D.}, {Ishino}, {Jeong}, {Kanai}, {Kaneko}, {Kashima}, {Katayama},
  {Kawasaki}, {Kernasovskiy}, {Keskitalo}, {Kibayashi}, {Kida}, {Kimura},
  {Kisner}, {Kohri}, {Komatsu}, {Komatsu}, {Kuo}, {Kurinsky}, {Kusaka},
  {Lazarian}, {Lee}, {Li}, {Linder}, {Maffei}, {Mangilli}, {Maki}, {Matsumura},
  {Matsuura}, {Meilhan}, {Mima}, {Minami}, {Mitsuda}, {Montier}, {Nagai},
  {Nagasaki}, {Nagata}, {Nakajima}, {Nakamura}, {Namikawa}, {Naruse},
  {Nishino}, {Nitta}, {Noguchi}, {Ogawa}, {Oguri}, {Okada}, {Okamoto},
  {Okamura}, {Otani}, {Patanchon}, {Pisano}, {Rebeiz}, {Remazeilles},
  {Richards}, {Sakai}, {Sakurai}, {Sato}, {Sato}, {Sawada}, {Segawa},
  {Sekimoto}, {Seljak}, {Sherwin}, {Shimizu}, {Shinozaki}, {Stompor}, {Sugai},
  {Sugita}, {Suzuki}, {Tajima}, {Takada}, {Takaku}, {Takakura}, {Takatori},
  {Tanabe}, {Taylor}, {Thompson}, {Thorne}, {Tomaru}, {Tomida}, {Tomita},
  {Tristram}, {Tucker}, {Turin}, {Tsujimoto}, {Uozumi}, {Utsunomiya}, {Uzawa},
  {Vansyngel}, {Wehus}, {Westbrook}, {Willer}, {Whitehorn}, {Yamada},
  {Yamamoto}, {Yamasaki}, {Yamashita}, \& {Yoshida}}]{Litebird2018}
{Suzuki} A. {et~al.}, 2018, ArXiv:1801.06987

\bibitem[{{Tristram} {et~al}\mbox{.}(2005){Tristram},
  {Mac{\'{\i}}as-P{\'e}rez}, {Renault}, \& {Santos}}]{Tristram2005}
{Tristram} M., {Mac{\'{\i}}as-P{\'e}rez} J.~F., {Renault} C., {Santos} D.,
  2005, \mnras, 358, 833

\bibitem[{{Zeldovich} \& {Sunyaev}(1969)}]{Zeldovich1969}
{Zeldovich} Y.~B., {Sunyaev} R.~A., 1969, \apss, 4, 301

\end{thebibliography}

%%%%%%%%%%%%%%%%%%%%%%%%%%%%%%%%%%%%%%%%%%%%%%%%%%%%
%%%%%%%%%%%%%%%%%%%%%%%%%%%%%%%%%%%%%%%%%%%%%%%%%%%%
%%%%%%%%%%%%%%%%%%%%%%%%%%%%%%%%%%%%%%%%%%%%%%%%%%%%
%%%
%%% Appendix
%%%
%%%\newpage

\end{document}